\documentclass[10pt,aps,prc,floatfix,twocolumn,nofootinbib]{revtex4-1}

\usepackage[labelfont={small},subrefformat=parens,caption=false]{subfig}
\usepackage{graphicx,amsmath,amssymb,bm}
\usepackage{amsfonts}
\usepackage[utf8]{inputenc}

\usepackage{verbatim}
\usepackage{float}
\usepackage{cancel}
\usepackage{multirow}
\usepackage{array}
\usepackage{xparse}
\usepackage{makecell}

\usepackage{physics}
\usepackage{color}
\usepackage{dsfont}
\usepackage{tabularx}

\usepackage[pdfencoding=auto, pdfpagelabels]{hyperref}

\usepackage[overload]{textcase}

\definecolor{linkcolor}{rgb}{0,0,0.40} 
\hypersetup{%
    pdfsubject=Paper,
    pdfkeywords={nuclear physics} {Bayesian} {chiral EFT},
    unicode = true,
    breaklinks = true,
    colorlinks = true,
    linkcolor = linkcolor,
    citecolor = linkcolor,
    menucolor = linkcolor,
    urlcolor = linkcolor
}

\usepackage[linesnumbered,algoruled,lined,commentsnumbered]{algorithm2e}

\usepackage{cellspace}
\graphicspath{{./figures/}}

\usepackage{silence}
\usepackage[T1]{fontenc}

\newcommand{\phantomsublabel}[3]{%
\unitlength=1in%
\put(#1,#2){%
    \subfloat[]{%
        \label{#3}%
    }}%
}

\makeatletter
\newcommand\newsubcommand[3]{\newcommand#1{#2\sc@sub{#3}}}
\def\sc@sub#1{\def\sc@thesub{#1}\@ifnextchar_{\sc@mergesubs}{_{\sc@thesub}}}
\def\sc@mergesubs_#1{_{\sc@thesub#1}}

\newcommand\newsupcommand[3]{\newcommand#1{#2\sc@sup{#3}}}
\def\sc@sup#1{\def\sc@thesup{#1}\@ifnextchar^{\sc@mergesups}{^{\sc@thesup}}}
\def\sc@mergesups^#1{^{\sc@thesup#1}}
\makeatother

\DeclareMathAlphabet{\mathbcal}{OMS}{cmsy}{b}{n}

\newcommand{\ordervec}{\vec}

\newcommand{\inputvec}{\mathbf}

\newsubcommand{\ckvec}{\ordervec{c}}{k}

\newsubcommand{\bkvec}{\ordervec{b}}{k}

\newsubcommand{\ckvecset}{\ordervec{\inputvec{c}}}{k}

\newsubcommand{\ckvecapprox}{\mathbf{c}'}{k}
\newsubcommand{\ckvecapproxset}{\mathbf{C}'}{k}

\newsubcommand{\bkvecapprox}{\mathbf{b}'}{k}
\newsubcommand{\bkvecset}{\mathbf{B}}{k}
\newsubcommand{\bkvecapproxset}{\mathbf{B}'}{k}

\newcommand{\genobs}{y}

\newsubcommand{\genobsvec}{\ordervec{\genobs}}{k}
\newsubcommand{\genobsvecset}{\ordervec{\inputvec{\genobs}}}{k}

\newsubcommand{\akvec}{\mathbf{a}}{k}

\newsubcommand{\akvecapprox}{\mathbf{a}'}{k}
\newsubcommand{\akvecset}{\mathbf{A}}{k}
\newsubcommand{\akvecapproxset}{\mathbf{A}'}{k}

{}  

\def\diffd{\mathrm{d}}  

\DeclareDocumentCommand\differential{ o g d() }{ 
    \IfNoValueTF{#2}{
        \IfNoValueTF{#3}
            {\diffd\IfNoValueTF{#1}{}{^{#1}}}
            {\mathinner{\diffd\IfNoValueTF{#1}{}{^{#1}}\argopen(#3\argclose)}}
        }
        {\mathinner{\diffd\IfNoValueTF{#1}{}{^{#1}}#2} \IfNoValueTF{#3}{}{(#3)}}
    }

\newcommand{\pathd}{\mathcal{D}}  

\DeclareDocumentCommand\pathdifferential{ o g d() }{ 
    \IfNoValueTF{#2}{
        \IfNoValueTF{#3}
            {\pathd\IfNoValueTF{#1}{}{^{#1}}}
            {\mathinner{\pathd\IfNoValueTF{#1}{}{^{#1}}\argopen(#3\argclose)}}
        }
        {\mathinner{\pathd\IfNoValueTF{#1}{}{^{#1}}#2} \IfNoValueTF{#3}{}{(#3)}}
    }

\newcommand{\params}{\theta}
\newcommand{\datasetD}{\mathcal{D}}
\newcommand{\midp}{\,|\,}
\newcommand{\pzCondD}[2]{p\bigl(z_{#1}^{(#2)} \midp \datasetD \bigr)}
\newcommand{\pCondZ}[4]{%
  p\bigl(z_{#1}^{(#3)} \midp z_{#2}^{(#4)}\bigr)}

\begin{document}


\title{Criticality analysis of nuclear binding energy neural networks }

\author{S.~A. Sundberg}
\email{sundberg.24@osu.edu}
\affiliation{Department of Physics, The Ohio State University, Columbus, OH 43210, USA}

\author{R.~J. Furnstahl}
\email{furnstahl.1@osu.edu}
\affiliation{Department of Physics, The Ohio State University, Columbus, OH 43210, USA}

\date{\today}

\begin{abstract}

 Machine learning methods, in particular deep learning methods such as artificial neural networks (ANNs) with many layers, have become widespread and useful tools in nuclear physics.
However, these ANNs are typically treated as ``black boxes'', with their architecture (width, depth, and weight/bias initialization) and the training algorithm and parameters  chosen empirically by optimizing learning based on limited exploration.
We test a non-empirical approach to understanding and optimizing nuclear physics ANNs by adapting a criticality analysis based on renormalization group flows in terms of the hyperparameters for weight/bias initialization, training rates, and the ratio of depth to width. 
This treatment utilizes the statistical properties of neural network initialization to find a generating functional for network outputs at any layer, allowing for a path integral formulation of the ANN outputs as a Euclidean statistical field theory.  
We use a prototypical example to test the applicability of this approach: a simple ANN for nuclear binding energies.
We find that with training using a stochastic gradient descent optimizer, the predicted criticality behavior is realized, and optimal performance is found with critical tuning.
However, the use of an adaptive learning algorithm leads to somewhat superior results without concern for tuning and thus obscures the analysis.
Nevertheless, the criticality analysis offers
a way to look within the black box of ANNs,
which is a first step towards potential improvements in network performance beyond using adaptive optimizers.
\end{abstract}

\maketitle


\section{Introduction}
\label{sec:introduction}

Machine learning methods such as artificial neural networks (ANNs) offer nuclear physicists new ways to explore and understand nuclear systems, as well as improve existing solution methods~\cite{Boehnlein:2021eym,Carleo_2019,Deiana_2022}.
Examples of ANN applications include the extrapolation of no-core shell model (NCSM) observables from smaller model space calculations~\cite{Wolfgruber:2023ehw}, variational Monte Carlo with neural network quantum states~\cite{Gnech:2023prs}, and learning nuclear mass models from data~\cite{Zeng:2022azv}.
ANNs are commonly treated as black boxes that are empirically optimized. Given their growing prominence as a tool for nuclear physics research, it is desirable to have a framework that allows for a more structured analysis based on an understanding of how they work. 
Several authors have proposed a field theory approach to analyze and optimize neural networks~\cite{Roberts:2021lll,Roberts:2021fes,Halverson:2020trp,Halverson:2021aot,schoenholz2017correspondence,PhysRevResearch.3.023034,Bachtis:2021xoh,Erdmenger:2021sot,Erbin:2021kqf,Erbin:2022lls}, using a combination of methods from quantum field theory and Bayesian statistics.
Here we assess whether this approach works in practice for the test case of a feed-forward ANN trained to predict binding energies in the nuclear landscape.

\begin{figure}[htb!]
    \centering
    \includegraphics[width=0.98\linewidth]{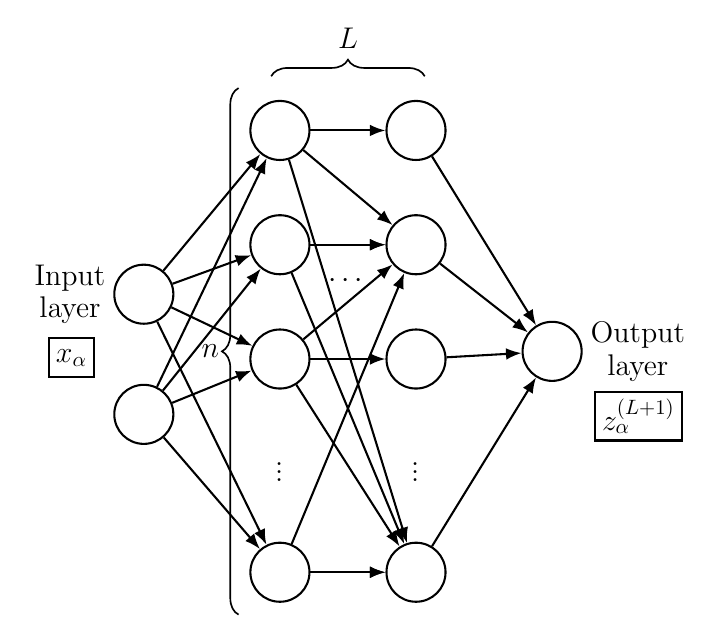}
    \caption{Schematic of a feed-forward neural network with two inputs $x_{\alpha}$ and one output $z_{\alpha}^{(L+1)}$, where $\alpha$ is an index running over the dataset $\mathcal{D}$. The width is $n$ and the depth is $\ell_{out} = L+1$, with $L$ being the number of hidden layers. The neurons are totally connected (some lines are omitted here for clarity).}
    \label{fig:schematic_ANN}
\end{figure}

\begin{figure*}[tbh!]
    \centering
    \includegraphics[width=0.95\textwidth]{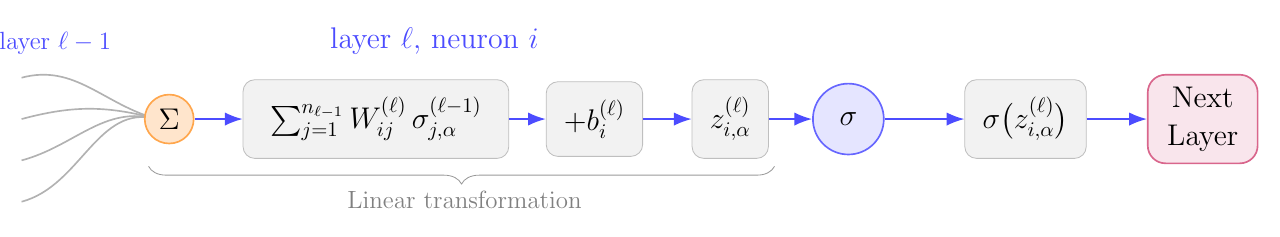}
    \caption{Schematic of a neuron within a neural network. In the layer $\ell$, the product between the weight matrix $W^{(\ell)}$ and the previous layers' output $\sigma^{(\ell-1)}$ is summed over the width of the prior layer $n_{\ell-1}$, and a bias vector $b^{(\ell)}$ is added, forming the preactivation $z^{(\ell)}$. The activation function $\sigma$ acts on the preactivation to yield the output $\sigma \bigl(z^{(\ell)}\bigr)$ (also notated $\sigma^{(\ell)}$) to be passed to the next layer.}
   \label{fig:schematic_Neuron}
\end{figure*}

An ANN uses connected neurons arranged in layers that, when trained, give a functional relationship between the input data vector $x$ and the known output data vector $y$, which are entries in a data set $\datasetD$~\cite{bishop2006pattern,AbuMostafa2012,Murphy:2012abc,mehta2019high}. 
We use the notation $x_\alpha$ and $y_\alpha$ to refer to a particular entry $\alpha$ in $\datasetD$, while $x_{i,\alpha}$ and
$y_{i,\alpha}$ are components of these vectors.
In Fig.~\ref{fig:schematic_ANN} we show a schematic of a feed-forward ANN with two inputs (so $i$ in $x_{i,\alpha}$ is 1 or 2) and one output (so $i$ in $y_{i,\alpha}$ is 1).
The outputs for a neural network are determined by the weights $W_{ij}$ and biases $b_i$, denoted collectively
by the parameters $\params$. 
On a more ``macroscopic level'', outputs are determined by the hyperparameters of the network such as the depth $\ell_{out}$, the width $n$, the activation function $\sigma$ (these three hyperparameters are defined here as the architecture hyperparameters) and the initialization hyperparameters that determine the pre-training distributions for the weights and biases. 
To simplify the analysis,
weights and biases will be initialized according to mean-zero Gaussians, so only the variances will be used as initialization hyperparameters. In addition, the width $n_{\ell}$ in a layer $\ell$ will be uniform across all hidden layers; the length of $x$ will be $n_0$ and the length of $y$ will be $n_{\ell_{out}}$($\ell_{out}\equiv L+1$).

In Fig.~\ref{fig:schematic_Neuron}, the process by which layer outputs are passed between layers is outlined. 
Going from the input layer to the first hidden layer ($\ell=1$) we define $\sigma^{(0)}_{i,\alpha} \equiv x_{i,\alpha}$. 
For all subsequent layers, a linear transformation maps the outputs from layer $\ell-1$ into the preactivations $z^{(\ell)}_{i,\alpha}$ in layer $\ell$ followed by a transformation by the activation function $\sigma$. 
For the final hidden layer, the output $z^{(\ell_{out})}_{\alpha} \equiv f(x_\alpha,\params)$ is produced with only a linear transformation. 

For a given initialization of the network parameters $\params$, the preactivations and the output function $f$ are uniquely determined.
However, repeated random initializations of $\params$ induces a probability distribution on preactivations and $f$.
Explicit expressions for these probability distributions in the limit of large width $n$ (from Refs.~\cite{Roberts:2021lll,Roberts:2021fes,Halverson:2020trp,Halverson:2021aot}) will be reviewed in Sec.~\ref{sec:ANNFTintro}, with the ratio of depth to width revealed as an expansion parameter for the distribution's action. 
The output distribution in a given layer $\ell$ refers to the distribution of preactivations $z^{(\ell)}_{j,\alpha}$. 
This choice allows for parallel treatment of all other layers with the distribution of functions $f$ in the output layer. 

To achieve the desired functional relationship between $x_{i,\alpha}$ and $y_{j,\alpha}$, a neural network is trained by adjusting the weights and biases in each layer according to the gradients of the loss function $\mathcal{L}\bigl(f(x,\params),y\bigr)$. 
The loss function quantifies the difference between the network output $f(x,\params)$ and the output data $y$, and the goal of training is to minimize $\mathcal{L}\bigl(f(x,\params),y\bigr)$. 
The gradients of $\mathcal{L}\bigl(f(x,\params),y\bigr)$ with respect to the network parameters $\params$ depend on the size of the output, as well as the architecture hyperparameters. 
It is known empirically that poor choices of architecture and/or initialization lead to exploding/vanishing gradients in the absence of adaptive learning algorithms, and this was problematic for training networks for many years~\cite{Bishop_Bishop_2024,Murphy:2012abc,Roberts:2021fes}.

In modern practice, standard adaptive learning algorithms, plus experience in which activation functions and initializations offer good network training performance, have led routinely to satisfactory results for ANNs using standard libraries.
This leaves experimentation with widths and depths of networks as the remaining knobs to adjust for many practitioners~\cite{math11112466,liu2021varianceadaptivelearningrate,Bishop_Bishop_2024}.
However, a more general theory of network behavior offers a way to determine what works and does not work ab initio, to see what network behaviors arise from which hyperparameter, and to identify what improvements can be made.
We follow the lead of Refs.~\cite{Roberts:2021lll,Roberts:2021fes,Halverson:2020trp,Halverson:2021aot}, who construct a theory of neural networks by considering the statistics of the network, calculating a generating functional distribution for moments of the network's output distribution, and analyzing this distribution in analogy to quantum field theory.
While this is highly promising, it is also largely theoretical and we seek to test whether it will be of practical use for problems in nuclear physics. 

The plan for this paper is as follows.
The properties and methods of artificial neural network field theory will be reviewed
in Sec.~\ref{sec:ANNFTintro}, then central claims on initialization and learning are validated in Sec.~\ref{sec:ANNFTvalidation} and Sec.~\ref{sec:binding_energy_results}, respectively.
As a test case, we examine two-input ANNs trained using AME2020 binding energy data.
Our conclusions and plans for future studies are summarized in Sec.~\ref{sec:conclusion}.

\section{Basics of ANNFT}
\label{sec:ANNFTintro}

\begin{figure*}[tbh!]
    \centering
    \includegraphics[width=0.7\linewidth]{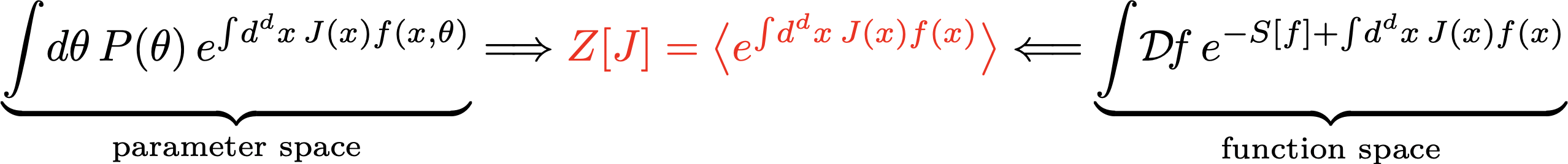}
    \caption{An ensemble of ANNs and a path integral in field theory are both distributions over random functions. 
    The partition function $Z[J]$ is a generating functional for the moments of functions $f$.
}
    \label{fig:function_parameter_space}
\end{figure*}

\subsection{Overview}

The goal of artificial neural network field theory (ANNFT) is to analyze networks using the tools of effective field theory and the renormalization group. 
This entails working with degrees of freedom based on the output probability distributions of the network in a controlled expansion in the ratio of the depth to width of the ANN.

The ANN output function $f(x,\params)$ for any fixed $\params$ (that is, specified weights and biases) is a deterministic function.
But, as already stressed, with repeated initializations of the ANN with $\params$ drawn from random distributions, the network acquires a probability distribution over functions $p\left(f(x)\midp\datasetD\right)$. 
Again, $\datasetD$ is used to represent the data set of inputs and outputs that the network is finding a relationship between.
In the pre-training case, this quantity simply refers to the inputs $x_\alpha$ for the network.
The outputs $y_\alpha$ will become relevant later during training.
As we illustrate in Secs.~\ref{subsec:first_hidden} and \ref{subsec:second_hidden}, the output probability distribution in any layer for any initialization can be calculated by marginalizing over the weights and biases of the network.
The network's dependence on the weights and biases is then transformed into a dependence on the preactivations $z^{(\ell)}$.

In this context, the weights and biases are independently and identically distributed (i.i.d.) random variables, which means the sum defining preactivations $z_{i,\alpha}^{(\ell)}$ in Fig.~\ref{fig:schematic_Neuron} as the hidden layer width $n$ is taken to be large satisfies the conditions for the central limit theorem (CLT) (assuming mean-zero distributions with appropriate normalization of the variances to keep the sum finite).
This implies that the preactivations in each layer and the function output will tend toward a (correlated) Gaussian distribution.
In short, the distribution over functions becomes a Gaussian process (GP) in the $n\rightarrow \infty$ limit~\cite{Neal1996,Williams1996ComputingWI,lee2018deep,matthews2018gaussian}.
Note that the activation functions do not disturb this conclusion because a nonlinear function of an i.i.d.\ random variable is still i.i.d.

At a large but finite width, the distribution over output functions acquires non-Gaussian components, but the near-CLT averaging enables these correlations to be treated in a controlled manner.
At the same time, increasing depth $\ell_{out}$ 
will be seen to introduce increasing correlations.
These opposing tendencies lead to the identification of the ratio $r \equiv \ell_{out}/n$ being significant as both $n$ and $\ell_{out}$ become large. 
$r$ acts as an expansion parameter that suppresses higher-order, non-Gaussian correlations in the limit of small $r$.
The nature of width and depth, their ratio $r$ as an expansion parameter, and perturbative non-Gaussianity of distributions will be explored Secs.~\ref{subsec:first_hidden},~\ref{subsec:second_hidden},~\ref{subsec:recursion}, and~\ref{sec:ANNFTvalidation}.

A complete description of the probability distributions in ANNFT can be formulated in terms of their moments, which characterize the shape of a distribution.
The $n^{\text{th}}$-order moment of a distribution of the (untrained) function $f$, $\langle f(x_1) \cdots f(x_n) \rangle$, is the expectation value of this product of functions at specified inputs $x_i$.
Examples of moments include the mean, variance, skewness, and kurtosis of a univariate probability distribution (as well as their multivariate distribution counterparts), which are, respectively, the first, second, third, and fourth moments of the distribution.
The partition function $Z[J]$ defined in the center of Fig.~\ref{fig:function_parameter_space} is the generating function for the moments, for example,
\begin{equation}
    \langle f(x_1) f(x_2) \rangle
    = \left .\frac{\delta}{\delta J(x_1)}\frac{\delta}{\delta J(x_2)}
    Z[J] \right|_{J=0} ,
\end{equation}
so $Z[J]$ is a fundamental object to calculate.
In the limit of infinite $n$ (when the distribution becomes Gaussian) the only independent moments are the mean (which will be zero before training if the initialization distributions are mean zero, along with all other odd-ordered moments) and the covariance. 
In particular, moments of order greater than two are dominantly given by factorized products of the covariance, whereas non-Gaussian contributions will be suppressed by powers of $1/n$. 
We will take advantage of this simplification and expand about this limit.

Because the distribution of $f$ is determined by the distributions of the weights and biases in $\params$, one way to calculate the partition function is to evaluate the integral over this parameter space as on the left of Fig.~\ref{fig:function_parameter_space}, where $P(\params)$ is the joint probability of all the weights and biases.
We will review how to formulate and evaluate this integral by considering successive layers of the ANN in Secs.~\ref{subsec:first_hidden}, \ref{subsec:second_hidden}, and \ref{subsec:recursion}.
The formulation of $Z[J]$ in these parameter-space degrees of freedom (dofs) would seem at first to suggest that the complexity of the network should become very high in the large width and depth limit.
However, one finds that the smoothing effect from the approach to the CLT implies that this is not the correct characterization of complexity for the ANN~\cite{Roberts:2021fes}.

An alternative is to express the statistical properties of ensembles of neural networks in a path-integral-like formulation, where the integration is over the distribution of random functions (see the right equation in Fig.~\ref{fig:function_parameter_space}).
If applied to a given layer, this generating functional can be used to calculate any correlation function for that layer; if applied to the output layer, this yields correlation functions in terms of the network outputs.
By choosing the initialization distributions to be mean-zero Gaussians, the probability distribution for the preactivations at any layer $\ell$ (including the output) takes the form of an exponentiated action: 
\begin{align}
    \pzCondD{}{\ell} \propto \exp(-S(z^{(\ell)})) ,
\end{align}
where we have omitted the normalization constant.
Derivations of the moments of the output function (and more generally the preactivations for each layer) can be simplified with Wick’s theorem and diagrammatics. 
Observables in a given layer can be calculated just like in quantum field theory with perturbation theory. 
This depends on having a small ratio of depth to width $r$, and using that as an expansion parameter for the network (see Chapters 0 and $\epsilon$ in Ref~\cite{Roberts:2021fes} for a broader overview).

In the $n\rightarrow\infty$ limit, the non-Gaussian contributions are suppressed, and we get a Gaussian free field theory (i.e., a quadratic action).
By backing off this limit, the output distribution in a given layer becomes nearly Gaussian. 
This introduces connected (non-Gaussian) contributions from moments higher than second order, which can be treated perturbatively. 
The diagrams mentioned before can now include vertices representing the connected contributions, and observables for any layer can be calculated.
All correlations of interest are included in an action $S(z)$, each with data-dependent couplings.
These couplings now provide a measure of complexity for the network.

The couplings will be shown in Sec.~\ref{subsec:recursion} to evolve under recursion relations with depth, which act like a renormalization group flow.
Increasing depth introduces correlations via this RG flow, while increasing width reduces them through CLT averaging.
As such, the terms in the action will use the ratio of depth to width $r = \ell_{out}/n$ as an expansion parameter, controlling which couplings contribute meaningfully to the output distribution of a network.
By analyzing these distributions, values of the initialization widths $C_W$ and $C_b$ (defined in Sec.~\ref{subsec:first_hidden}) can be found that make the variance constant under the RG flow with depth and lead to higher-order interactions with increasing powers of $r$. 
These values of the initialization hyperparameters are known as critical values, and by tuning the initialization to criticality, the network output magnitude will not explode or vanish with depth.
This in turn prevents the gradients of the network from exploding or vanishing, providing stability in training through the criticality analysis of the networks with ANNFT.

So far we have only considered the initialization of the neural network, with no training with respect to the $y$'s in $\datasetD$.
In carrying out such training, the original parameters are adjusted to a new set of parameters $\params^*$ that minimize the loss function $\mathcal{L}\bigl(f(x,\params),y\bigr)$ for each initialization of $\params$.
This implies a distribution $p\left(f(x,\params^*)\midp\datasetD\right)$ for trained functions $f(x,\params^*)$, which provides an ensemble 
of solutions for a given problem and enables uncertainty quantification.

The strategy will be to Taylor expand about initialization with large (but finite) width. 
If the trained parameters $\params^*$ are close to the initialized values $\params$, then we can truncate
\begin{align}
        f(x; \params^\star) 
        = f(x; \params) 
        + (\params^\star - \params) \frac{df}{d\params} 
        + \frac{1}{2} (\params^\star - \params)^2 \frac{d^2 f}{d\params^2} 
        + \cdots \label{eq:ANN_taylor_expansion}
    \end{align}
to good approximation.
This gives a way to understand solutions in terms of the initialization, but if the Taylor series has a large number of relevant terms dependent on derivatives of $f(x,\theta)$ to high order, the high dimensionality of the parameters makes such an analysis intractable. 
However, Ref.~\cite{Roberts:2021fes} shows that the Taylor series only needs up to the third derivative to describe network outputs with large $n$.
The higher derivatives are suppressed by powers of $r = \ell_{out}/n$ in this limit. 
With these higher-order terms suppressed, the critical initialization helps to ensure that the covariance of the output in any layer does not exponentially grow or vanish with depth, and therefore the magnitudes of the loss function gradients used to update $\params$ are not too large or small, as they depend on the magnitude of $z^{(l)}$. 
By regulating the magnitude of the gradients and, therefore, the difference between the initial and trained parameters, critical initialization partially ensures the validity of the Taylor series. 

To fully ensure the validity, critical training must be implemented to provide additional regulation of the gradients and the size of the trained parameters. 
The dependence of the Taylor series on the derivatives, as well as the information about the trained parameters, can be folded into what is called the neural tangent kernel (NTK) of a given network~\cite{Jacot2018NeuralTK,hanin2019finite,Roberts:2021fes}. 
The NTK is a matrix that governs how the outputs in each layer evolve during training, and will be discussed in more detail in Sec.~\ref{subsec:training}.
Like the couplings of the output distribution, the NTK also evolves under an RG flow with depth.
Training is tuned to criticality by scaling the learning rates of the network so that the NTK evolves in a controllable, non-exponential fashion like the moments of the distribution.
By having the trained output expressed in terms of its dependence on the statistical moments of the initialized output distribution and the NTK,
the Taylor series is cast into a form dependent on $z^{(L+1)}(T)$ (the final layer output after $T$ training epochs) and the NTK $H^{(\ell)}$ rather than $f(x,\params)$ and its many derivatives.
The Taylor series can then be truncated by the dependence of the stochastic variables $z^{(\ell)}$ and $H^{(\ell)}$ on the width $n$ and depth $\ell_{out}$. 

Significantly, the output distribution after training also becomes a Gaussian distribution in the infinite-width limit, and nearly Gaussian with connected correlations that depend on $r$ in the large-but-finite limit.
It can be shown in the limit of small $r$ that the Taylor series can be truncated to a few terms due to this dependence, allowing for predictions of trained network behavior through tractable perturbative calculations.

In the following sections, we fill in selected details of this overview.
For much more extensive treatments, the reader is directed to Refs.~\cite{Roberts:2021fes,Halverson:2020trp,Halverson:2021aot}.

\subsection{First hidden layer} \label{subsec:first_hidden}
    
To derive the statistical field theory for network outputs, we first consider the statistics of the network parameters. 
A network's weights $W_{ij}$ and biases $b_i$ are initialized from probability distributions. 
For the sake of simpler calculations, these are chosen to be mean-zero Gaussian distributions for the weights and biases:%
\footnote{The initialization distributions need not be mean zero, or even Gaussian at all. This choice makes calculations easier to perform, and enhances the analogy to field theories in physics later.}
\begin{align}
  p(W^{(\ell)}_{ij} \midp I) &= \sqrt{\frac{n}{2\pi C_W}}\exp({-\frac{n(W^{(\ell)}_{ij})^2}{2C_W}}) , 
  \label{eq:prob_W}\\
  p(b^{(\ell)}_i \midp I) &= \sqrt{\frac{1}{2\pi C_b}}\exp({-\frac{(b^{(\ell)}_i)^2}{2C_b}})  ,
  \label{eq:prob_b}
\end{align} 
where $n$ is the common width of the neural network hidden layers (previously the width $n_{\ell}$ was used to describe the width in a layer $\ell$), and the quantities $C_W$ and $C_b$ are the variances of the distributions. 
To keep the formulas for parameter distributions compact, we use ``I'' (for information) to represent these contingent quantities in probability density functions (pdfs). 
Note that while the true variance of the weight distribution is ${C_W}/{n}$,  the unscaled variance $C_W$ and the bias variance $C_b$ will be the more relevant quantities for calculation and understanding network behavior. 
The factor of $1/n$ serves to cancel the $n$ terms summed in the previous layer, so that we will have a useful large $n$ limit (as can be seen from the expressions for the preactivations in Fig.~\ref{fig:schematic_Neuron} or Eq.~\eqref{eq:ANNOutput}). 
Correspondingly, the variance for the weights between the input layer and the first hidden layer is scaled by a factor $1/n_0$ rather than $1/n$.

For a neural network of a given architecture, the probability distribution of preactivations $z^{(\ell)}_{i,\alpha}$ at a given layer $\ell$ from the initialization distributions \eqref{eq:prob_W} and \eqref{eq:prob_b} can be calculated directly. 
The preactivations $z^{(\ell)}$ at layer $\ell$ are (see Fig.~\ref{fig:schematic_Neuron})
\begin{equation}
z_{i,\alpha}^{(\ell)} 
= \sum_{j=1}^{n_{\ell-1}} W_{ij}^{(\ell)} \, \sigma_{j,\alpha}^{(\ell-1)} 
+ b_i^{(\ell)}   , 
\label{eq:ANNOutput}  
\end{equation}
where
\begin{equation}
 \sigma^{(\ell)}_{i,\alpha} 
= \sigma(z^{(\ell)}_{i,\alpha}),
\end{equation}
and
\begin{equation}
    \sigma_{i,\alpha}^{(0)} = x_{i,\alpha}  .
\end{equation}
In the first hidden layer ($\ell=1$), the joint probability distribution $\pzCondD{}{1}$ can be calculated by integrating over the model parameters from $-\infty$ to $\infty$,%
\footnote{Equation~\eqref{eq:DistLayer1Int} follows because the weights and biases are independent and $p(z^{(\ell)}_{i,\alpha} \midp b^{(\ell)}_i, W^{(\ell)}_{i,j})$ is a delta function for any $\ell$. }
\begin{align}
 \pzCondD{}{1} &= 
\int^{\infty}_{-\infty} \biggl[ \prod_{i} db^{(1)}_{i} \, p(b^{(1)}_{i}) \biggr] \biggl[ \prod_{i,j} dW^{(1)}_{ij} \, p(W^{(1)}_{ij}) \biggr] 
  \notag \\
  & \qquad \null \times
    \prod_{i,\alpha} \delta \Bigl( z^{(1)}_{i;\alpha} - b^{(1)}_{i} - \sum_{j} W^{(1)}_{ij} x_{j;\alpha} \Bigr) .
    \label{eq:DistLayer1Int}
\end{align}
A strategy for evaluating this integral is to use the integral representation of the delta function%
\footnote{The introduction of $\Lambda$ is the analog of the Hubbard-Stratonovich method used in quantum field theory~\cite{Roberts:2021fes}.}
\begin{align}
    \delta(z - a) = \int^\infty_{-\infty} \frac{d\Lambda}{2\pi} \, e^{i\Lambda(z-a)}.
\end{align} 
Applying this to all indices $\delta \bigl( z_{i;\alpha} - b_{i} - \sum_{j} W_{ij} x_{j;\alpha}\bigr)$ 
results in Eq.~\eqref{eq:DistLayer1Int} becoming 
\begin{widetext}
\begin{align}
 \pzCondD{}{1} &= \int^{\infty}_{-\infty} \biggl[\prod_i \frac{db^{(1)}_i}{\sqrt{2\pi C_b}}\biggr] \biggl[\prod_{i,j} \frac{dW^{(1)}_{ij}}{\sqrt{2\pi C_W/n_0}}\biggr] \biggl[\prod_{i,\alpha} \frac{d\Lambda_i^\alpha}{2\pi}\biggr] 
  \nonumber \\
 &\qquad\qquad \times 
\exp\biggl[-\sum_i \frac{\bigl(b^{(1)}_i\bigr)^2}{2C_b} - n_0\sum_{i,j} \frac{\bigl(W^{(1)}_{ij}\bigr)^2}{2C_W} + i\sum_{i,\alpha} \Lambda_i^\alpha \Bigl(z_{i;\alpha} - b^{(1)}_i - \sum_j W^{(1)}_{ij}x_{j;\alpha}\Bigr)\biggr].
\end{align}
\end{widetext}
Evaluating the integrals by completing the square for $W_{ij}$ and $b_i$ yields the distribution 
\begin{align}
    \pzCondD{}{1} = \frac{1}{|2\pi G^{(1)}|^{\frac{n_1}{2}}} 
    e^{-\frac{1}{2}
    \sum\limits_{i=1}^{n_1} 
    \sum\limits_{\alpha_1,\alpha_2 \in \mathcal{D}} 
    z_{i;\alpha_1}^{(1)} G_{(1)}^{\alpha_1\alpha_2} 
    z_{i;\alpha_2}^{(1)}} ,
\end{align}
where $G^{(1)}_{\alpha_1,\alpha_2}$ is the covariance matrix for the distribution, defined by 
\begin{align}
    G_{\alpha_1\alpha_2}^{(1)} = C_b + C_W \sum_j \frac{x_{j;\alpha_1} x_{j;\alpha_2}}{n_0}  ,
\end{align}
and the notation $G^{\alpha_1\alpha_2}_{(1)}$ with raised indices denotes the inverse of the covariance matrix. 

\subsection{Second hidden layer}\label{subsec:second_hidden}

In the first layer, all statistical correlations between neurons are governed entirely by the covariance matrix, making it a multivariate Gaussian distribution. 
Distributions in further layers develop non-Gaussianity for finite $n$. 
To see this explicitly, the second layer output distribution $p\bigl(z^{(2)}\midp\datasetD\bigr)$ can be calculated, and the method of calculation can be extrapolated for all deeper layers. 
By application of the product rule, and then the sum rule for probability distributions,  $z^{(1)}$ can be marginalized over.
From this, $p\bigl(z^{(2)}\midp\datasetD\bigr)$ can be expressed as 
\begin{align}
    \pzCondD{}{2} 
    = \int^{\infty}_{-\infty} \biggl[\prod_{j,\alpha} dz_{j;\alpha}^{(1)}\biggr] 
    \pCondZ{}{}{2}{1}
    \pzCondD{}{1},\label{eq:2nd_layer_distribution}
\end{align}
with the conditional distribution 
\begin{align}
    \pCondZ{}{}{2}{1} 
    &= \int^{\infty}_{-\infty} \biggl[\prod_i db_i^{(2)} p\bigl(b_i^{(2)}\bigr)\biggr]\! \biggl[\prod_{i,j} dW_{ij}^{(2)} p\bigl(W_{ij}^{(2)}\bigr)\biggr] 
    \notag \\
    & \quad \null \times 
    \prod_{i,\alpha} \delta\Bigl(
    z_{i;\alpha}^{(2)} 
    - b_i^{(2)} 
    - \sum_j W_{ij}^{(2)} 
    \sigma_{j;\alpha}^{(1)}\Bigr). \label{eq:conditional_dist}
\end{align}
The conditional distribution can be evaluated the same way as Eq.~\eqref{eq:DistLayer1Int}, resulting in the equation 
\begin{align}
    \pCondZ{}{}{2}{1} = \frac{1}{|2\pi \widehat{G}^{(2)}|^{\frac{n_2}{2}}} e^{-\frac{1}{2}\sum\limits_{i=1}^{n_2} \sum\limits_{\alpha_1,\alpha_2 \in \mathcal{D}} z_{i;\alpha_1}^{(2)} \widehat{G}_{(2)}^{\alpha_1\alpha_2}z_{i;\alpha_2}^{(2)}},
\end{align} 
with the covariance matrix 
\begin{align}
    \widehat{G}_{\alpha_1\alpha_2}^{(2)} \equiv C_b^{(2)} + C_W^{(2)} \frac{1}{n_1} \sum_{j=1}^{n_1} \sigma_{j;\alpha_1}^{(1)} \sigma_{j;\alpha_2}^{(1)}.
\end{align}

Significantly, the covariance matrix in the second layer conditional distribution and all deeper layer conditional distributions is stochastic, as the activation function $\sigma^{(1)}\equiv\sigma^{(1)}\bigl(z^{(1)}\bigr)$ is a function of the random variable $z^{(1)}$ from the previous layer.
The use of a hat on a quantity, such as $\widehat{G}_{\alpha_1\alpha_2}^{(2)}$, indicates that it is a function of random variables.
The mean of the conditional distribution's covariance $\widehat{G}_{\alpha_1\alpha_2}^{(2)}$ is
\begin{align}
    G_{\alpha_1\alpha_2}^{(2)} \equiv C_b^{(2)} + C_W^{(2)} \langle \sigma_{\alpha_1} \sigma_{\alpha_2} \rangle_{G^{(1)}}.
    \label{eq:mean_covariance}
\end{align} 
$\langle g(z) \rangle_{G}$ indicates a Gaussian expectation value of the function $g(z)$ for a mean zero Gaussian distribution of $z$ with a covariance matrix $G$. As such, $G_{\alpha_1\alpha_2}^{(2)}$ in Eq.~\eqref{eq:mean_covariance} does not have a hat as it is the mean of the stochastic covariance $\widehat{G}_{\alpha_1\alpha_2}^{(2)}$. 
Due to the stochastic nature of the covariance, there will be fluctuations in value around $G_{\alpha_1\alpha_2}^{(2)}$. These fluctuations can be expressed as
\begin{align}
    \Delta\widehat{G}_{\alpha_1\alpha_2}^{(2)} &\equiv \widehat{G}_{\alpha_1\alpha_2}^{(2)} 
    - G_{\alpha_1\alpha_2}^{(2)} 
    \notag \\
    &= C_W^{(2)} \frac{1}{n_1} 
    \sum_{j=1}^{n_1} \left(
    \sigma_{j;\alpha_1}^{(1)} \sigma_{j;\alpha_2}^{(1)} 
    - \langle\sigma_{\alpha_1} \sigma_{\alpha_2}\rangle_{G^{(1)}}\right). \label{eq:layer2fluctuations}
\end{align}
Calculating the mean fluctuations of the covariance $\ev{\Delta\widehat{G}_{\alpha_1\alpha_2}^{(2)}}_{G^{(1)}}$ results in the first term in Eq.~\eqref{eq:layer2fluctuations} to be equal to the second, 
so that the mean fluctuations of the conditional distribution's covariance are zero.
However, the variance of these fluctuations is non-zero,
\begin{align}
    \langle\Delta\widehat{G}_{\alpha_1\alpha_2}^{(2)}\Delta\widehat{ G}_{\alpha_3\alpha_4}^{(2)}\rangle_{G^{(1)}} 
    &= \frac{1}{n_1} 
    \Bigl(C_W^{(2)}\Bigr)^2 
    \bigl[\langle\sigma_{\alpha_1} \sigma_{\alpha_2} \sigma_{\alpha_3} \sigma_{\alpha_4}\rangle_{G^{(1)}} 
    \notag \\
    & \qquad \null
    - \langle\sigma_{\alpha_1} \sigma_{\alpha_2}\rangle_{G^{(1)}} \langle\sigma_{\alpha_3} \sigma_{\alpha_4}\rangle_{G^{(1)}}\bigr]
    \notag \\
    &\equiv \frac{1}{n_1} V_{(\alpha_1\alpha_2)(\alpha_3\alpha_4)}^{(2)}.
\end{align}
The quantity $V_{(\alpha_1\alpha_2)(\alpha_3\alpha_4)}^{(2)}$ is the non-Gaussian, or connected contribution to the fourth-order moment, and in analogy with field theory it serves as the four-point vertex for neuron interactions. 
As can be seen, the fluctuations in the covariance matrix of a Gaussian distribution depend on the width of the prior layer $n_1$. 
In the large-width limit, $\frac{1}{n_1}V_{(\alpha_1\alpha_2)(\alpha_3\alpha_4)}^{(2)}$ is suppressed. 
This serves as a clear example of simplification in the limit of large width, and allows for the vertex $V_{(\alpha_1\alpha_2)(\alpha_3\alpha_4)}^{(2)}$ to be treated perturbatively. 

At infinite width,  the fluctuations of $p\left(z^{(2)}\midp z^{(1)}\right)$'s covariance matrix vanish, and it becomes a deterministic Gaussian distribution. 
With the simplifications from large $n$, the full second-layer distribution $p\left(z^{(2)}\midp\mathcal{D}\right)$ can be calculated. 
The $\ell=2$ stochastic covariance matrix $\widehat{G}_{\alpha_1\alpha_2}^{(2)} = G^{(2)}_{\alpha_1\alpha_2} + \Delta\widehat{G}_{\alpha_1\alpha_2}^{(2)}$ can be inverted to second order in fluctuations about the mean covariance matrix: 
\begin{align}
    \widehat{G}_{(2)}^{\alpha_1\alpha_2} &= G_{(2)}^{\alpha_1\alpha_2} - \sum_{\beta_1,\beta_2 \in \mathcal{D}} G_{(2)}^{\alpha_1\beta_1} \Delta\widehat{G}_{\beta_1\beta_2}^{(2)} G_{(2)}^{\beta_2\alpha_2}
    \notag \\
   & \quad \null + \sum_{\beta_1,...,\beta_4 \in \mathcal{D}} G_{(2)}^{\alpha_1\beta_1} \Delta\widehat{G}_{\beta_1\beta_2}^{(2)} G_{(2)}^{\beta_2\beta_3} \Delta\widehat{G}_{\beta_3\beta_4}^{(2)} G_{(2)}^{\beta_4\alpha_2}
    \notag \\
    & \qquad\qquad \null + O\left(\Delta^3\right).
\end{align} 
This can then be plugged into the argument for the conditional distribution in Eq.~\eqref{eq:conditional_dist}, and after Taylor expanding in the fluctuations $\Delta\widehat{G}_{\alpha_1\alpha_2}^{(2)}$, the integral in Eq.~\eqref{eq:2nd_layer_distribution} evaluates to
\begin{widetext}
\begin{align}
  \pzCondD{}{2} &= \frac{1}{\sqrt{|2\pi G^{(2)}|^{n_2}}} \exp\biggl(-\frac{1}{2}\sum_{j=1}^{n_2} \sum_{\alpha_1,\alpha_2\in\mathcal{D}} G_{(2)}^{\alpha_1\alpha_2} z_{j;\alpha_1}^{(2)} z_{j;\alpha_2}^{(2)}\biggr) 
\biggl\{ \biggl[1 + O\biggl(\frac{1}{n_1}\biggr)\biggr] + \sum_{i=1}^{n_2} \sum_{\alpha_1,\alpha_2\in\mathcal{D}} \biggl[O\biggl(\frac{1}{n_1}\biggr)\biggr] z_{i1;\alpha_1}^{(2)} z_{i1;\alpha_2}^{(2)} 
  \notag\\
& \qquad\null + \frac{1}{8n_1} \sum_{i_1,i_2=1}^{n_2} \sum_{\alpha_1,...,\alpha_4\in\mathcal{D}} V_{(2)}^{(\alpha_1\alpha_2)(\alpha_3\alpha_4)} z_{i1;\alpha_1}^{(2)} z_{i1;\alpha_2}^{(2)} z_{i2;\alpha_3}^{(2)} z_{i2;\alpha_4}^{(2)} \biggr\} + O\biggr(\frac{1}{n_1^2}\biggl).
\end{align} 
A Gaussian is then multiplied by terms organized in powers of $1/n_1$. The contributions to zeroth and first order are contained within curved brackets, and the $O(1/n_1)$ in square brackets within the second term indicates subleading corrections to $G^{(2)}_{\alpha_1\alpha_2}$.
By taking the logarithm of the right hand side, the action $S\left(z^{(\ell)}\right)$ is isolated, and constant terms can be absorbed into the partition function, and the action can be expressed as 
\begin{align}
  S(z^{(2)}) &= \frac{1}{2} \sum_{\alpha_1,\alpha_2\in\mathcal{D}} \left[G_{(2)}^{\alpha_1\alpha_2} + O\left(\frac{1}{n_1}\right)\right]\sum_{i=1}^{n_2} z^{(2)}_{i;\alpha_1} z^{(2)}_{i;\alpha_2} 
  \nonumber \\
  & \quad \null 
  - \frac{1}{8} \sum_{\alpha_1,...,\alpha_4\in\mathcal{D}} \frac{1}{n_1}V_{(2)}^{(\alpha_1\alpha_2)(\alpha_3\alpha_4)} 
  \sum_{i_1,i_2=1}^{n_2} z^{(2)}_{i1;\alpha_1} z^{(2)}_{i1;\alpha_2} z^{(2)}_{i2;\alpha_3} z^{(2)}_{i2;\alpha_4}
   + O\left(\frac{1}{n_1^2}\right).
\end{align}
\end{widetext}
The higher-order connected contributions to the moments can also be calculated as terms in the action, and each term will have factors of 1/$n_1$ with increasing order.
In the large $n_1$ limit, only the lowest-order terms will contribute to the action, and in the infinite limit, the output distribution will become Gaussian, reflecting the behavior predicted by the central limit theorem. 

\subsection{Recursion relation} \label{subsec:recursion}

The method of derivation from the first hidden layer to the second can be applied to find an output distribution $p\bigl(z^{(\ell+1)}\midp\mathcal{D}\bigr)$ given the knowledge of $p\left(z^{(\ell)}\midp\mathcal{D}\right)$. 
The output distribution in layer $\ell+1$ can be calculated from $p\bigl(z^{(\ell)}\midp\mathcal{D}\bigr)$ through the integral 
\begin{align}
    \pzCondD{}{\ell+1} = \int^{\infty}_{-\infty} \biggl[\prod_{j,\alpha} dz_{j;\alpha}^{(\ell)}\biggr] 
    \pCondZ{}{}{\ell+1}{\ell} 
    \pzCondD{}{\ell}.
\end{align} 
The calculation proceeds identically to the one performed to obtain $\pzCondD{}{2}$. The conditional distribution $\pCondZ{}{}{\ell+1}{\ell}$ is defined by 
\begin{align}
  &  \pCondZ{}{}{\ell+1}{\ell} = \frac{1}{\sqrt{\bigl|2\pi\widehat{G}^{(\ell+1)}\Bigr|^{n_{\ell+1}}}}
    \notag \\
    & \quad\null\times
    \exp\left(-\frac{1}{2}\sum_{i=1}^{n_{\ell+1}} \sum_{\alpha_1,\alpha_2\in\mathcal{D}} z_{i;\alpha_1}^{(\ell+1)} \widehat{G}_{(\ell+1)}^{\alpha_1\alpha_2} z_{i;\alpha_2}^{(\ell+1)}\right),
\end{align}
where the stochastic metric in the $(\ell+1)$-th layer $\widehat{G}_{\alpha_1\alpha_2}^{(\ell+1)} $ is given by 
\begin{align}
    \widehat{G}_{\alpha_1\alpha_2}^{(\ell+1)} \equiv C_b^{(\ell+1)} + C_W^{(\ell+1)} \frac{1}{n_\ell} \sum_{j=1}^{n_\ell} \sigma_{j;\alpha_1}^{(\ell)} \sigma_{j;\alpha_2}^{(\ell)}.
\end{align} 

Again, the metric can be split into a mean and fluctuations, and then the various expansions and the integration can take place to produce distribution $\pzCondD{}{\ell+1}$ from $\pzCondD{}{\ell}$, as desired. 
This generalization of calculating the output distributions also provides a generalization of the action $S\bigl(z^{(\ell)}\bigr)$: 
\begin{align}
S\left(z^{(\ell)}\right) &\equiv \frac{1}{2}\sum_{i=1}^{n_\ell} \sum_{\alpha_1,\alpha_2\in\mathcal{D}} g_{(\ell)}^{\alpha_1\alpha_2}z_{i;\alpha_1}^{(\ell)} z_{i;\alpha_2}^{(\ell)} 
   \notag \\
  & \quad \null - \frac{1}{8} \sum_{i_1,i_2=1}^{n_\ell} \sum_{\alpha_1,...,\alpha_4\in\mathcal{D}} v_{(\ell)}^{(\alpha_1\alpha_2)(\alpha_3\alpha_4)}
  \notag \\
  & \qquad\quad\null\times 
  z_{i1;\alpha_1}^{(\ell)} z_{i1;\alpha_2}^{(\ell)} z_{i2;\alpha_3}^{(\ell)} z_{i2;\alpha_4}^{(\ell)} + \ldots,
\end{align}
where
\begin{align}
g_{(\ell)}^{\alpha_1\alpha_2} &= G_{(\ell)}^{\alpha_1\alpha_2} + O\left(\frac{1}{n_1},\ldots\right), \label{eq:2point}\\
v_{(\ell)}^{(\alpha_1\alpha_2)(\alpha_3\alpha_4)} &= \frac{1}{n_{\ell-1}}V_{(\ell)}^{(\alpha_1\alpha_2)(\alpha_3\alpha_4)} + O\left(\frac{1}{n_1^2},\ldots\right).\label{eq:4point}
\end{align}
Terms such as $g^{(\ell)}_{\alpha_1\alpha_2}$ and $v^{(\ell)}_{(\alpha_1\alpha_2)(\alpha_3\alpha_4)}$ can be thought of as data-dependent couplings in the action. $G^{(\ell)}_{\alpha_1\alpha_2}$ and $\frac{1}{n_{\ell-1}}V_{(\ell)}^{(\alpha_1\alpha_2)(\alpha_3\alpha_4)}$ are the leading-order contributions to the quadratic and quartic couplings. In the limit of large width, these will be the contributing couplings within the action, and the higher-order moments can be neglected. By setting the hidden layer widths $n_1,n_2,\dots,n_{L},n_{\ell_{out}}$ to a common value $n\gg 1$, the leading-order couplings in any layer can be calculated according to the recursion relations 
%
\begin{align}
    G_{\alpha_1\alpha_2}^{(\ell+1)} = C_b^{(\ell+1)} + C_W^{(\ell+1)} \langle\sigma_{\alpha_1}\sigma_{\alpha_2}\rangle_{G^{(\ell)}} + O\biggl(\frac{1}{n}\biggr) , \label{eq:2PointRecursion}  
 \end{align}   
    
\begin{align}
    \frac{1}{n}V_{(\alpha_1\alpha_2)(\alpha_3\alpha_4)}^{(\ell+1)} &= \frac{1}{n} \bigl(C_W^{(\ell+1)}\bigr)^2 [\langle\sigma_{\alpha_1}\sigma_{\alpha_2}\sigma_{\alpha_3}\sigma_{\alpha_4}\rangle_{G^{(\ell)}} 
        \notag \\ & \qquad\qquad\null 
    -\langle\sigma_{\alpha_1}\sigma_{\alpha_2}\rangle_{G^{(\ell)}} \langle\sigma_{\alpha_3}\sigma_{\alpha_4}\rangle_{G^{(\ell)}}]  
    \notag \\ & \null 
    + \frac{1}{n} \frac{\bigl(C_W^{(\ell+1)}\bigr)^2}{4} 
   \sum_{\beta_1,...,\beta_4\in\mathcal{D}} V_{(\ell)}^{(\beta_1\beta_2)(\beta_3\beta_4)} 
    \notag \\
    & \qquad \null\times
    \langle\sigma_{\alpha_1}\sigma_{\alpha_2} (z_{\beta_1} z_{\beta_2} - g_{\beta_1\beta_2})\rangle_{G^{(\ell)}} \notag \\
    &\qquad \null \times
    \langle\sigma_{\alpha_3}\sigma_{\alpha_4} (z_{\beta_3} z_{\beta_4} - g_{\beta_3\beta_4})\rangle_{G^{(\ell)}} 
    \notag \\ & \null 
    + O\biggl(\frac{1}{n^2}\biggr) .\label{eq:4PointRecursion}
\end{align}
%

In evaluating the Gaussian expectation values in these recursion relations, the couplings in the action will exponentially grow or vanish as the depth $\ell$ increases. 
To regulate this behavior, a neural network is subjected to a criticality analysis, where values of the initialization hyperparameters $C_W$ and $C_b$ are chosen to facilitate an action where the leading-order quadratic coupling $G^{(\ell)}_{\alpha_1\alpha_2}$ value is fixed with depth in the recursion relation.
The values of $C_W$ and $C_b$ may have layer dependence in general, but in satisfying the recursion relations in a network with uniform width $n$, they can be made layer-independent.
When this criticality condition is met, the leading-order contributions for higher-order moments gain a factor of the depth $\ell$ to a power equal to that of the order $1/n$ in the expansion. The expansion in $1/n$ then becomes an expansion in $r = \ell_{out}/n$. As an expansion parameter, $r$ governs which connected correlations contribute to the output distribution for the network. 

Depending on the application, ANNs perform better as they grow deeper or wider, and so practitioners will increase their network's depth, width or both to suit their problem~\cite{Fan_Lai_Wang_2022,Álvarez-López_Slimane_Zuazua_2024,Radhakrishnan_Belkin_Uhler_2023,2202.03841,2010.15327}. 
However, problems like exploding/vanishing gradients and overfitting/underfitting also become more prevalent as these parameters are increased. 

From Eqs.~\eqref{eq:2point} and~\eqref{eq:4point}, it can be seen that the width $n$ suppresses connected contributions to moments as $n\rightarrow\infty$. 
When the width is infinite, the distributions in each layer become multivariate Gaussian distributions. 
This is again a manifestation of the central limit theorem (CLT), as the sum over all elements in Eq.~\eqref{eq:ANNOutput} combined with the factor of ${1}/{n}$ from the distribution of weights results in the preactivations $z^{(\ell)}_{i,\alpha}$ being averaged over all elements in the prior layer. 
For most neural networks that can be tuned to criticality, the biases will be initially set to zero, and will not spoil the Gaussian behavior in the infinite-width limit where the CLT kicks in. 

While increasing width averages out correlations between neurons, the depth $\ell$ introduces new ones through the recursion relations in Eqs.~\eqref{eq:2PointRecursion} and~\eqref{eq:4PointRecursion}. 
In analogy to field theory, these moment recursion relations with depth can be thought of as a renormalization group flow for the different ``couplings'' in the action. 
Tuning $C_W$ and $C_b$ to criticality describes the values of these initialization hyperparameters that result in the variance being at a fixed point, meaning the coupling is constant with depth.

In order to accomplish the goal of a limited number of degrees of freedom that still describe a realistic ANN, a small ratio of depth to width is desirable. 
In this limit, the width averages out the higher-order connected contributions from the moments in the distribution, while the depth keeps the lowest-order connected contributions present. 
This allows the distributions to be ``nearly Gaussian'', possessing only a few contributing connected moments that can be easily understood and analyzed.

As an expansion parameter, $r$ quantifies the degree of correlation between neurons in a network. This results in three regimes describing the initialized output distribution: 
\begin{itemize}
    \item $r\rightarrow0$, all terms in the action dependent on $r$ vanish, and the output distribution becomes Gaussian (effectively infinite width, CLT kicks in). These networks have turned off their correlations.
    \item $0<r\ll1$, the moments are controllable, truncated, and nontrivial, as is desired in our effective theory approach. These networks are in what is known as the ``effectively deep'' regime.
    \item $r\geq1$, the moments are strongly coupled, and every term in the r expansion contributes to the action. In this regime, the theory becomes highly non-perturbative, and an effective description becomes impossible~\cite{Roberts:2021fes}.
\end{itemize}
With these considerations, ANNFT offers prescriptions for architecture and initialization hyperparameters. 
To have a network that can be treated as an effective field theory expansion in $r$, the value of $r$ must lie within the effectively deep regime, dictating that the width in the hidden layers must be much larger than the depth. 
In the case of practitioners increasing width or depth depending on the application, these hyperparameters should be increased together so that $r$ does not enter the non-perturbative or Gaussian regime. 
Furthermore, the activation function should be chosen such that it admits criticality, with the initialization hyperparameters tuned to critical values. 
These conditions partially stabilize the network output, and further stabilization comes from training considerations.

\subsection{Training}\label{subsec:training}

All discussion so far has been focused on the initialization of the network, but of course training must also be considered.

The neural tangent kernel (NTK) is the central object for the analysis of network training in ANNFT.
The NTK governs the evolution of network outputs during training for ANNs and can depend on the outputs $z_{i,\alpha}$ in not just the final layer, but all layers prior due to the chain rule derivatives present in loss updates during training.
The NTK is defined as
\begin{align}
    H^{(\ell)}_{i_1 i_2; \alpha_1 \alpha_2} 
    = \sum_{\mu, \nu} 
    \lambda_{\mu \nu} \frac{dz^{(\ell)}_{i_1; \alpha_1}}{d\params_\mu} \frac{dz^{(\ell)}_{i_2; \alpha_2}}{d\params_\nu}, \label{eq:NTK}
\end{align} 
where $\lambda_{\mu \nu}$ is the learning rate tensor~\cite{Roberts:2021fes}.
$\lambda_{\mu \nu}$ is chosen to be a diagonal matrix with separate learning rates for the weights and biases: 
\begin{align}
    \lambda_{b_{i_1}^{(\ell)} b_{i_2}^{(\ell)}} = \delta_{i_1 i_2} \lambda_b^{(\ell)}, \quad \lambda_{W_{i_1 j_1}^{(\ell)} W_{i_2 j_2}^{(\ell)}} = \delta_{i_1 i_2} \delta_{j_1 j_2} \frac{\lambda_W^{(\ell)}}{n_{\ell-1}}.
\end{align} 
This allows for an analysis of the training using separate hyperparameters for weights and biases in a manner similar to $C_W$ and $C_b$ in the analysis of the initialization. 
Applying this choice to Eq.~\eqref{eq:NTK}, the NTK becomes 
\begin{align}
    \widehat{H}_{i_1 i_2; \alpha_1 \alpha_2}^{(\ell)} &= \sum_{\ell'=1}^{\ell} \biggl[ \sum_{j=1}^{n_{\ell'}} \biggl( \lambda_b^{(\ell')} \frac{dz_{i_1; \alpha_1}^{(\ell')}}{db_j^{(\ell')}} \frac{dz_{i_2; \alpha_2}^{(\ell')}}{db_j^{(\ell')}} 
    \notag \\
    & \quad\qquad \null
    + \frac{\lambda_W^{(\ell')}}{n_{\ell'-1}} \sum_{k=1}^{n_{\ell'-1}} \frac{dz_{i_1; \alpha_1}^{(\ell')}}{dW_{jk}^{(\ell')}} \frac{dz_{i_2; \alpha_2}^{(\ell')}}{dW_{jk}^{(\ell')}} \biggr) \biggr]. \label{eq:NTK_diagonal_lr}
\end{align}
It should be noted that the NTK is a stochastic quantity for our network, and so a hat is added to $H^{(\ell)}_{i_1 i_2; \alpha_1 \alpha_2}$ to indicate this, like it was for $\widehat{G}^{(\ell)}_{\alpha_1\alpha_2}$. 
The terms in square brackets refer to the contribution to the NTK for each layer $\ell'=1$ to $\ell'=\ell$.

Exploiting the similarity between this quantity and the moments of the pre-training distribution, the NTK's dependence on width and depth is analyzed in the same way.
The infinite-width limit for ANNs yielding Gaussian processes results in them being subject to kernel methods.
The NTK's training dynamics become governed by a linear differential equation in the infinite-width limit~\cite{Jacot2018NeuralTK,Jacot_Gabriel_Hongler_2020,Roberts:2021fes,Prince_2024,Weng_2022}. 
As we back off the infinite-width limit, algorithm-dependent non-linear training dynamics become present, like connected contributions of higher-order moments of the pre-training distribution~\cite{Roberts:2021fes,Novak_Sohl-Dickstein_Schoenholz_2022,Jacot_Gabriel_Hongler_2020}.
As such, the finite-width kernel $H^{(\ell)}$ contains corrections to the linear infinite-width kernel $H_{\alpha_1\alpha_2}^{\{0\}(\ell)}$ suppressed by powers of $1/n$: \begin{align}
H_{\alpha_1\alpha_2}^{(\ell)} 
&= H_{\alpha_1\alpha_2}^{\{0\}(\ell)} 
+ \frac{1}{n_{\ell-1}} H_{\alpha_1\alpha_2}^{\{1\}(\ell)} 
\notag \\
& \quad \null
+ \frac{1}{n_{\ell-1}^2} H_{\alpha_1\alpha_2}^{\{2\}(\ell)} 
+ O\left(\frac{1}{n^3}\right).
\end{align}

The NTK also features depth dependence similar to the couplings of the pre-training distribution.
By comparing the expressions for $\widehat{H}_{i_1 i_2; \alpha_1 \alpha_2}^{(\ell)}$ and $\widehat{H}_{i_1 i_2; \alpha_1 \alpha_2}^{(\ell+1)}$ from Eq.~\eqref{eq:NTK_diagonal_lr}, the recursion relation for the NTK can be derived: 
\begin{align}
    \widehat{H}_{i_1 i_2; \alpha_1 \alpha_2}^{(\ell+1)} 
    = &\sum_{j=1}^{n_{\ell+1}} 
    \biggl( \lambda_b^{(\ell+1)} 
    \frac{dz_{i_1; \alpha_1}^{(\ell+1)}}{db_j^{(\ell+1)}} 
    \frac{dz_{i_2; \alpha_2}^{(\ell+1)}}{db_j^{(\ell+1)}} 
    \notag \\
    & \qquad \null
    + \frac{\lambda_W^{(\ell+1)}}{n_{\ell}} \sum_{k=1}^{n_{\ell}} 
    \frac{dz_{i_1; \alpha_1}^{(\ell+1)}}{dW_{jk}^{(\ell+1)}} 
    \frac{dz_{i_2; \alpha_2}^{(\ell+1)}}{dW_{jk}^{(\ell+1)}} \biggr) \notag \\
    &+ \sum_{j_1, j_2=1}^{n_{\ell}} 
    \frac{dz_{i_1; \alpha_1}^{(\ell+1)}}{dz_{j_1; \alpha_1}^{(\ell)}} 
    \frac{dz_{i_2; \alpha_2}^{(\ell+1)}}{dz_{j_2; \alpha_2}^{(\ell)}} \widehat{H}_{j_1 j_2; \alpha_1 \alpha_2}^{(\ell)}.\label{eq:NTK_recursion}
\end{align}

The NTK analysis then follows a route similar to that of $\widehat{G}^{(\ell)}_{\alpha_1\alpha_2}$. 
In the first hidden layer, the NTK $H^{(1)}_{i_1i_2\alpha_1\alpha_2}$ is deterministic like $G^{(1)}_{\alpha_1\alpha_2}$.
This stems from the fact that in the recursion relation Eq.~\eqref{eq:NTK_recursion}, $\widehat{H}^{(0)}_{j_1j_2;\alpha_1\alpha_2}=0$, as the inputs to the network are not being altered by training.
Calculating the derivatives in the recursion relation yields
\begin{align}
    H^{(1)}_{i_1i_2\alpha_1\alpha_2}
    = \delta_{i_1i_2}\biggl[\lambda_b^{(1)} 
    + \lambda^{(1)}_W\biggl(\frac{1}{n_0}\sum_{j=1}^{n_0}x_{j;\alpha_1}x_{j;\alpha_2}\biggr)\biggr],
\end{align}
which does not depend on any stochastic variables.

The NTK in all other layers will have stochastic properties that require it to be split into its mean and fluctuations $\widehat{H}^{(\ell)}_{i_1i_2;\alpha_1\alpha_2} = H^{(\ell)}_{i_1i_2;\alpha_1\alpha_2} + \Delta \widehat{H}^{(\ell)}_{i_1i_2;\alpha_1\alpha_2}$.
The recursion relations for the different parts of the NTK are calculated in full in chapters 8 and 9 of Ref.~\cite{Roberts:2021fes}, but all of the relations depend on architecture hyperparameters such as $C_W$ and $C_b$, and the choice of activation function in addition to the weight and bias learning rates $\lambda^{(\ell)}_W$ and $\lambda^{(\ell)}_b$.
Like the couplings of the pre-training distribution, the NTK will exhibit exponential behavior without specific treatment of the hyperparameters.
As the architecture hyperparameters are already tuned to criticality, the remaining tuneable hyperparameters are the learning rates. 

To avoid this exponential behavior that contributes to the exploding and vanishing gradients, the learning rates are scaled by factors of the width and depth that lead to corrections to the infinite-width NTK $H_{\alpha_1\alpha_2}^{\{0\}(\ell)}$ controlled by powers of $r$.
As an example, the critically scaled learning rates for the ReLU and Tanh activation functions are 
\begin{alignat}{2}
 \lambda_{b,\text{ReLU}}^{(\ell)} &= \frac{\Tilde{\lambda}_{b,\text{ReLU}}}{\ell_{out}} \,, \quad
 \lambda^{(\ell)}_{W,\text{ReLU}} 
 &= \frac{\Tilde{\lambda}_{W,\text{ReLU}}}{\ell_{out}}\,,\\
        \lambda_{b,\text{Tanh}}^{(\ell)} &= \frac{\Tilde{\lambda}_{b,\text{Tanh}}}{\ell} \,, \quad
 \lambda^{(\ell)}_{W,\text{Tanh}} 
 &= \Tilde{\lambda}_{W,\text{Tanh}}.\label{eq:critical_lr}
    \end{alignat} 
The learning rates for the weights used in practice also include a factor of $1/n$ (and $1/n_0$ in the first layer) to cancel sums over the width in the recursion relations similar to the scaling of $C_W$.
These scalings adjust the learning to provide stability to training, and alter the size of the ``raw'' rates $\Tilde{\lambda}_W$ and $\Tilde{\lambda}_b$ to allow for equivalence of learning rates across the different layers and between different network architectures.

This prescription is applied to networks using stochastic gradient descent learning algorithms, which feature a fixed learning rate.
In conjunction with critically tuned initialization, this is claimed to result in stable networks that avoid the exploding and vanishing gradient problem.
In addition, these networks and their behavior before and after training can be analyzed and understood using techniques from perturbation theory in QFT and the renormalization group, as promised by ANNFT.

In the next two sections, many of these claims will be validated using two-input (the neutron number $N$ and the proton number $Z$) ANNs trained by data from the AME2020 binding energy dataset.

\section{Validating ANNFT initialization distribution behavior}
\label{sec:ANNFTvalidation}

\begin{figure*}[ptbh!]
    \centering
    \includegraphics[width=0.82\columnwidth]{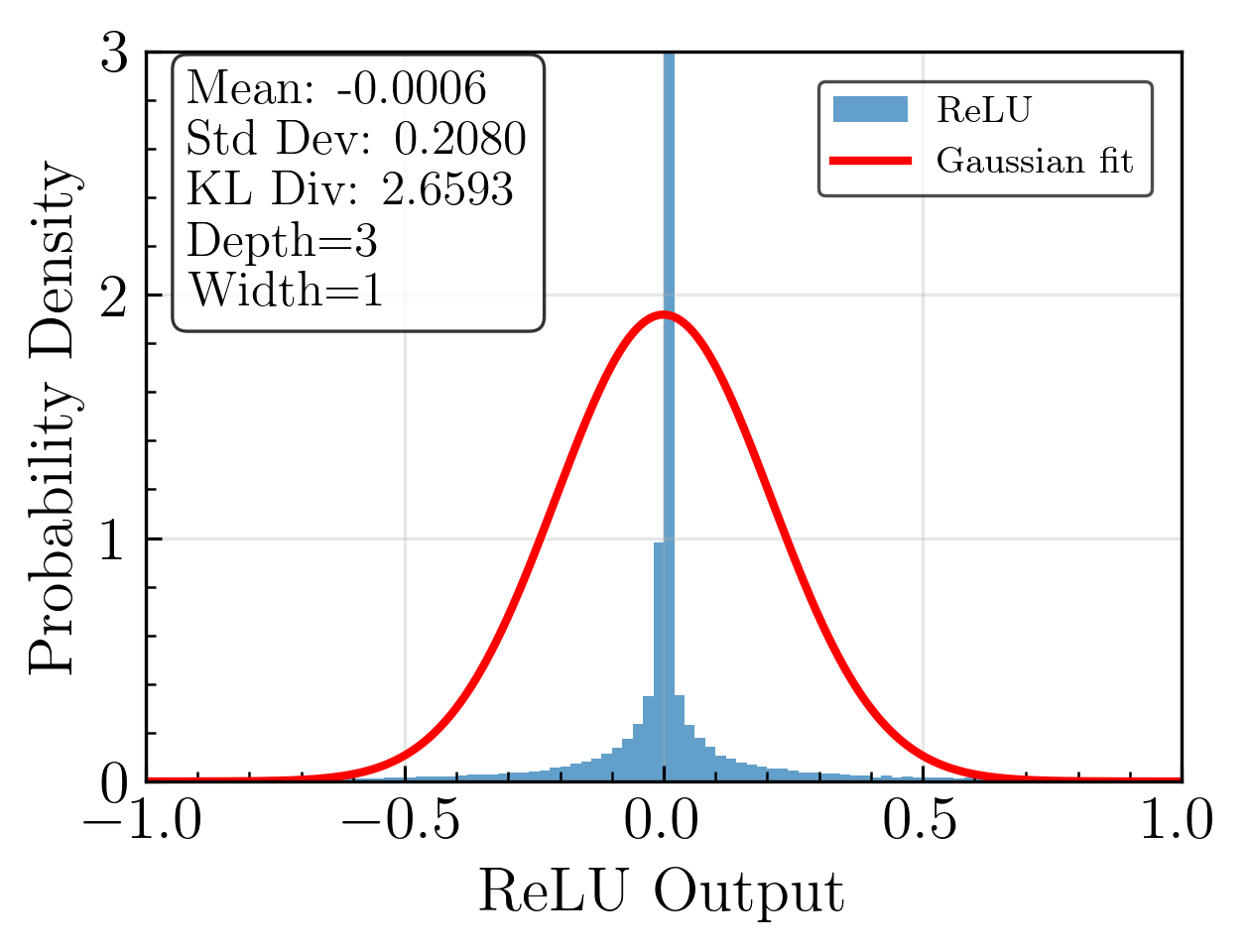}
        \phantomsublabel{-2.4}{0.4}{fig:dist_scaled_depth3_width1}
    \qquad
    \includegraphics[width=0.82\columnwidth]{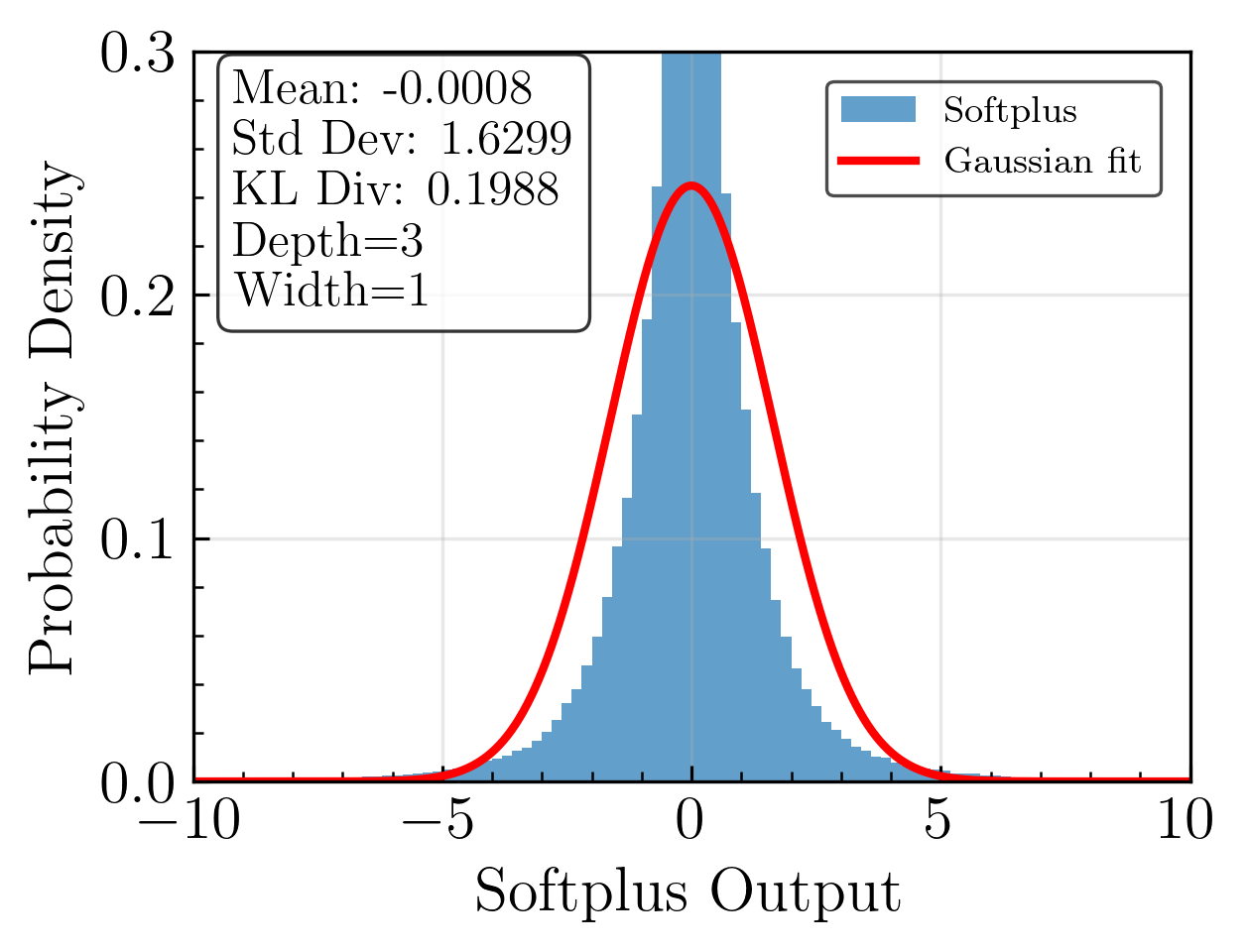}
        \phantomsublabel{-2.3}{0.4}{fig:dist_scaled_depth3_width1_softplus}
    
    \includegraphics[width=0.82\columnwidth]{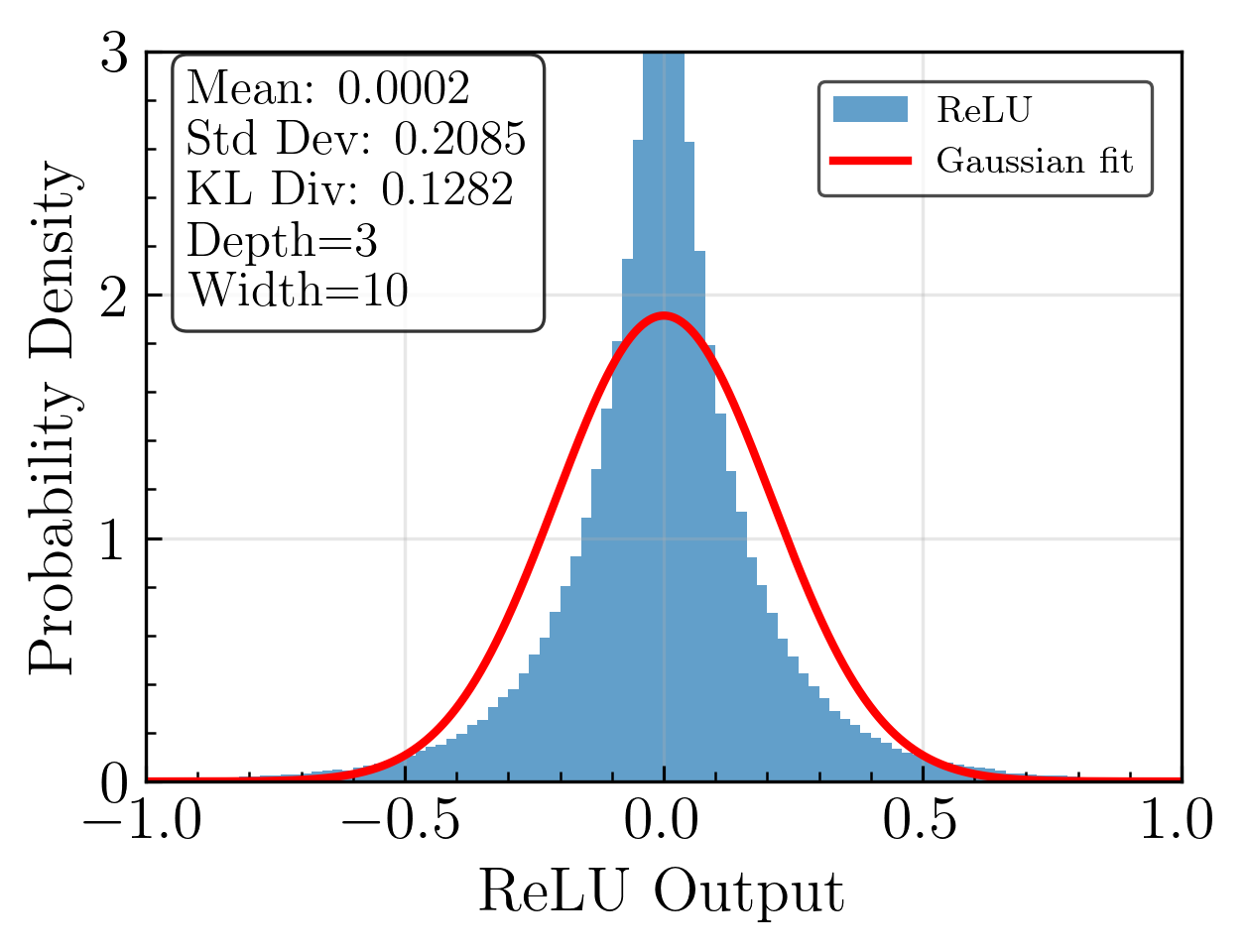}
        \phantomsublabel{-2.4}{0.4}{fig:dist_scaled_depth3_width10}
    \qquad
    \includegraphics[width=0.82\columnwidth]{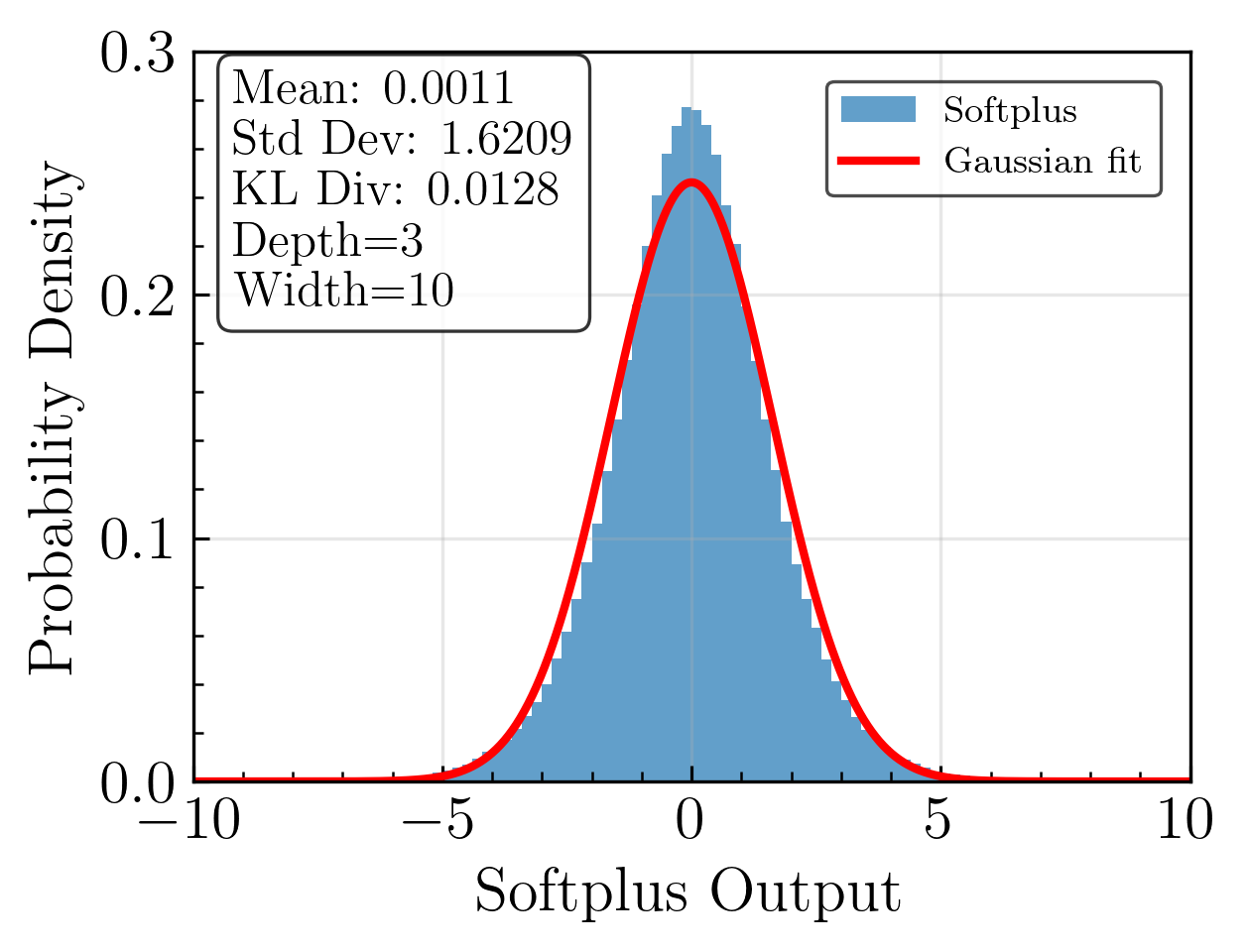}
        \phantomsublabel{-2.3}{0.4}{fig:dist_scaled_depth3_width10_softplus}
        
    \includegraphics[width=0.82\columnwidth]{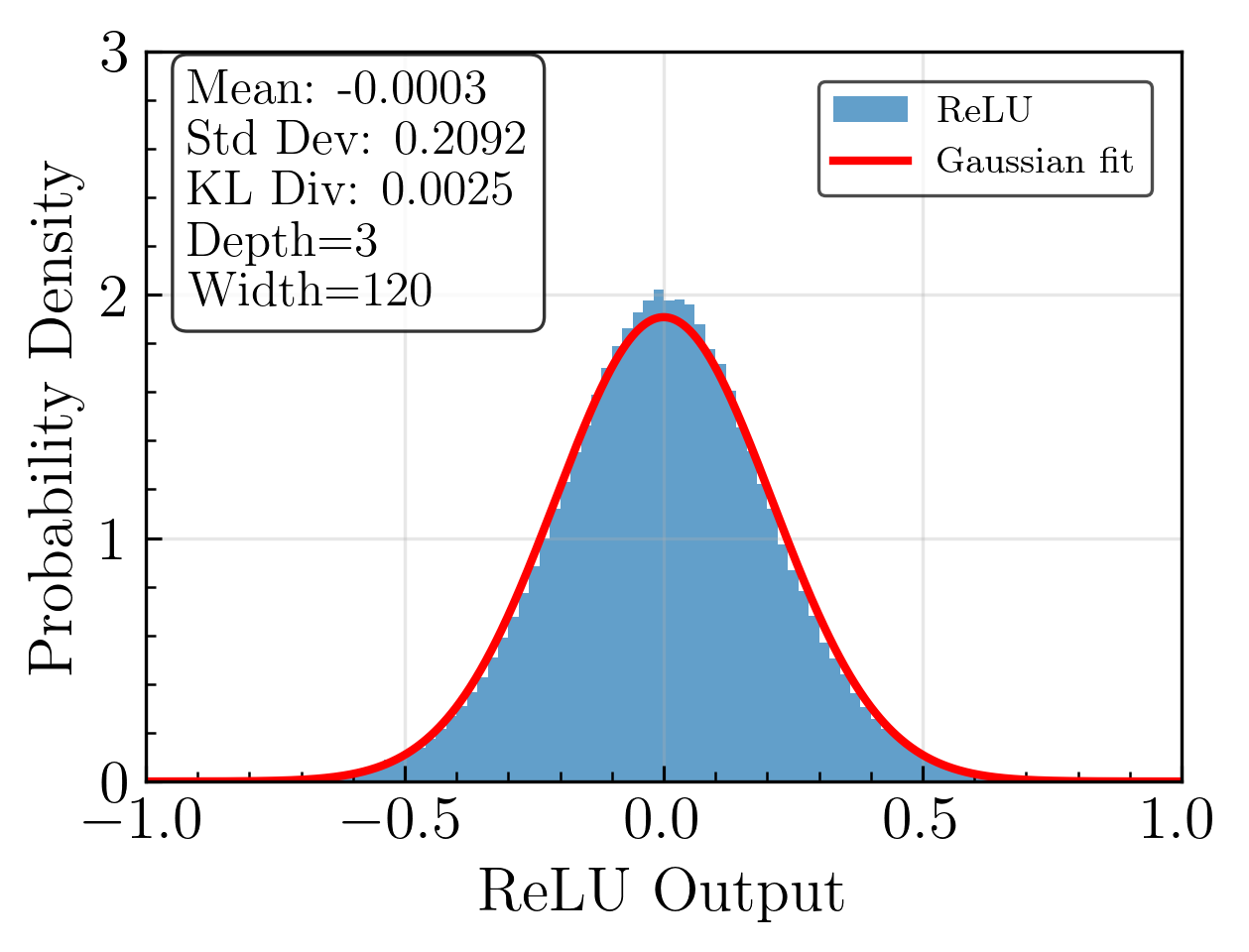}
        \phantomsublabel{-2.4}{0.4}{fig:dist_scaled_depth3_width120}
    \qquad
    \includegraphics[width=0.82\columnwidth]{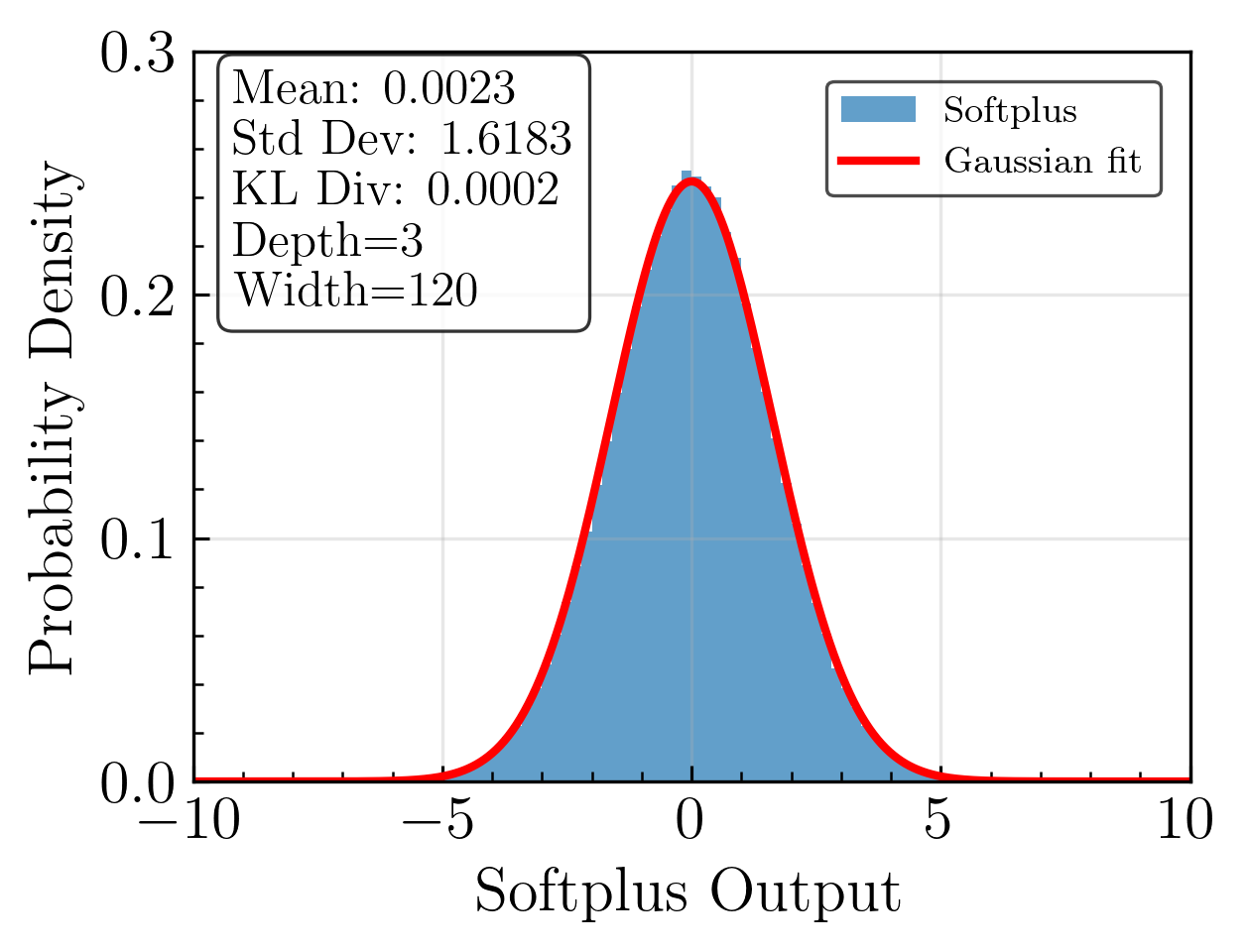}
        \phantomsublabel{-2.3}{0.4}{fig:dist_scaled_depth3_width120_softplus}

    \includegraphics[width=0.82\columnwidth]{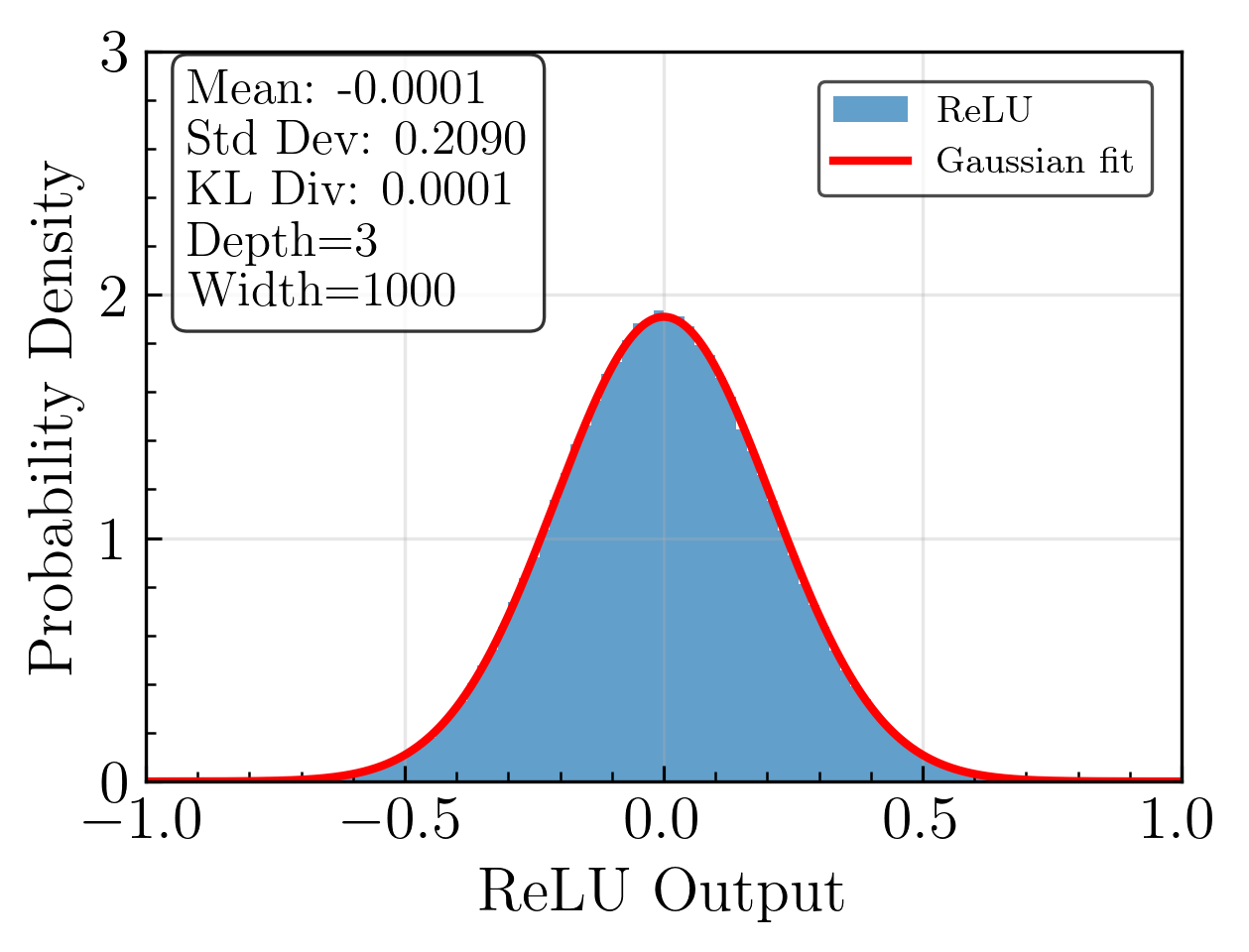}
        \phantomsublabel{-2.4}{0.4}{fig:dist_scaled_depth3_width1000}
    \qquad   \includegraphics[width=0.82\columnwidth]{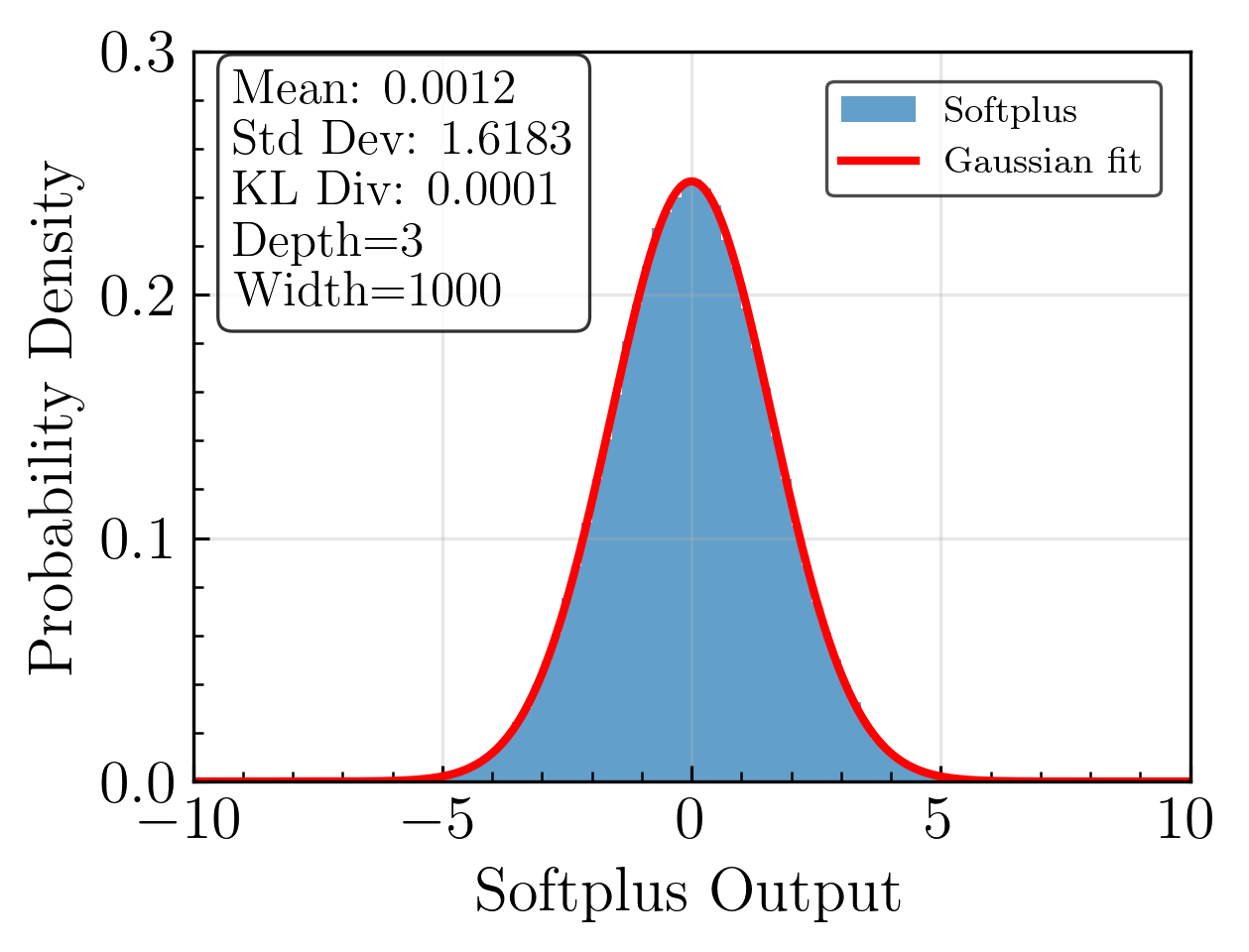}
        \phantomsublabel{-2.3}{0.4}{fig:dist_scaled_depth3_width1000_softplus}
 
    \caption{Final layer ANN output distributions for the scaled $^{56}$Fe input ($Z=26\rightarrow0.13$ and $N=30\rightarrow0.16$)  with two hidden layers. 
    Results are given for a ReLU activation function with widths (a) 1, (c) 10, (e) 120, and (g) 1000 neurons per layer and for a Softplus activation function with widths (b) 1, (d) 10, (f) 120, and (h) 1000 neurons per layer.
\label{fig:dist_10_ReLU}}
\end{figure*}

The first claim we consider is that the large-width limit of the neural networks tends toward a GP prior for the output distribution of functions.
We know mathematically that this must happen eventually; the relevant question is how rapidly the limit is approached.
We test this first by initializing the network at various widths,
with a fixed depth and fixed input. 
Two different activation functions were used when generating these distributions, ReLU and Softplus.
The ReLU activation function is one of the most popular activation functions used in neural networks.
ReLU is zero for negative inputs, and linear for inputs greater than or equal to zero.
The Softplus activation function is designed to be a smoothed ReLU, but has been observed to be less stable during training~\cite{pmlr-v15-glorot11a,Szandała2021,dubey2022activationfunctionsdeeplearning}.
These two activation functions were chosen for their similarity and because ReLU possesses a fixed point in its variance, whereas Softplus does not.

Figure~\ref{fig:dist_10_ReLU} shows the histograms of $500,000$ total outputs from different initializations for the ReLU and Softplus activation functions with a fixed depth of two hidden layers and widths from 1 to 1000 neurons per layer. 
$^{56}$Fe was used as the test input to generate all these distributions, and all inputs to the network were scaled according to a MinMax scaler.
In applying ANNFT, inputs to the network should be scaled to be approximately $\mathcal{O}(1)$~\cite{Roberts:2021fes}. 
 MinMax scaling converts the inputs to be within a range $0$ to $1$ using the transformation
\begin{align}
X_{\text{scaled}} = \frac{X - X_{\min}}{X_{\max} - X_{\min}},
\end{align}
on both the number of neutrons $N$ and protons $Z$, with the min and max terms referring to the largest and smallest values of the dataset.
The red line in Fig.~\ref{fig:dist_10_ReLU} shows a normal distribution using the mean and variance of the given histogram to compare the output distribution to what the distribution would look like if it were Gaussian.  
As the width increases (and $r$ decreases), the output distributions approach Gaussian distributions at about a width of 120 neurons as the large width averages out connected contributions from higher-order moments in the distribution. 
Both activation functions' output distributions converge to Gaussian distributions in the infinite-width limit, regardless of whether the activation function has a fixed point in its variance or not.
Indeed, the Softplus distributions approach the Gaussian limit somewhat faster than the ReLU distributions.

\begin{figure*}[ptbh!]
    \centering
    \includegraphics[width=0.99\columnwidth]{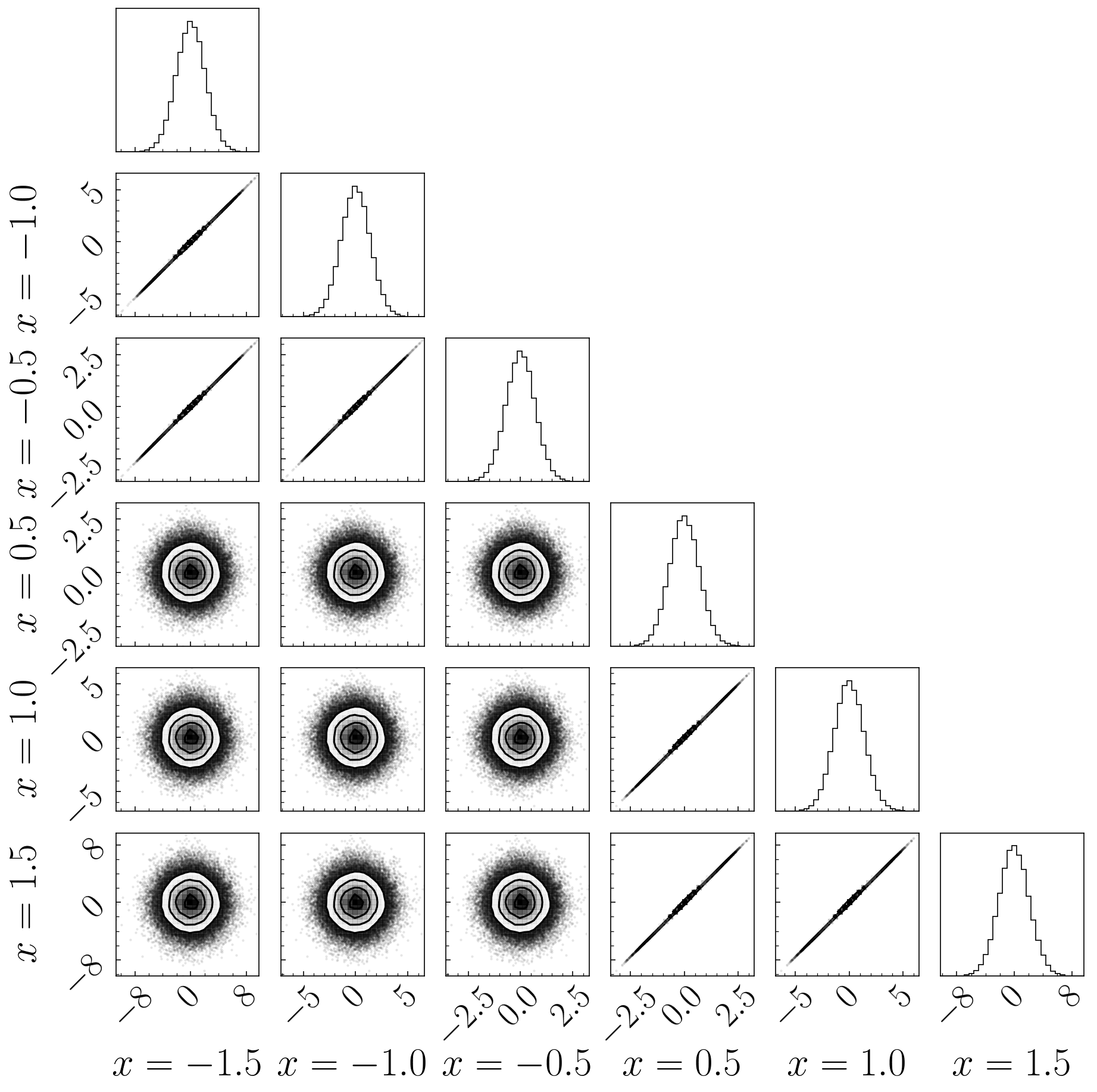}
        \phantomsublabel{-1.5}{3.0}{fig:corner_plot_1_layer_ReLU}
    \includegraphics[width=0.98\columnwidth]{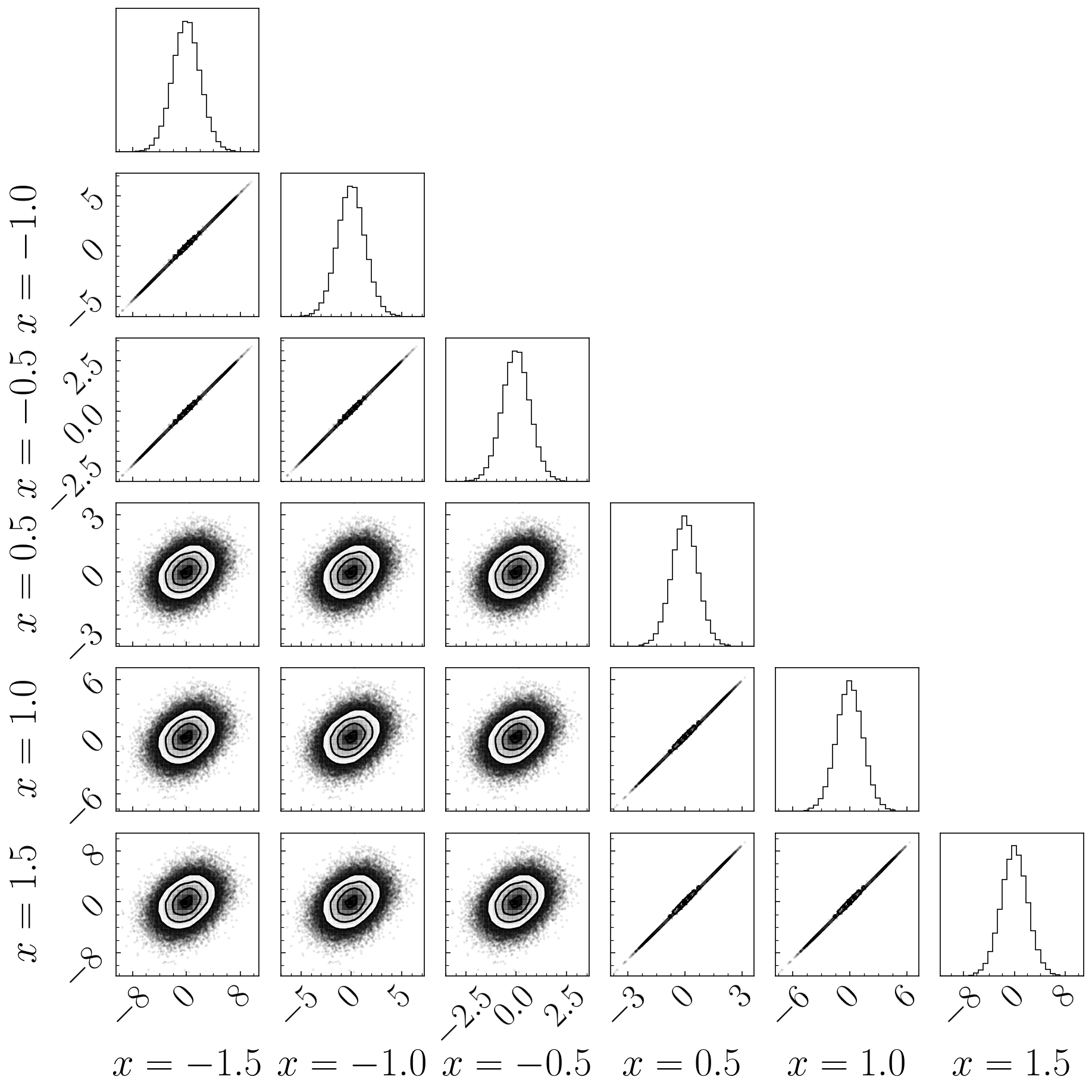}
        \phantomsublabel{-1.5}{3.0}{fig:corner_plot_2_layer_ReLU}
    
    \includegraphics[width=0.98\columnwidth]{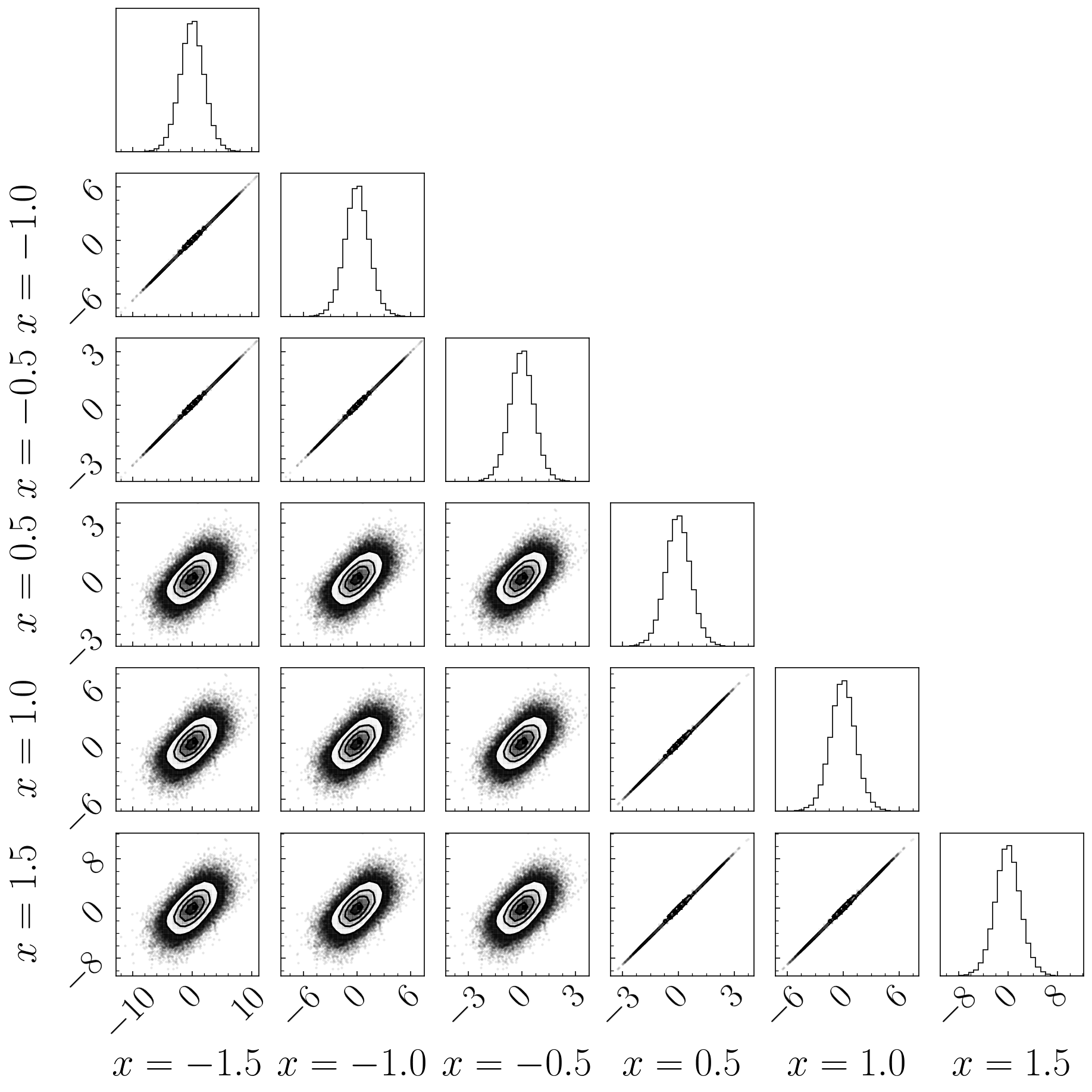}
        \phantomsublabel{-1.5}{3.0}{fig:corner_plot_4_layer_ReLU}
    \includegraphics[width=0.98\columnwidth]{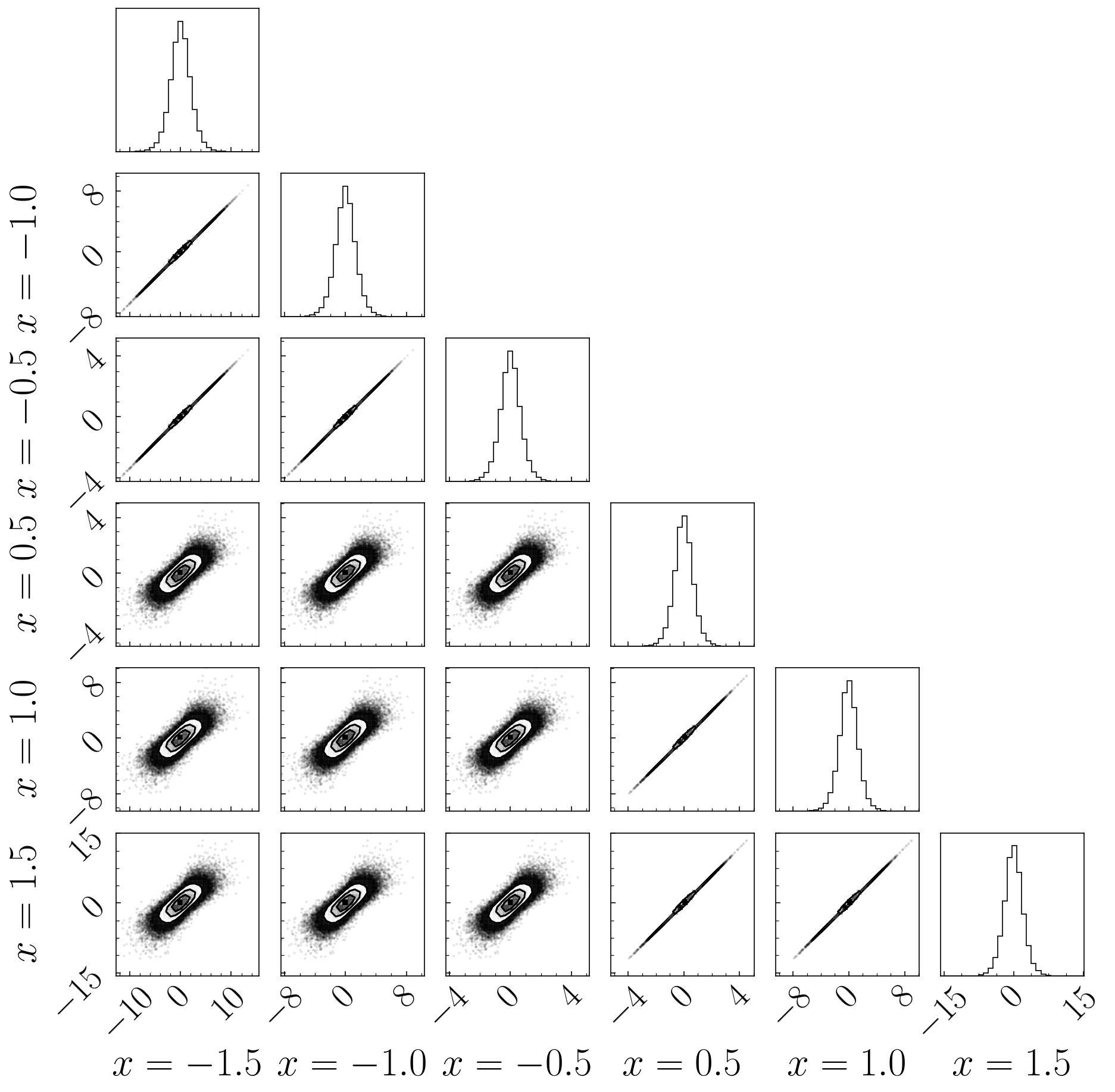}
        \phantomsublabel{-1.5}{3.0}{fig:corner_plot_8_layer_ReLU}

    \caption{Corner plots for ReLU activation function outputs. The corner plots all feature ReLU networks of a fixed width of 120 neurons, and depths of (a) 1, (b) 2, (c) 4, and (d) 8. The addition of hidden layers introduces correlations to the network output distributions. \label{fig:corner_plot_ReLU}}
    
\end{figure*}

To fully show that neural network output distributions become GPs  as the ratio of depth to width becomes small, we must also look at correlations between inputs.
Figure~\ref{fig:corner_plot_ReLU} compares corner plots for outputs from a pre-training ReLU-based ANN with scalar inputs ranging from $-1.5$ to $+1.5$. 
These inputs, being $\mathcal{O}(1)$ or less, are left untreated by the MinMax scaler.
The width for all plots is fixed at 120 neurons while the depth increases from one hidden layer up to eight hidden layers.
The diagonal histograms are consistent with the approach to Gaussian outputs seen in Fig.~\ref{fig:dist_10_ReLU} (e.g., a clear Gaussian shape for one layer with $r = 1/60$ while a clear non-Gaussian shape for eight layers with $r = 1/15$).

\begin{figure*}[ptbh!]
    \centering
    \includegraphics[width=0.9\columnwidth]{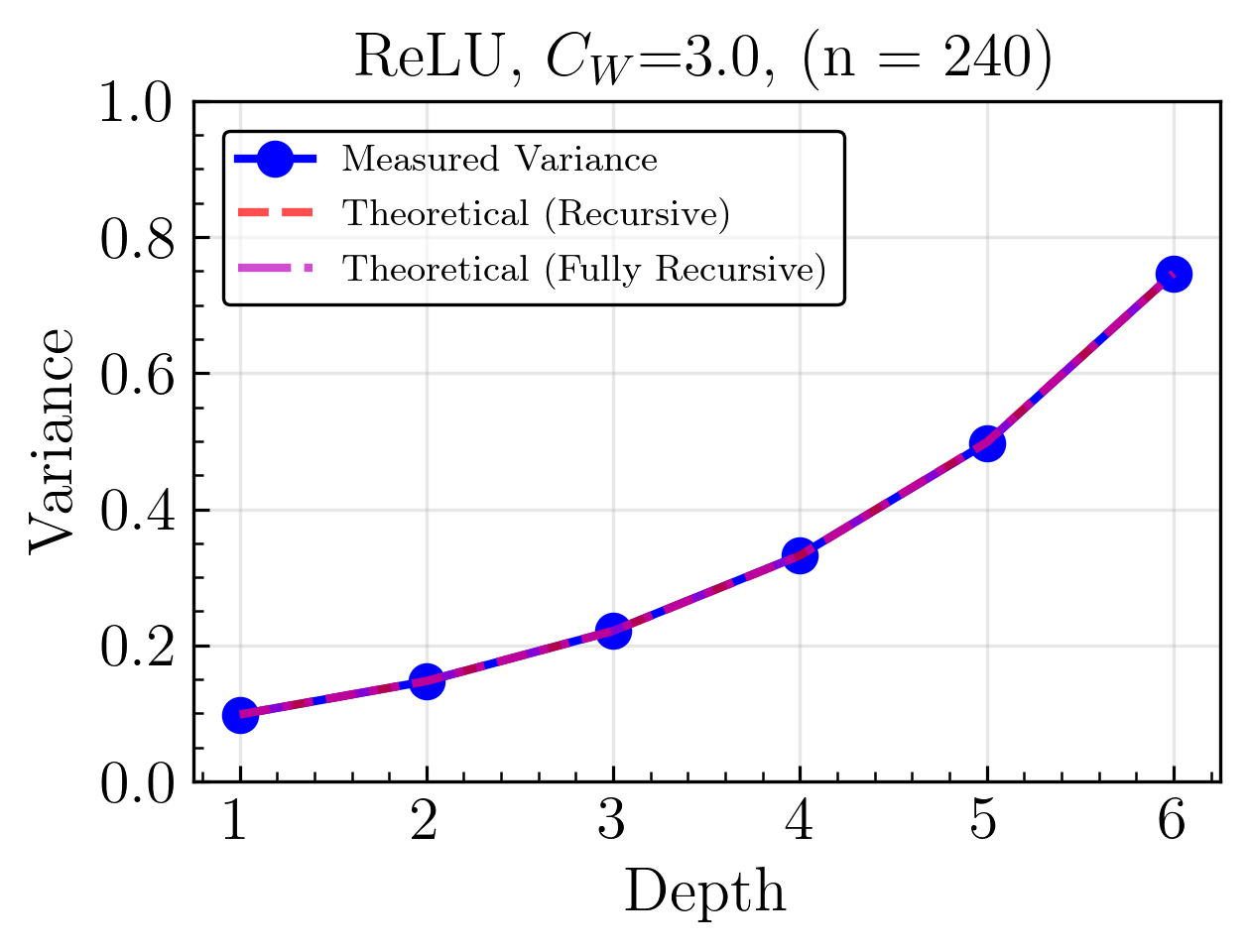}
        \phantomsublabel{-0.5}{0.5}{fig:variance_vs_depth_width_240_ReLU_above_critical}
    \qquad
    \includegraphics[width=0.9\columnwidth]{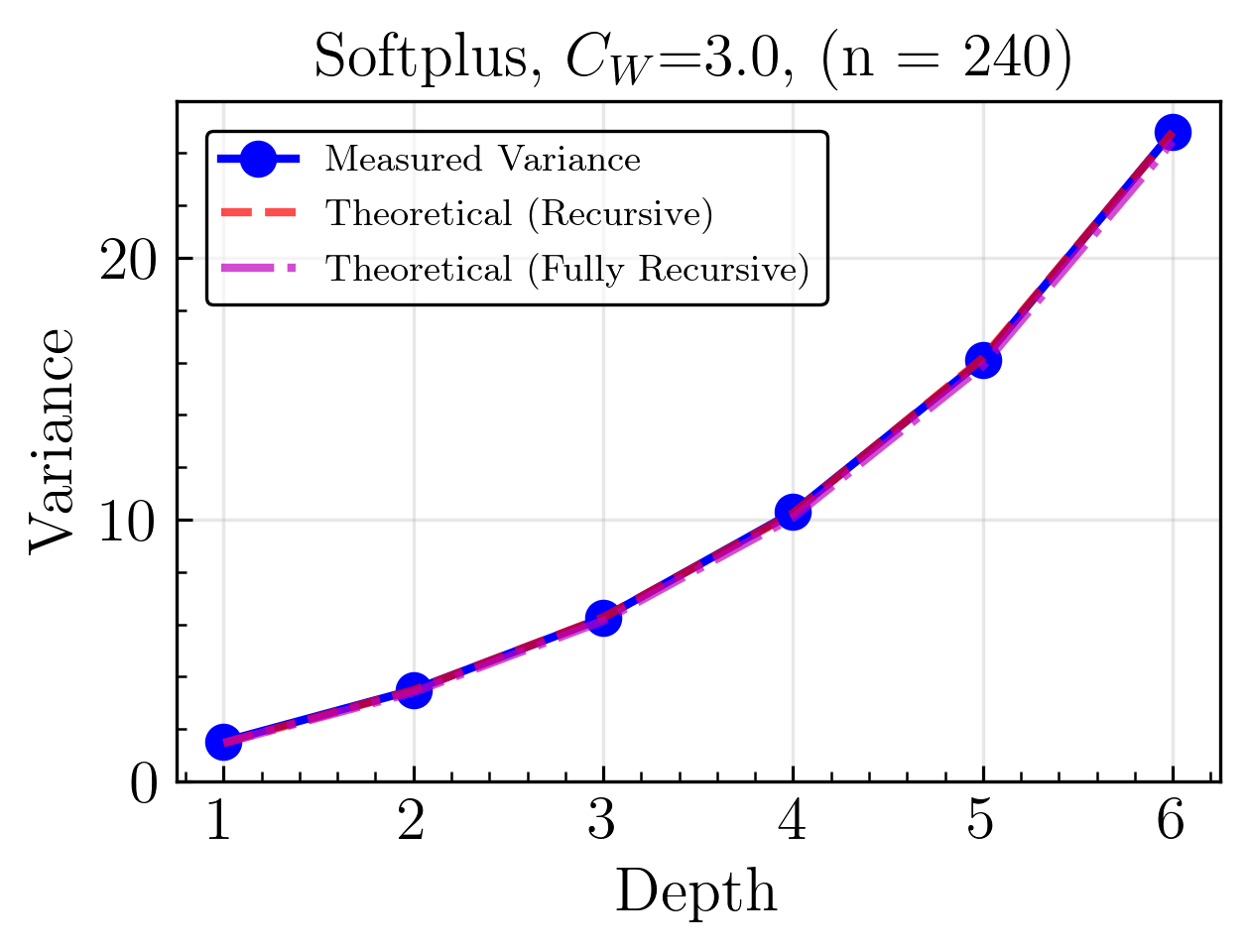}
        \phantomsublabel{-0.5}{0.5}{fig:variance_vs_depth_width_240_Softplus_above_critical}
    
    \includegraphics[width=0.9\columnwidth]{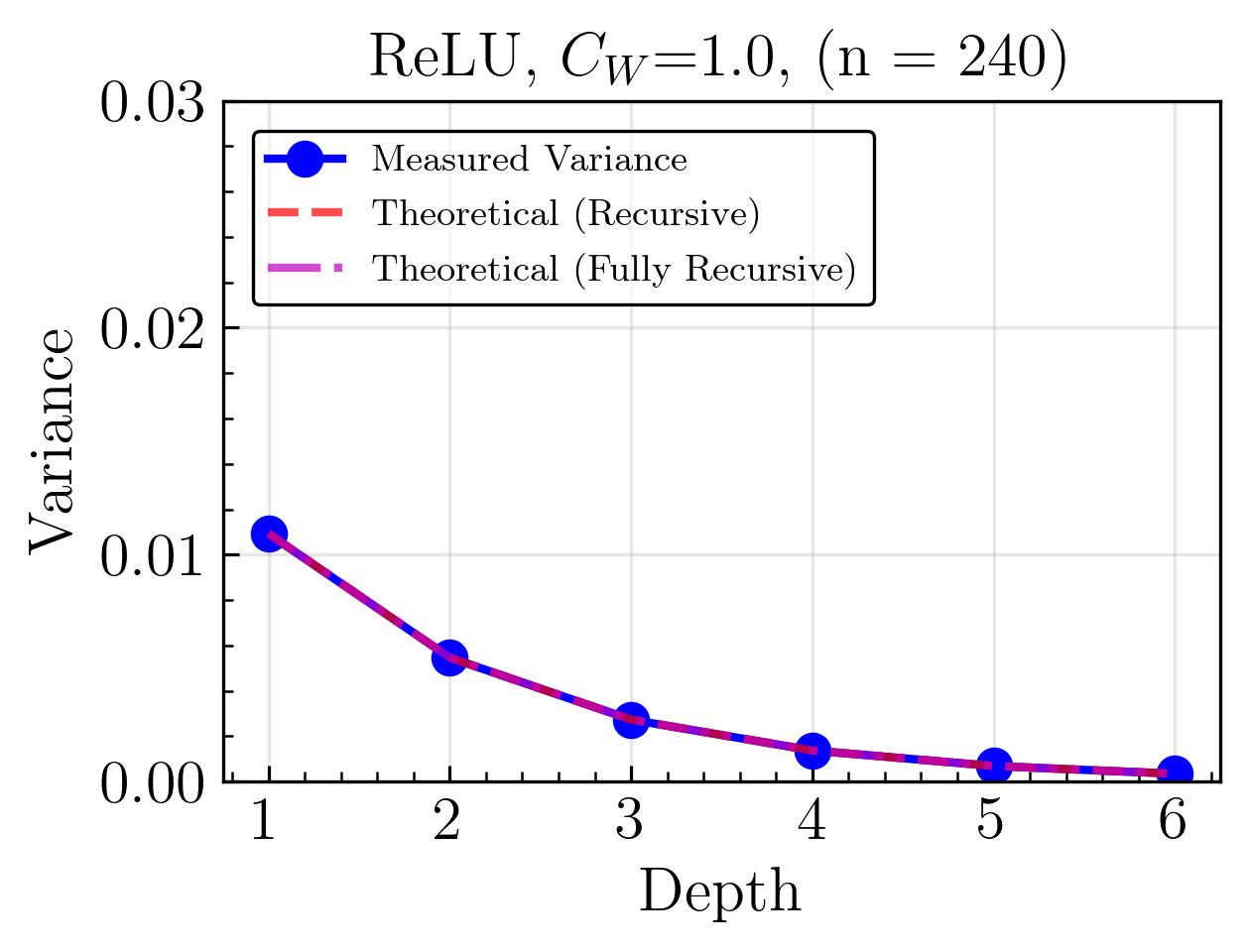}
        \phantomsublabel{-0.5}{0.5}{fig:variance_vs_depth_width_240_ReLU_below_critical}
    \qquad
    \includegraphics[width=0.9\columnwidth]{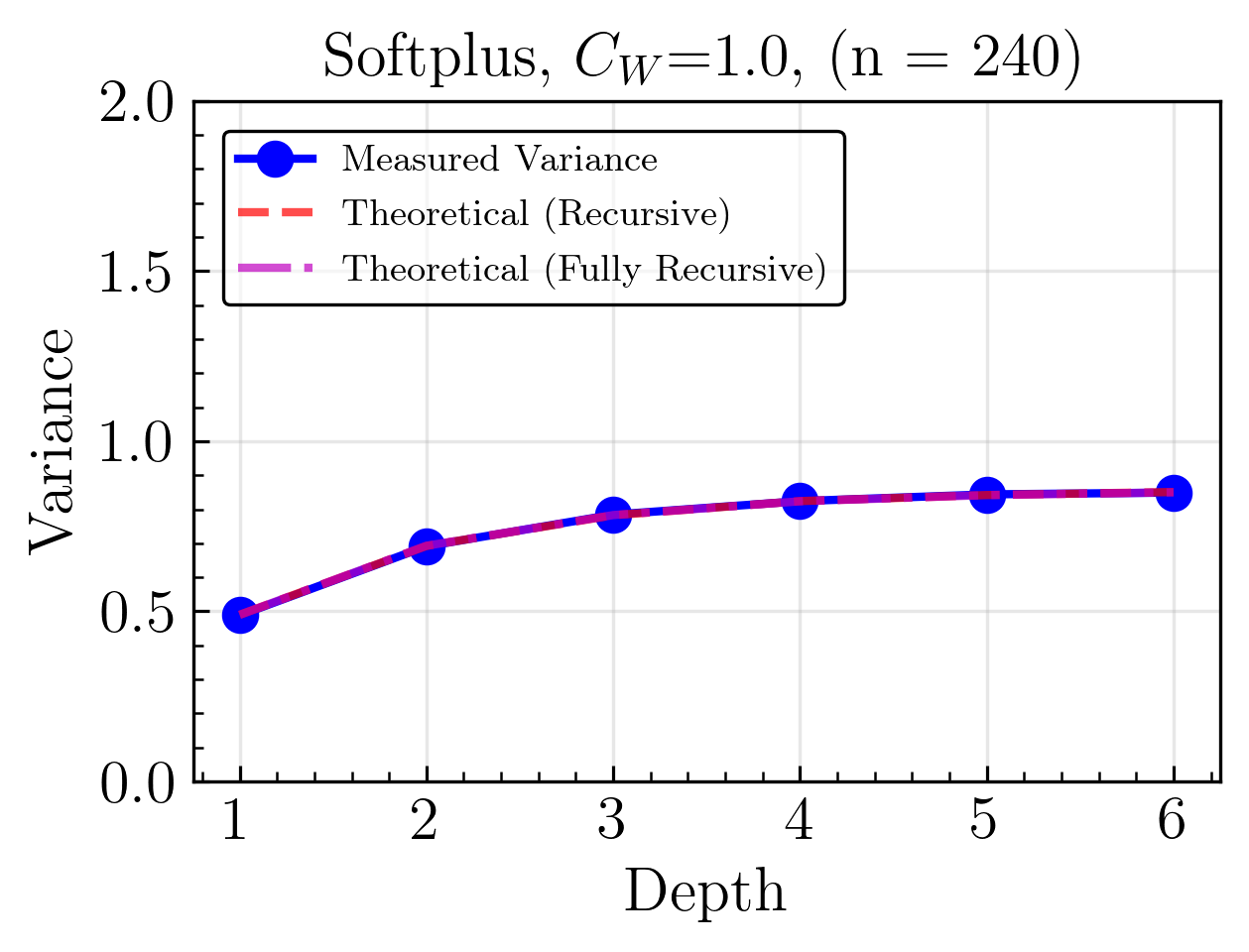}
        \phantomsublabel{-0.5}{0.5}{fig:variance_vs_depth_width_240_Softplus_below_critical}

     \includegraphics[width=0.9\columnwidth]{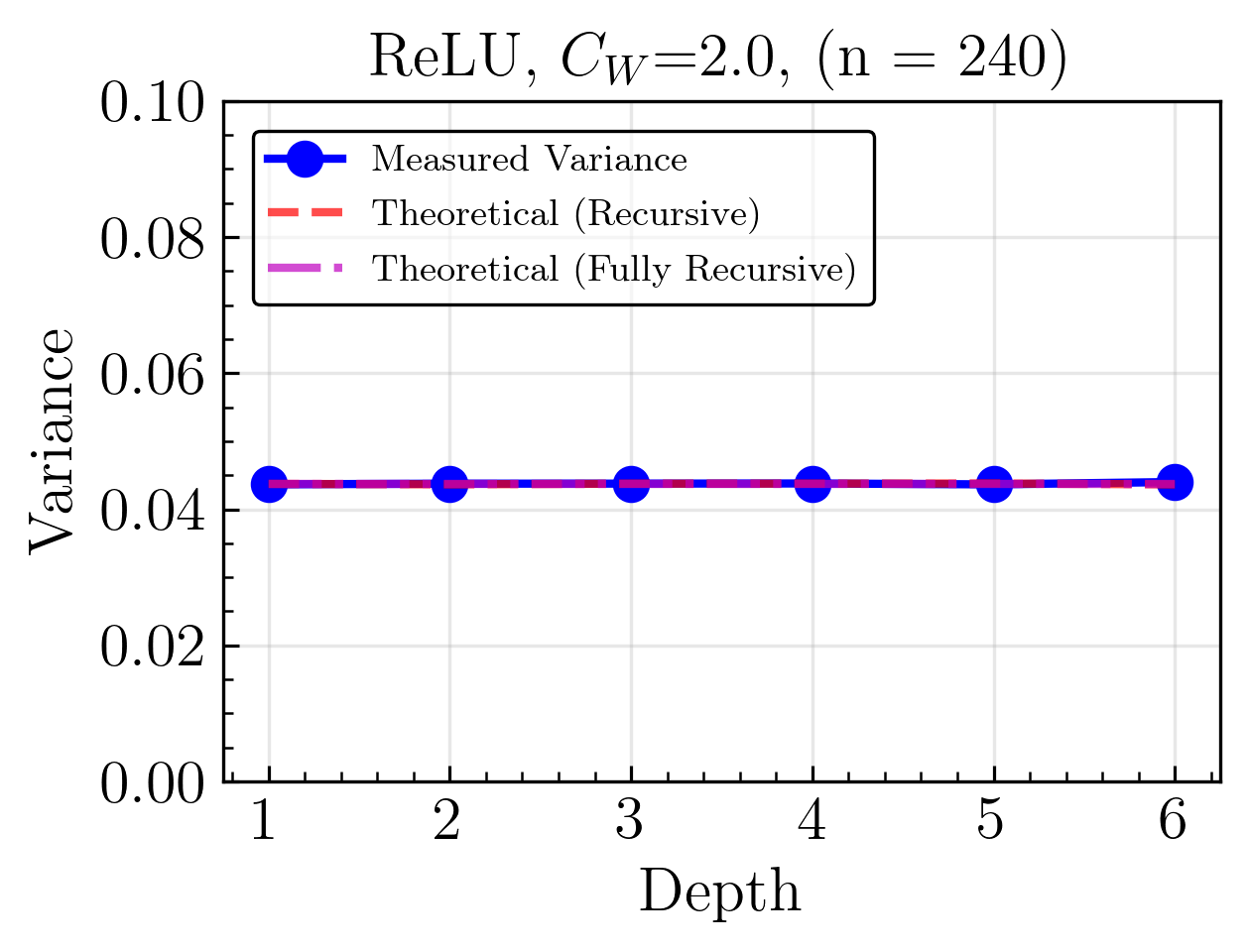}
        \phantomsublabel{-0.5}{0.5}{fig:variance_vs_depth_width_240_ReLU_critical}
     \qquad  
     \includegraphics[width=0.9\columnwidth]{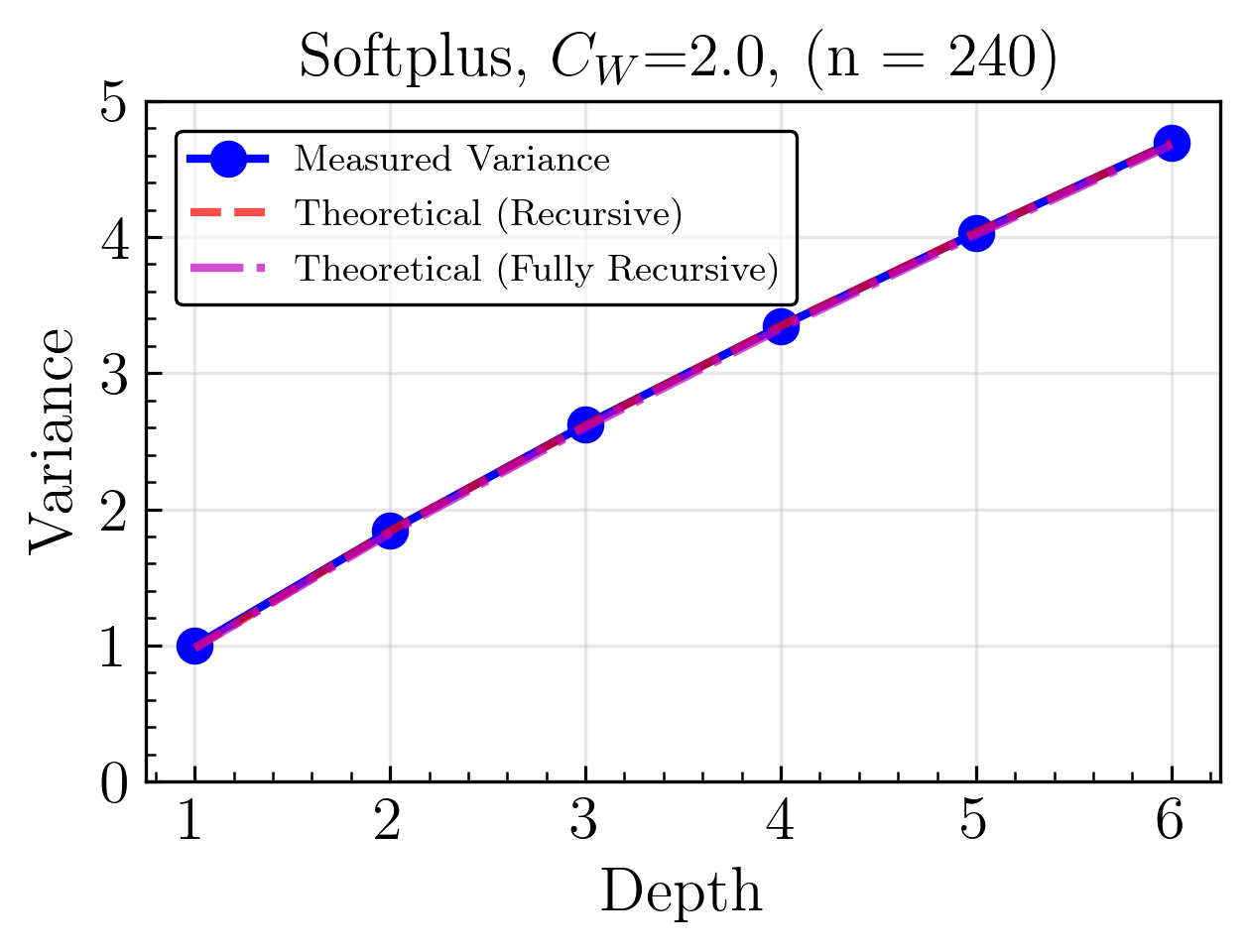}
        \phantomsublabel{-0.5}{0.5}{fig:variance_vs_depth_width_240_Softplus_critical}

    \caption{The final layer pre-training output variance as a function of neural network depth with a fixed width of 240. 
    Results are given for a ReLU activation function tuned (a) above ($C_W = 3.0$, $C_b = 0.0$) (c) below ($C_W = 1.0$, $C_b = 0.0$), and (e) at the critical point ($C_W = 2.0$, $C_b = 0.0$), and for a Softplus activation function with (b) highest, (d) lowest, and (f) intermediate initialization widths for the weights by using the same initialization hyperparameters as the ReLU.
    The measured variance is plotted in blue, a recursive calculation using empirical values of the variance is plotted as a dashed red line, and a recursive calculation using only theoretically calculated variances is plotted as a dashed pink line.
    When above/below the critical values of the initialization width, the variance explodes/vanishes with depth. When the ReLU network is tuned to critical initialization, the variance is fixed with depth. The Softplus activations do not have a critical point for initialization. As such, the Softplus variance either grows or asymptotes towards a constant value with depth, and does not have initialization hyperparameters that give a fixed variance with depth.
}
    \label{fig:variance_above_ReLU}
\end{figure*}

\begin{figure*}[ptbh]
    \centering
    \includegraphics[width=0.95\columnwidth]{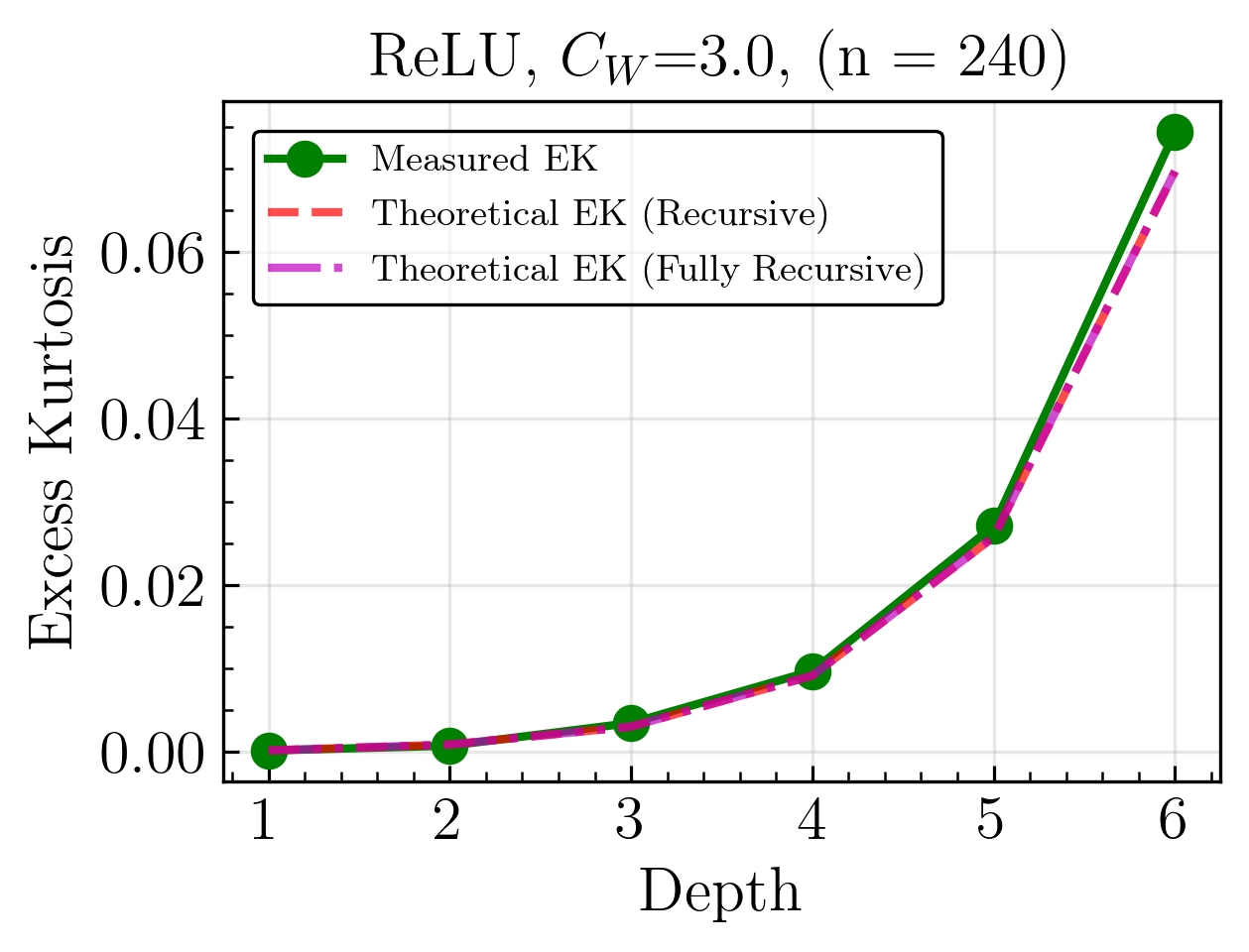}
        \phantomsublabel{-0.4}{0.6}{fig:kurtosis_vs_depth_width_240_ReLU_above_critical}
    \qquad
    \includegraphics[width=0.95\columnwidth]{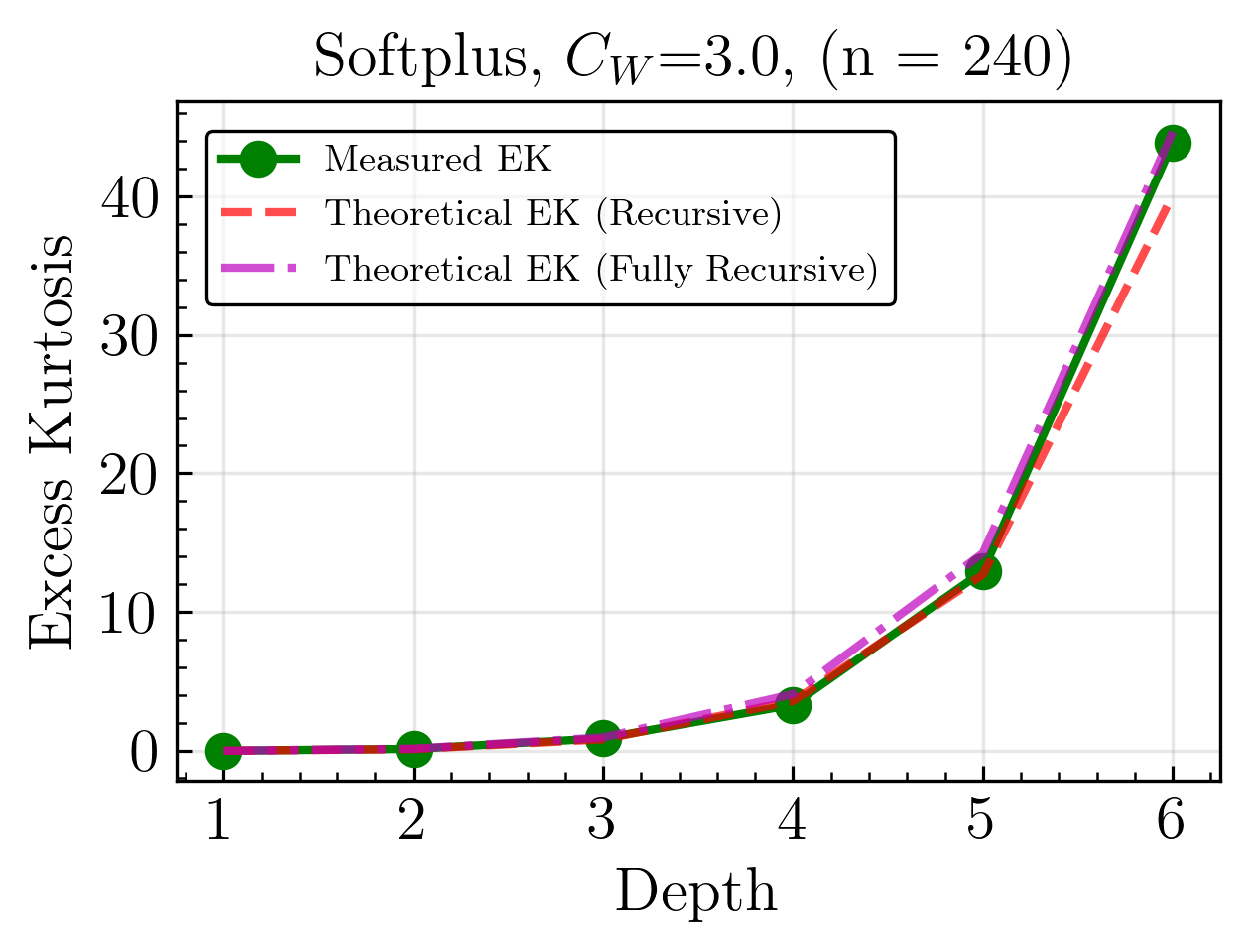}
        \phantomsublabel{-0.4}{0.6}{fig:kurtosis_vs_depth_width_240_Softplus_above_critical}
    
    \includegraphics[width=0.95\columnwidth]{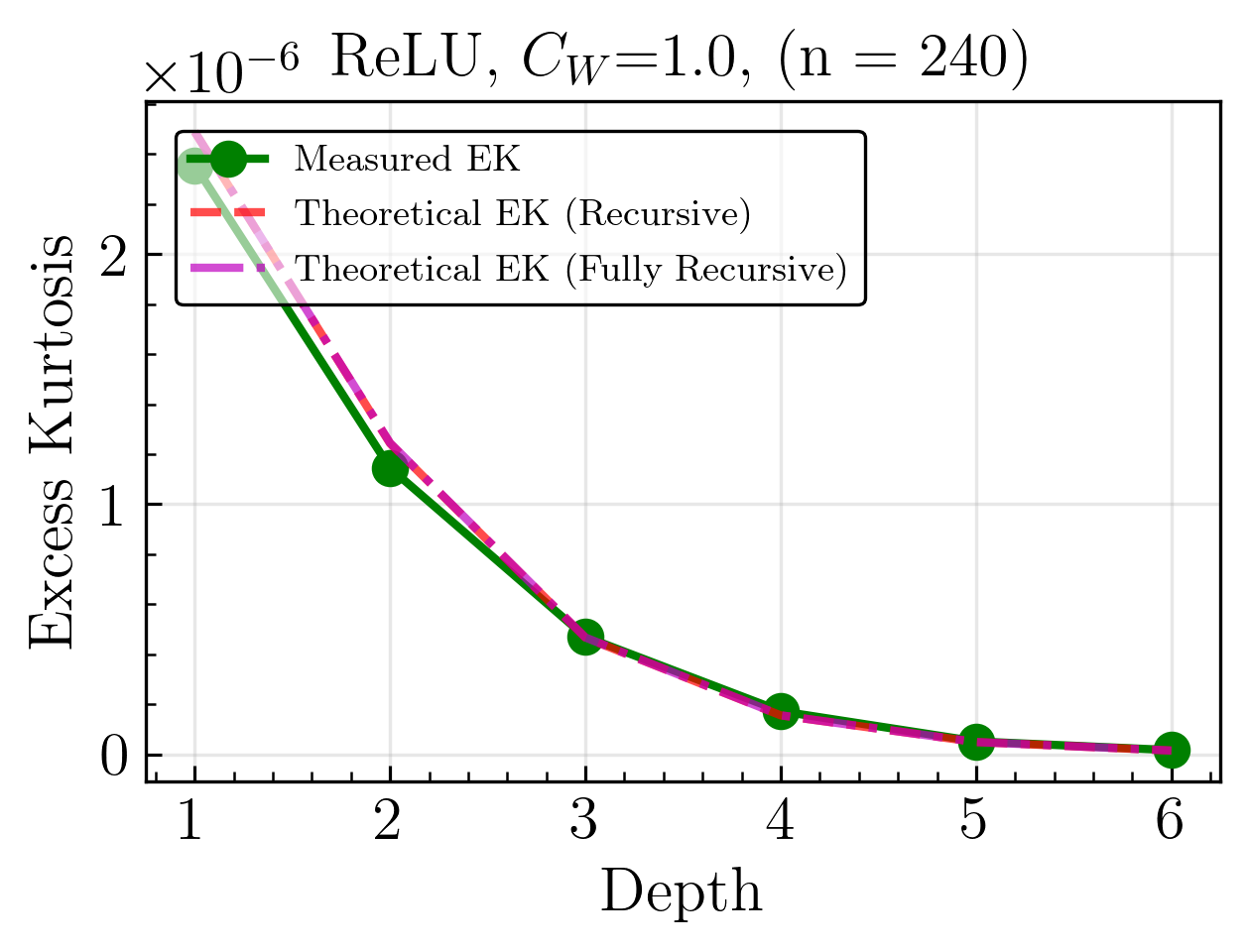}
        \phantomsublabel{-0.4}{0.6}{fig:kurtosis_vs_depth_width_240_ReLU_below_critical}
    \qquad
    \includegraphics[width=0.95\columnwidth]{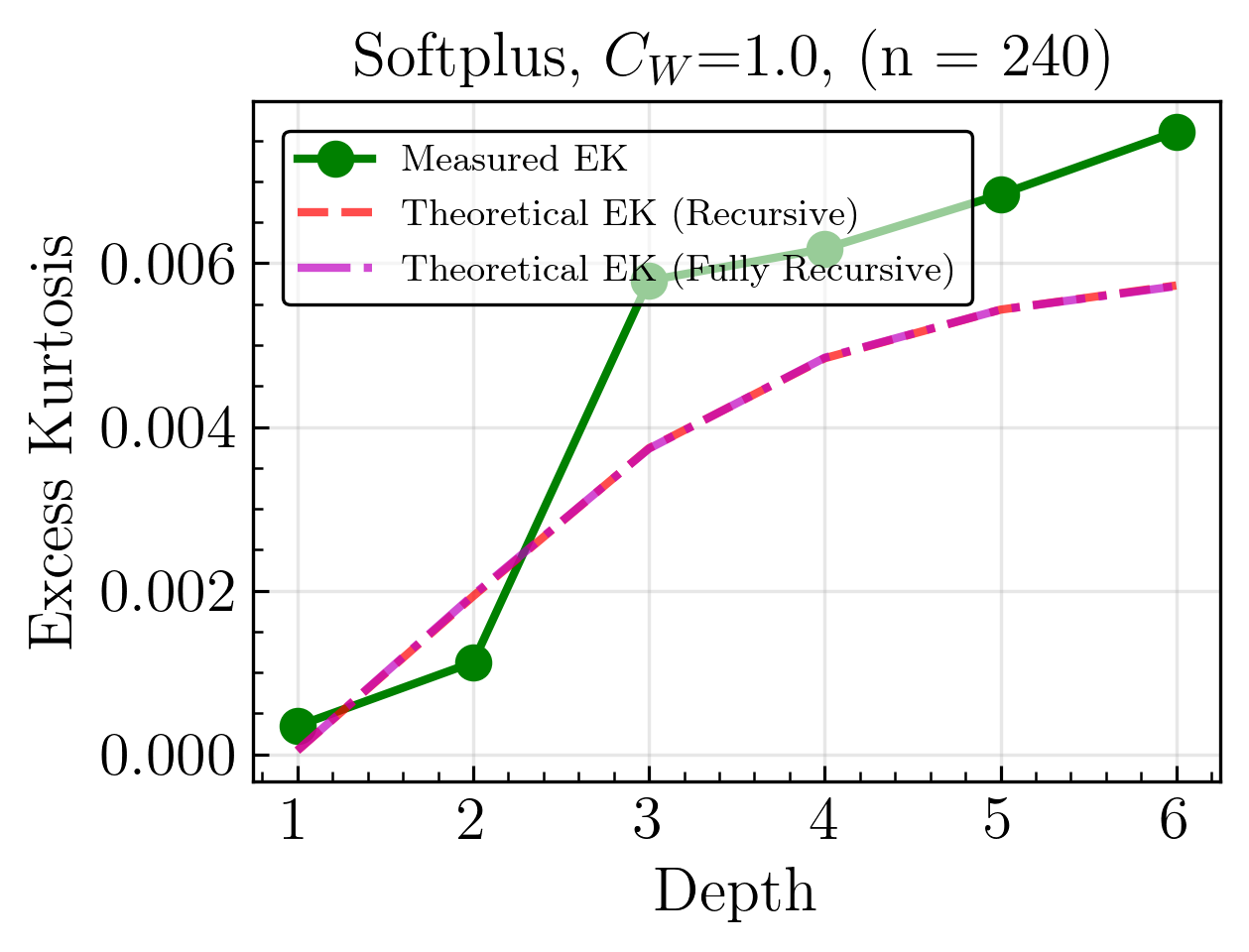}
        \phantomsublabel{-0.4}{0.6}{fig:kurtosis_vs_depth_width_240_Softplus_below_critical}

     \includegraphics[width=0.95\columnwidth]{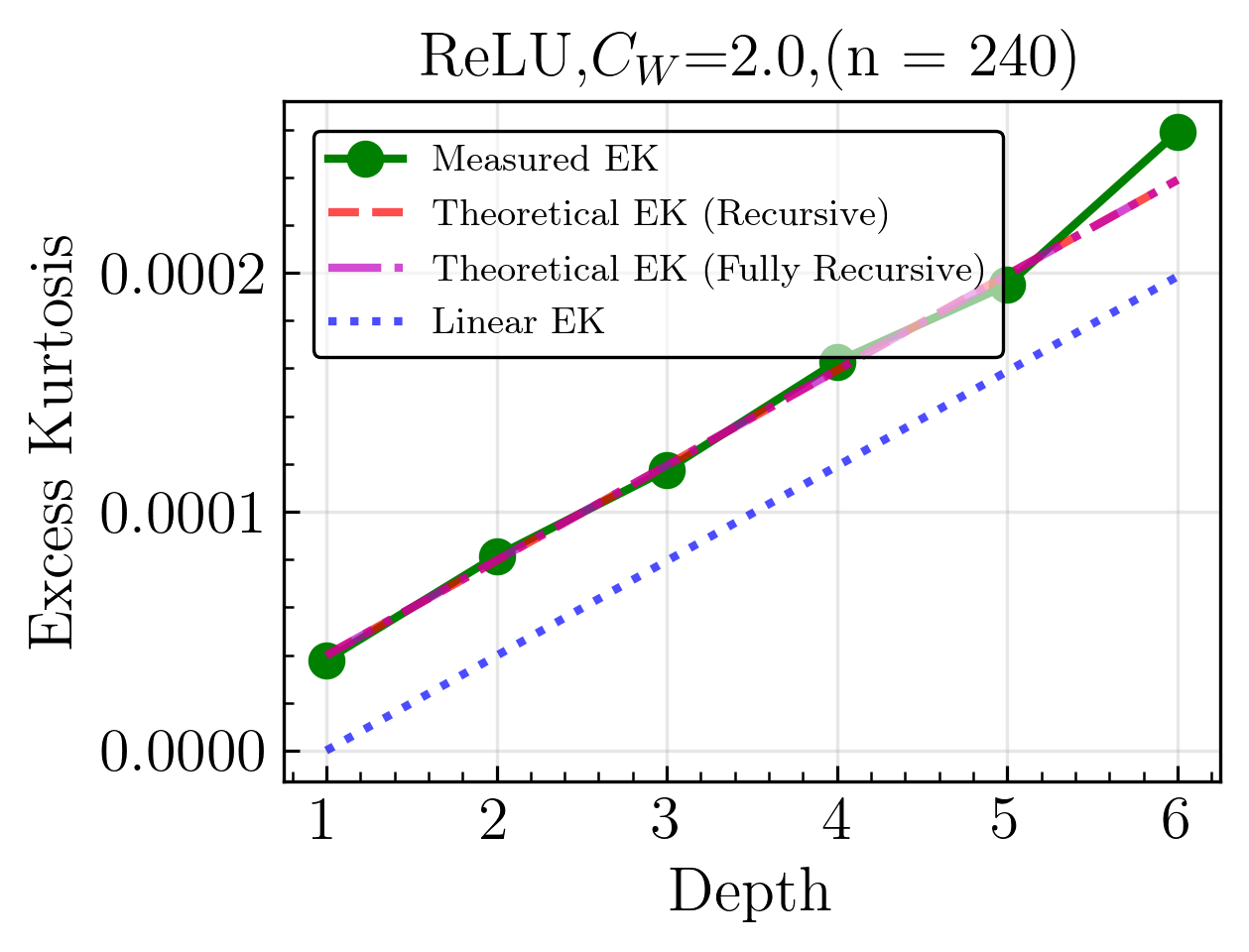}
     \qquad
        \phantomsublabel{-0.5}{0.6}{fig:kurtosis_vs_depth_width_240_ReLU_critical}
     \includegraphics[width=0.95\columnwidth]{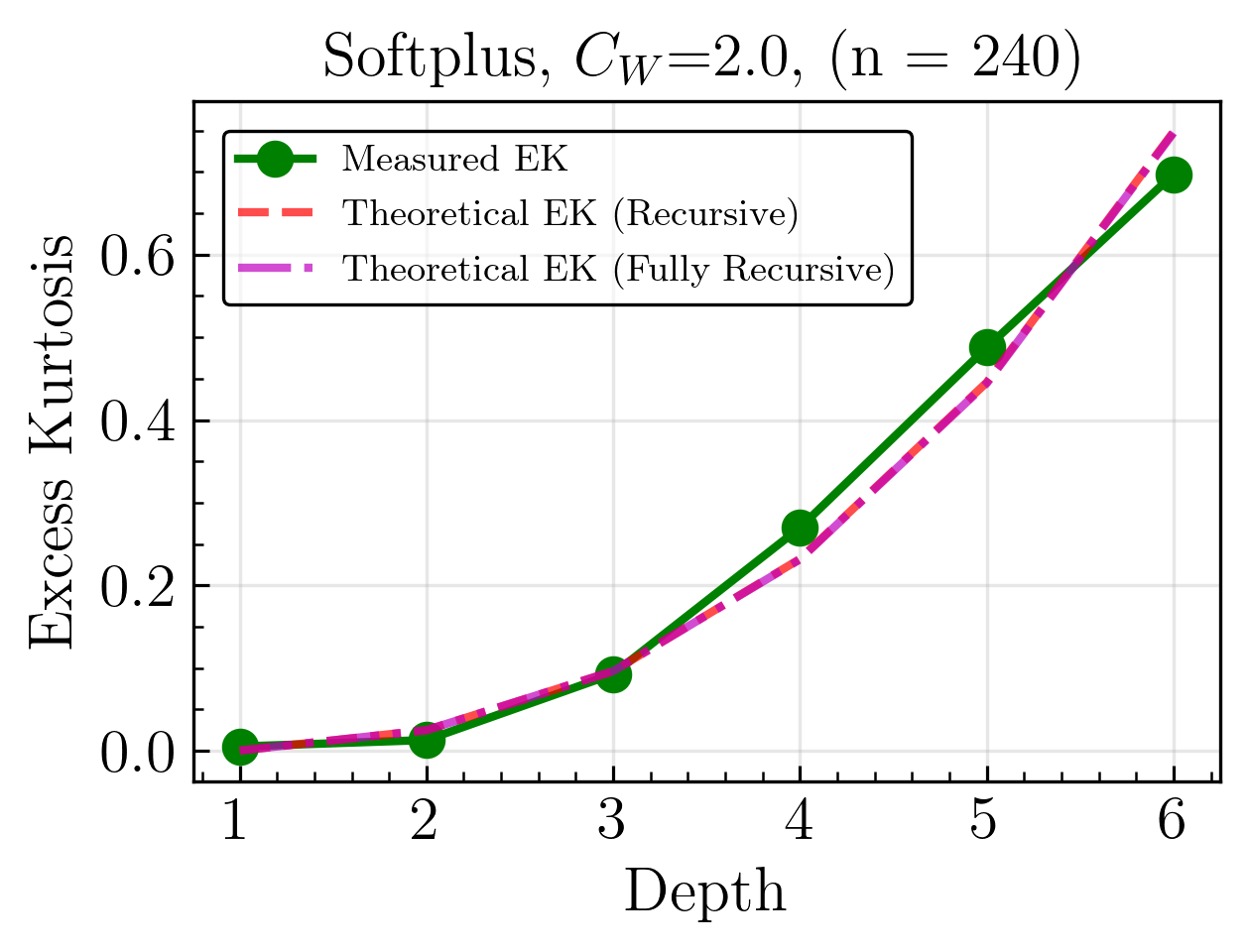}
        \phantomsublabel{-0.5}{0.6}{fig:kurtosis_vs_depth_width_240_Softplus_critical}

    \caption{The final-layer pre-training-output (unstandardized) excess kurtosis (abbreviated here as EK, and also known as the non-Gaussian 4-point correlations) as a function of neural network depth with a fixed width of 240. 
    Results are given for a ReLU activation function tuned (a) above ($C_W=3.0$, $C_b=0.0$), (c) below ($C_W=1.0$, $C_b=0.0$), and (e) at the critical point ($C_W=2.0$, $C_b=0.0$), and for a Softplus activation function with (b) highest, (d) lowest, and (f) intermediate initialization widths for the weights by using the same initialization hyperparameters as ReLU.
    The measured EK is plotted as a green line, a recursive calculation of EK using empirical values of the variance and initial EK is plotted as a dashed red line, and a recursive calculation using only theoretical values is plotted as a dashed pink line.
    An additional blue line is present in (e), representing an exact calculation of the critical EK without treating $1/n$ terms as subleading.
    When critically tuned, ReLU networks' EK behaves linearly with depth as predicted by ANNFT, demonstrating that the higher-order moments of the distribution are controlled by an expansion in $r$.
    Non-critical distributions are seen to grow non-linearly with depth, reflecting that criticality allows for a perturbative ANNFT that lets deeper network network behavior to be analyzed.
}
    \label{fig:kurtosis_above_ReLU}
\end{figure*}

The behavior of the off-diagonal joint distributions is likely unfamiliar to practitioners familiar with the most commonly used GPs (such as RBF or Matern kernels).
We can understand the pattern by applying the equations presented in Sec.~\ref{sec:ANNFTintro}.
Indeed, kernels for the limiting GPs can be calculated by recognizing that the output distributions are mean zero, and that the variance is the only surviving moment~\cite{Halverson:2020trp}. 
The variance $G^{(\ell)}$ of outputs $z^{(\ell)}$ for a given layer can be used to find the variance of the next using the equation.
\begin{align}
G^{(\ell+1)} 
= C_b 
+ C_W \int_{-\infty}^{\infty} dz^{(\ell)} \, \sigma^2(z^{(\ell)}) \, e^{-\frac{(z^{(\ell)})^2}{2G^{(\ell)}}}.
\end{align}
Using this expression, the final layer variance $G^{(\ell_{out})}$ can be calculated, and the output distribution for the GP prior completely determined. To acquire the GP kernel for the prior, all that is required is the covariance of the outputs in the final layer $\expval{z^{(\ell)}(x)z^{(\ell)}(x')}$, where $x$ and $x'$ are inputs to the network. 

The kernel for an infinitely wide neural network with one hidden layer and a critically initialized ReLU activation function can be calculated analytically~\cite{Halverson:2020trp}: 
\begin{align}
      \expval{z^{(\ell=1)}(x)z^{(\ell=1)}(x')} &= \frac{2}{\pi} |x \cdot x'| (\pi - \phi) \cos(\phi) , \\
    \phi &= \arccos(\frac{x \cdot x'}{|x\cdot x'|}) .
\end{align}
These equations impose that the correlation coefficients are either 0 or 1, depending on whether $x\cdot x'$ is less than or greater than zero.
This behavior, which is that of a manifestly non-stationary GP, is evident in Fig.~\subref*{fig:corner_plot_1_layer_ReLU}.
As the ratio of depth to width increases, correlations grow, as seen in the other panels.
    
Returning to the pre-training binding energy output distributions, it is important to see that the couplings in the distributions' actions evolve according to an RG flow with each layer of the ANN. 
As we have emphasized, finding a fixed point in the variance prevents gradient issues during training.
ReLU is an example of an activation function that has a fixed point. 
In Fig.~\ref{fig:variance_above_ReLU}, the variance of the final layer output distribution either explodes (Fig.~\subref*{fig:variance_vs_depth_width_240_ReLU_above_critical}) or vanishes (Fig.~\subref*{fig:variance_vs_depth_width_240_ReLU_below_critical}) with depth when the initialization widths are away from criticality, and remains at a fixed value when tuned to criticality (Fig.~\subref*{fig:variance_vs_depth_width_240_ReLU_critical}). 

The variance recursion relation given in Eq.~\eqref{eq:2PointRecursion} can be used to calculate the variance in any layer $\ell$. 
The variance value used to calculate the initial value for $G^{(1)}$ is $G^{(0)} = C_W(\frac{1}{n_0}\sum_{i=1}^{n_0}x_i^2) +C_b$.
Applying the MinMax scaling to a test input of $^{56}$Fe, two calculations of the variance were performed. 
The first used the measured value of $G^{(1)}$ from the data, then calculated the variance in all further layers.
The second used $G^{(0)}$ to calculate $G^{(1)}$ instead, recursively computing all variances in every layer with only theoretically predicted values of $G^{(\ell)}$.

As can be seen in Figs.~\subref*{fig:variance_vs_depth_width_240_ReLU_above_critical}, ~\subref*{fig:variance_vs_depth_width_240_ReLU_below_critical}, and ~\subref*{fig:variance_vs_depth_width_240_ReLU_critical}, there is agreement between the data and both ANNFT approaches for calculating the variance in all layers.
ANNFT accurately calculates the variance for neural networks using multiple different initialization hyperparameter values. 
In addition, the fixed point that ANNFT predicts for the ReLU activation function is manifested in \subref*{fig:variance_vs_depth_width_240_ReLU_critical} with a value of $G^{(0)} = 4.37 \times 10^{-2}$.
The presence of this fixed point is significant, as it demonstrates that there exists an initialization that stabilizes the variance with depth.
The other initializations lead to the variance of initial neural network outputs growing or shrinking exponentially, which in turn causes the initial ANN outputs to grow or shrink.
Effective ANN training is dependent on the size of the outputs, and the fixed point in the variance helps to stabilize gradients. 

By contrast, the Softplus activation function does not have critical initialization and thus no fixed point for its variance. 
In Fig.~\subref*{fig:variance_vs_depth_width_240_Softplus_above_critical}, the Softplus's output variance exhibits exponential growth like the ReLU does when $C_W=3.0$.
However, Softplus's variance grows to much larger values than ReLU, which in turn would cause more unstable gradients with increased network depth.
Figures~\subref*{fig:variance_vs_depth_width_240_Softplus_below_critical} and \subref*{fig:variance_vs_depth_width_240_Softplus_critical} show the variance as a function of depth becoming concave compared to the convex behavior in Fig.~\subref*{fig:variance_vs_depth_width_240_Softplus_above_critical}.
With $C_W=2.0$, the variance for Softplus is still growing with larger values than the ReLU distribution, but the growth is linear.
When $C_W=1.0$, the variance appears to be approaching a constant value with depth. 
To arrive at a true fixed point, $C_W$ must approach $0$, which is not possible as this choice of initialization would not break the permutation symmetry required to have a trainable neural network~\cite{Roberts:2021fes}.
For an activation function to have a fixed point in their variance they need to meet the conditions: 
\begin{align}
    \sigma(0) = 0 ,\\
    \sigma'(0) \neq0,
\end{align}
and Softplus does not meet the first condition.%
\footnote{Note that Softplus \emph{can} meet both conditions for criticality by adding a constant $C = -\log(2)$ so that $\sigma(0) =0$.
Not all activation functions without a fixed point in the variance can be adjusted in this way, but these criteria provide a way to either adjust an activation function so it can be tuned to criticality, or to disregard an activation function if it cannot be adjusted to meet these conditions.}
Despite being intended as a smooth replacement for the ReLU activation function, Softplus is much more prone to exploding/vanishing gradients and therefore inferior to ReLU~\cite{pmlr-v15-glorot11a,Szandała2021,dubey2022activationfunctionsdeeplearning}.

The next important moment for ANN output distributions is the fourth moment, or kurtosis.
The excess kurtosis (EK), which is the difference of the full kurtosis from the product of covariances (so EK is zero for a Gaussian distribution):
\begin{align}
    EK &= \ev{z_{i_1;\alpha_1}z_{i_2;\alpha_2}z_{i_3;\alpha_3}z_{i_4;\alpha_4}}
    \notag \\
   & \qquad \null - \ev{z_{i_1;\alpha_1}z_{i_2;\alpha_2}}\ev{z_{i_3;\alpha_3}z_{i_4;\alpha_4}} ,
\end{align}
is given by the connected four-point interaction in the ANNFT formulation.%
\footnote{Kurtosis is typically defined as a standardized fourth moment, meaning it is divided by the product of covariances. Our definition is not standardized.}
In the infinite-width limit, the connected contributions from higher-order moments are suppressed and the neural network becomes a Gaussian process with training determined by a linear differential equation. 
These higher-order moments' connected contributions allow for more complex correlations between neurons and network's abilities to represent complex features in data.
As noted in Sec~\ref{sec:ANNFTintro}, tuning to criticality allows for the action of a neural network to be treated through an expansion in the ratio of depth to width $r$, and the connected contributions to the distribution to be treated as perturbations to the infinite-width Gaussian distribution, or free field theory.

The second term in an action for an ANN with mean zero Gaussian initialization distributions is $V^{(\ell)}/n$, 
which is the EK of the distribution without the normalization by the variance squared.
The quantity $V^{(\ell)}$ satisfies a recursion relation
\begin{align}
    V^{(\ell+1)} 
    &= \frac{C_W}{G^{(\ell)}}\ev{z\sigma(z)\sigma'(z)}_{G^{(\ell))}}V^{(\ell)} \notag \\
    & \quad \null + C_W^2[\ev{\sigma^4(z)}_{G^{(\ell)}}
    -\ev{\sigma^2(z)}_{G^{(\ell)}}^2].
    \label{eq:kurtosis_recursion}
\end{align}
If left untreated, $V^{(\ell)}$ will cause the EK to grow and shrink exponentially with depth.
However, with a critically tuned variance $G^{(\ell)}$, the excess kurtosis grows linearly in $\ell$ (and therefore in $r$ as well).

This behavior can be seen for the ReLU activation function in Figs.~\subref*{fig:kurtosis_vs_depth_width_240_ReLU_above_critical}, \subref{fig:kurtosis_vs_depth_width_240_ReLU_below_critical}, and \subref{fig:kurtosis_vs_depth_width_240_ReLU_critical}.
Along with the empirical EK from the generated output distributions, two calculations of the kurtosis were performed using the recursion relation in Eq.~\eqref{eq:kurtosis_recursion}.
The first computed $V^{(1)}$ directly from the distribution using Scipy's stats.kurtosis function to find the excess kurtosis, and then scaling the value appropriately to acquire $V^{(1)}$~\cite{2020SciPy-NMeth}.
From there, the $V^{(\ell)}$s in all subsequent layers were computed recursively using the $G^{(\ell)}$ calculated directly from the output distributions.
The second calculation used initial values of $G^{(0)} = C_W(\frac{1}{n_0}\sum_{i=1}^{n_0}x_i^2) +C_b$ and $V^{(0)}=0$ (the output is Gaussian from Sec~\ref{sec:ANNFTintro}) to compute $V^{(\ell)}$ completely from theoretically predicted values.

Figures~\subref*{fig:kurtosis_vs_depth_width_240_ReLU_above_critical} and \subref{fig:kurtosis_vs_depth_width_240_ReLU_below_critical} show the excess kurtosis growing or shrinking exponentially in agreement with both versions of the recursive calculation.
Figure~\subref*{fig:kurtosis_vs_depth_width_240_ReLU_critical} shows the desired linear behavior from criticality, and closely matches both recursive calculations.
Figure~\subref*{fig:kurtosis_vs_depth_width_240_ReLU_critical} also features an additional line corresponding to an exact calculation of $V^{(\ell)}/n$ to leading-order in $1/n$ calculation from Eq.~\eqref{eq:kurtosis_recursion} for ReLU:
\begin{align}
    V^{(\ell)} 
    = 5(G^{(\ell)})^2\frac{\ell-1}{n} 
    +\mathcal{O}(\frac{1}{n^2}).
    \label{eq:linear_approx}
\end{align}
If this expression is rearranged such that the expansion is leading-order in $r$, and the $1/n$ contribution present becomes a subleading contribution to the expansion, then 
\begin{align}
    V^{(\ell)} = 5(G^{(\ell)})^2r +\mathcal{O}(r^2),\label{eq:linear_kurtosis_ReLU}
\end{align}
and the blue line matches the data and recursive calculations.
The recursive calculations show agreement with the data, with the exception of Fig.~\subref*{fig:kurtosis_vs_depth_width_240_Softplus_below_critical} (for which the excess kurtosis is very small and is subject to large relative fluctuations).

In the case of the Softplus activation, Fig.~\subref*{fig:kurtosis_vs_depth_width_240_Softplus_above_critical} and \subref{fig:kurtosis_vs_depth_width_240_Softplus_critical} demonstrate the excess kurtosis growing exponentially, while \subref*{fig:kurtosis_vs_depth_width_240_Softplus_below_critical} exhibits concave behavior with increasing depth. 
The excess kurtosis asymptotically approaching a constant value for $C_W = 1.0$ rather than being linear means these terms in the action are not being adjusted as depth increased.
The additional correlations that grow linearly with depth in an output distribution for a critically tuned activation function allow for more expressivity, and are controlled by the width of the network.
With a constant EK, the Softplus activation function becomes less expressive and capable of learning features of a dataset for $C_W=1.0$.
Having a variance that remains constant
when tuned to criticality, ReLU's excess kurtosis scales linearly with the ratio of depth to width $r$~\cite{Roberts:2021fes}. 
This roughly linear behavior can be seen in Fig.~\subref*{fig:kurtosis_vs_depth_width_240_ReLU_critical}, and away from criticality, the kurtosis explodes. The kurtosis being linear in $r$ demonstrates the presence of an expansion in $r$, and for small $r$ allows for control of statistical moments order by order in $r$.

\section{Validating ANNFT training behaviors}
\label{sec:binding_energy_results}

The authors of Ref.~\cite{Zeng:2022azv} trained their binding-energy neural networks with GeLU activation functions and the Adam optimizer~\cite{kingma2017adammethodstochasticoptimization}.
However, the critical prescription for training discussed in Sec.~\ref{subsec:training} is applicable in practice only to the stochastic gradient descent (SGD) optimizer, so we focus on that and return at the end to consider adaptive optimizers.
To test the ANNFT training behaviors for the two-input binding-energy ANN from Ref.~\cite{Zeng:2022azv},
we trained networks with ReLU and Tanh activation functions using SGD, as the critical analysis described in Secs.~\ref{subsec:recursion} and \ref{subsec:training} finds them to be the most stable for neural network training~\cite{Roberts:2021fes}.
For each activation function, different combinations of critical and non-critical initialization and training were considered.
In particular, we explored critical-initialization critical-training (CICT), critical-initialization non-critical-training (CINT), and non-critical-initialization non-critical-training (NINT).
All training used a mean absolute error (MAE) loss function to offer direct comparison to Ref.~\cite{Zeng:2022azv}.

\begin{figure*}[ptbh!]
    \centering
    \includegraphics[width=0.75\linewidth]{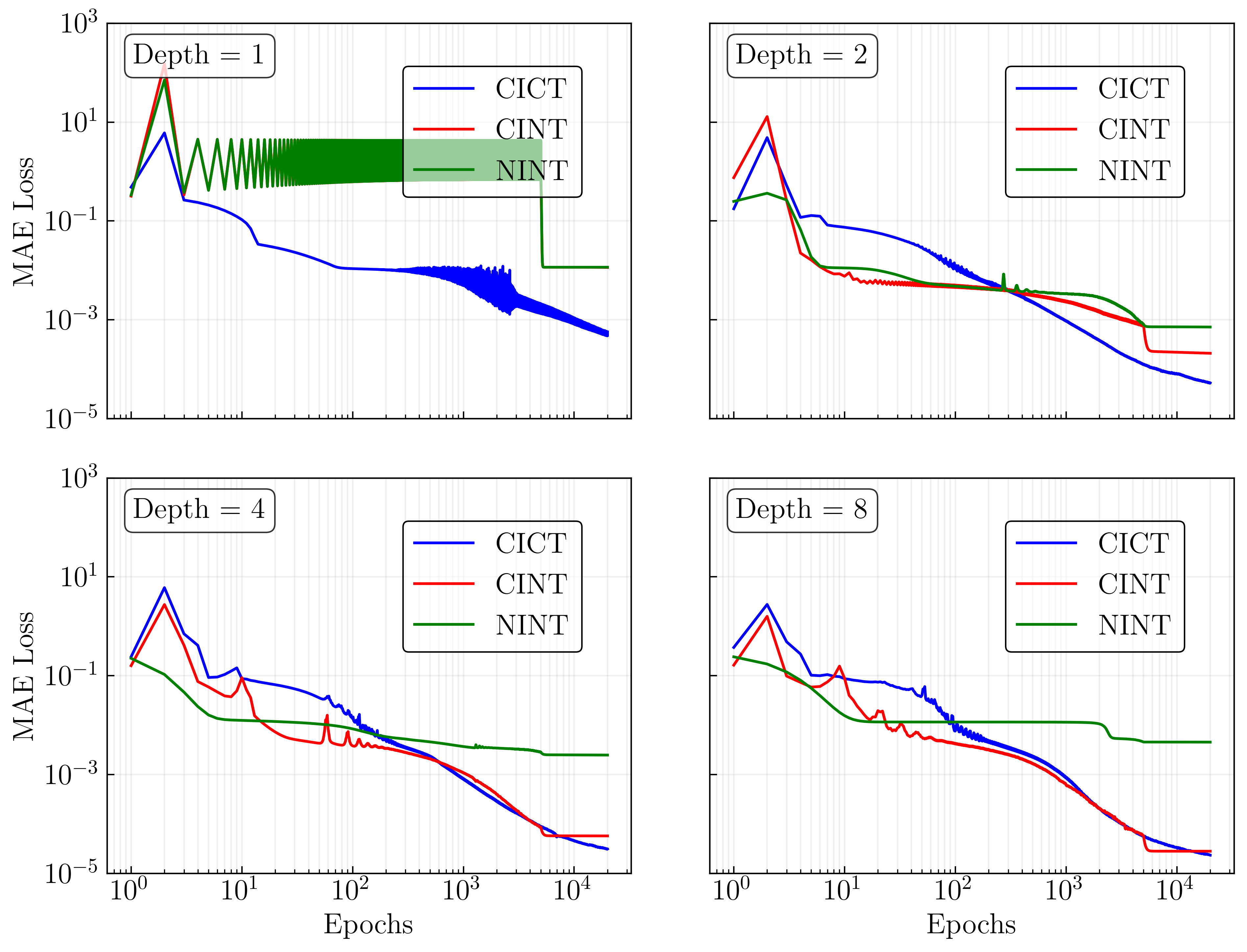}
    \phantomsublabel{-4.75}{2.3}{fig:loss_comparison_ReLU_width_100_a}
    \phantomsublabel{-2.2}{2.3}{fig:loss_comparison_ReLU_width_100_b}
    \phantomsublabel{-4.75}{0.37}{fig:loss_comparison_ReLU_width_100_c}
    \phantomsublabel{-2.2}{0.37}{fig:loss_comparison_ReLU_width_100_d}

    \caption{The mean absolute error loss vs.\ epochs for ReLU activation functions. Hidden layer widths are at 100 neurons, and depths are 1,2,4, and 8 hidden layers. The CICT and CINT architectures outperform the NINT, and their performance improves with depth, whereas the NINT networks stagnate in performance. \label{fig:BE_loss_ReLU}}
    

   
   \bigskip 
   \includegraphics[width=0.75\linewidth]{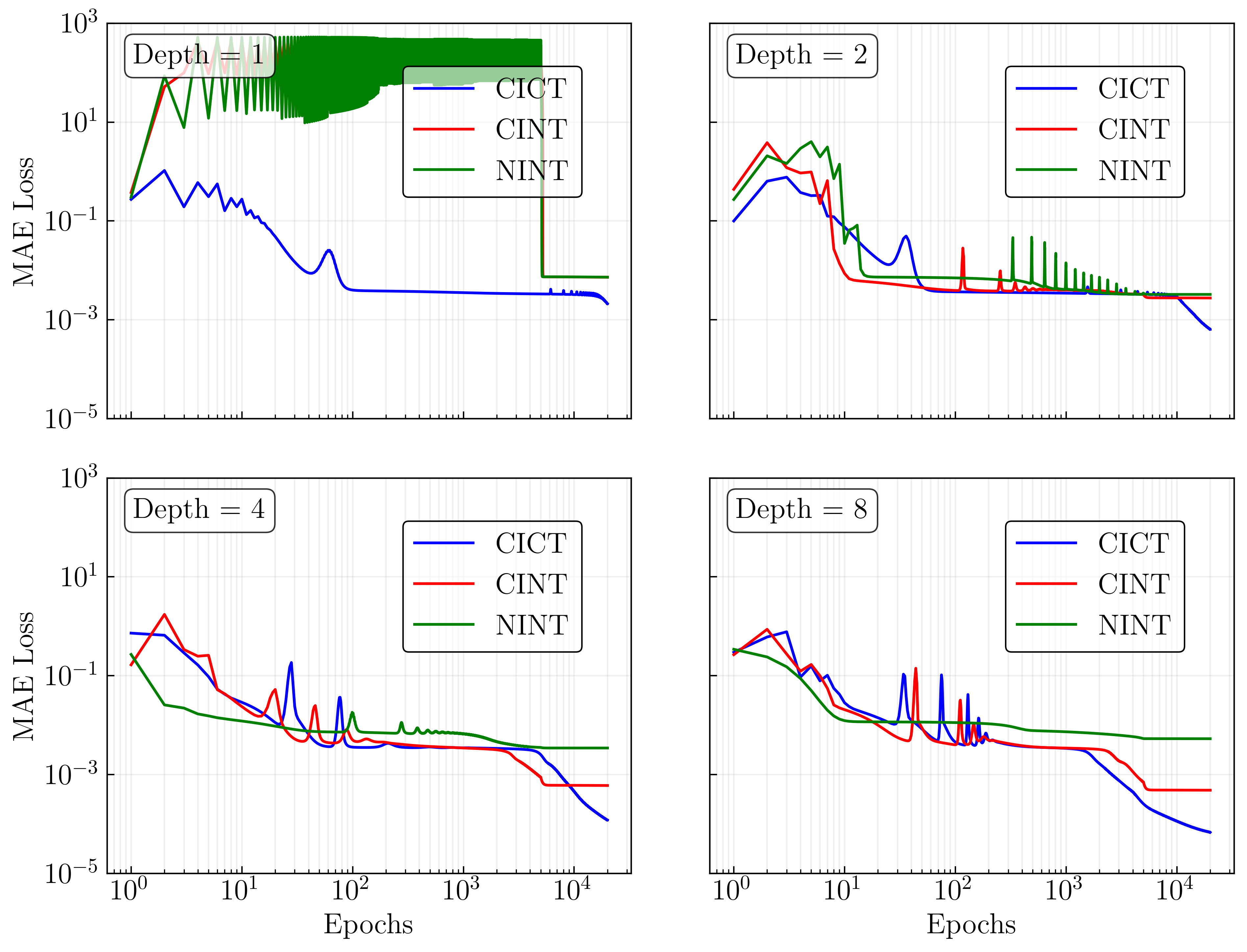}
    \phantomsublabel{-4.75}{2.3}{fig:loss_comparison_Tanh_width_100_a}
    \phantomsublabel{-2.2}{2.3}{fig:loss_comparison_Tanh_width_100_b}
    \phantomsublabel{-4.75}{0.37}{fig:loss_comparison_Tanh_width_100_c}
    \phantomsublabel{-2.2}{0.37}{fig:loss_comparison_Tanh_width_100_d}
    
    \caption{Same as Fig.~\ref{fig:BE_loss_ReLU} but for Tanh activation functions. \label{fig:BE_loss_Tanh}}
    
\end{figure*}

In training their two-input binding-energy networks, Ref.~\cite{Zeng:2022azv} used a learning rate for their Adam optimizer of 0.0001, and decay constants 0.9 and 0.999.
To make a fair comparison, the SGD-optimized networks were trained in two phases.
In the first phase, the unscaled learning rates for the weights $\tilde{\lambda}_W$ and biases $\tilde{\lambda}_b$ are adjusted to be much larger than 0.0001 to allow the network to initially take much larger steps toward a minimum in the parameters.
The second phase switches to a much smaller learning rate value after 5000 epochs.
The second learning rate's smaller size is intended to fine-tune the network parameters and match the $\mathcal{O}(10^{-4})$ value of the adaptive case.
The CICT networks feature the critically scaled learning rates from Eq.~\eqref{eq:critical_lr}.
As discussed in Sec.~\ref{subsec:training}, critical training scales the learning rates so they depend on the depth and the width to avoid exploding or vanishing gradients. 
As such, layers may feature different learning rates depending on the choice of activation function, whereas traditionally a global learning rate is used for training. 
For the SGD trained networks without layer dependence (CINT and NINT), a global learning rate is calculated by averaging over the layer-dependent learning rates of the CICT ReLU or Tanh network with the same depth, width, and activation function to be used in the CINT and NINT case.

Figures~\ref{fig:BE_loss_ReLU} and \ref{fig:BE_loss_Tanh} depict the CICT loss in blue, the CINT loss in red, and the NINT loss in green, all as a function of the number of epochs. 
For ReLU, the fully critical (CICT) network outperforms the half critical (CINT) and non-critical (NINT) network for all depths. 
Networks with criticality had loss values orders of magnitude lower than the fully non-critical network.
In all cases (except depth 1), CICT network outperforms the other two cases. Critical networks (CICT, CINT) improve with depth, while NINT appears to stagnate after two hidden layers.

\begin{figure*}[phtb!]
    \centering
    \includegraphics[width=0.95\linewidth]{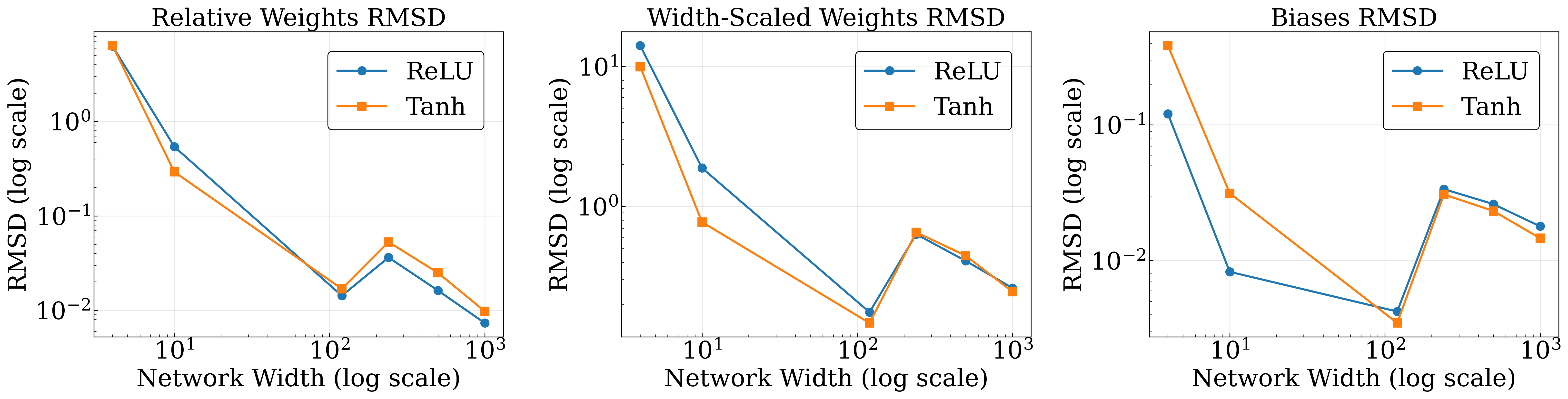}
        \phantomsublabel{-6.25}{0.28}{fig:parameter_change_RMSD_a}
        \phantomsublabel{-4.0}{0.28}{fig:parameter_change_RMSD_b}
        \phantomsublabel{-1.75}{0.28}{fig:parameter_change_RMSD_c}
    \caption{Plots of the mean RMSD between final and initial parameters for the average network hidden layer plotted versus multiple widths (4, 10, 120, 240, 500, 1000). (a) depicts the relative RMSD for the weights (compared to the initial weights), while (b) depicts the absolute RMSD with  the weight matrices rescaled by a factor of $n$ to remove width dependence, and (c) has the unscaled RMSD for the biases (which are initially zero). Within the regimes useful for training, it can be seen that the changes in parameters are small, a condition required for Taylor expanding around a solution.  \label{fig:param_change_RMSD}}
\end{figure*}

\begin{figure*}[phtb!]
    \centering
    \includegraphics[width=0.95\linewidth]{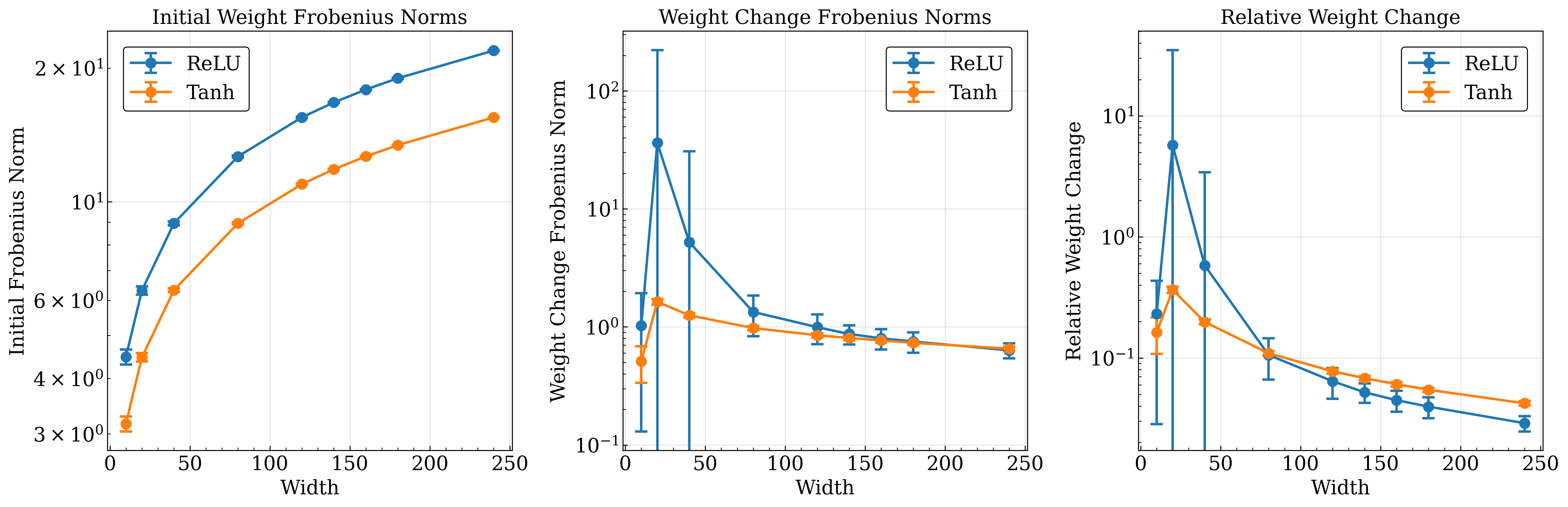}
            \phantomsublabel{-4.7}{0.28}{fig:norms_vs_width_depth4_epochs20000_a}
        \phantomsublabel{-2.51}{0.28}{fig:norms_vs_width_depth4_epochs20000_b}
        \phantomsublabel{-0.31}{0.28}{fig:norms_vs_width_depth4_epochs20000_c}
    \caption{(a) The mean hidden layer Frobenius norms of the pre-taining weight matrices, (b) the difference between pre and post training matrices, and (c) the relative difference matrices versus the width of the network containing the matrices. Each norm per width is computed by averaging the mean norm for 100 network trainings of 20000 epochs, with outliers outside of the interquartile range of the distribution of each norm being excluded. The depth of each network configuration was held fixed at 4 layers ($L=3$,$\ell_{out}=4$). The norms give an estimate of the size of the initial parameters, and their absolute and relative difference after training, and their dependence on the width of the network.  \label{fig:param_change_width}}
\end{figure*}

\begin{figure*}[phtb!]
    \centering
    \includegraphics[width=0.95\linewidth]{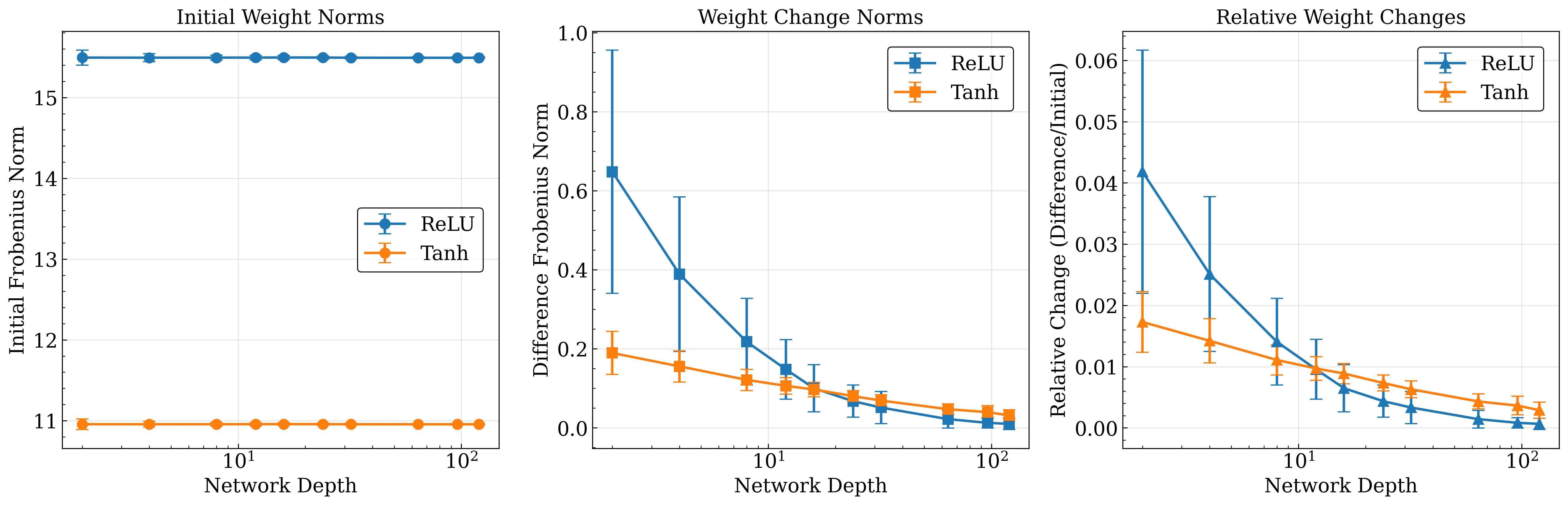}
                \phantomsublabel{-6.3}{0.34}{fig:norms_vs_depth_width120_epochs20000_a}
        \phantomsublabel{-4.1}{0.34}{fig:norms_vs_depth_width120_epochs20000_b}
        \phantomsublabel{-1.82}{0.34}{fig:norms_vs_depth_width120_epochs20000_c}
    \caption{Same as Fig.11, but now with fixed width instead of depth. (a) The mean hidden layer Frobenius norms of the pre-taining weight matrices, (b) the difference between pre and post training matrices, and (c) the relative difference matrices versus the depth of the network containing the matrices. Each norm per depth is computed by averaging the mean norm for 100 network trainings of 20000 epochs, with outliers outside of the interquartile range of the distribution of each norm being excluded. The width was held fixed at 120 neurons per hidden layer. The norms give an estimate of the size of the initial parameters, and their absolute and relative difference after training, and their dependence on the depth of the network..  \label{fig:param_change_depth}}
\end{figure*}

The controlled analysis of ANNs relies on the applicability of the truncated Taylor series of Eq.~\eqref{eq:ANN_taylor_expansion}, which is an expansion about the initialization parameters $\params$.
Therefore, it is important to verify that the difference between the initial and final parameters of the networks is small to ensure that the Taylor series is valid.
For critical ReLU and Tanh networks, we find in Figs.~\ref{fig:param_change_RMSD}, \ref{fig:param_change_width}, and \ref{fig:param_change_depth} that the difference in the initial and final parameters is indeed small in the regimes of interest for the ratio of depth to width $r$ (see Sec.~\ref{subsec:recursion}).

Figure~\ref{fig:param_change_RMSD} depicts the mean of the root-mean-square deviation (RMSD) of the parameter changes in the hidden layers for 100 trainings. 
The networks used had a fixed depth of $\ell_{out}=4$ with widths of $4$, $10$, $120$, $240$, $500$, and $1000$. 
In Fig.~\subref*{fig:parameter_change_RMSD_a} the mean relative (compared to the initial weights) RMSD of the weights declines rapidly with width $n$. 
In Fig.~\subref*{fig:parameter_change_RMSD_b} we show the absolute RMSD with weight matrices scaled by $n$ to normalize the comparison between initial and final weights. 
We find a similar pattern to the mean relative RMSD in the mean absolute RMSD, which follows because the initialization weights are of order unity.
The story for the biases in Fig.~\subref*{fig:parameter_change_RMSD_c} is similar. 
This verifies that as the network gets closer to $r\ll 1$, the change in parameters within the network decreases.

This conclusion is seen again in the mean hidden layer Frobenius norms versus widths in Fig.~\ref{fig:param_change_width}.
As the width increases, so does the Frobenius norm of the hidden layer matrices in Fig.~\subref*{fig:norms_vs_width_depth4_epochs20000_a}, as the matrix is $n\times n$, where $n$ is the width of the network's hidden layers.
The difference between initial and final parameters increases with width initially in Fig.~\subref*{fig:norms_vs_width_depth4_epochs20000_b}, but decreases for width values greater than 48 neurons.
The relative norm in Fig.~\subref*{fig:norms_vs_width_depth4_epochs20000_c} is very small in the region of interest, and its decrease with width shows the desired small parameter changes for fixed-depth networks.

Figure~\ref{fig:param_change_depth} shows the Frobenius norms for networks with a fixed width of $n=120$ and increasing depth.
At fixed width the Frobenius norms remains constant as depth is increased, as seen in Fig.~\subref*{fig:norms_vs_depth_width120_epochs20000_a}, as the increased depth will only include new matrices to be averaged over rather than increasing the size of the weight matrices like the fixed-depth case.
The absolute changes in difference norms are shown in Fig.~\subref*{fig:norms_vs_depth_width120_epochs20000_b} and the relative changes in Fig.~\subref*{fig:norms_vs_depth_width120_epochs20000_c}.
At any depth with this large width (up to $r = 1$), the relative changes are small, much smaller than one.
In summary, we find that networks with small $r$ values warrant the use of Taylor expansions of the network solutions around their initial parameters $\params$.

\begin{figure}[htb!]
    \centering
    \includegraphics[width=0.95\columnwidth]{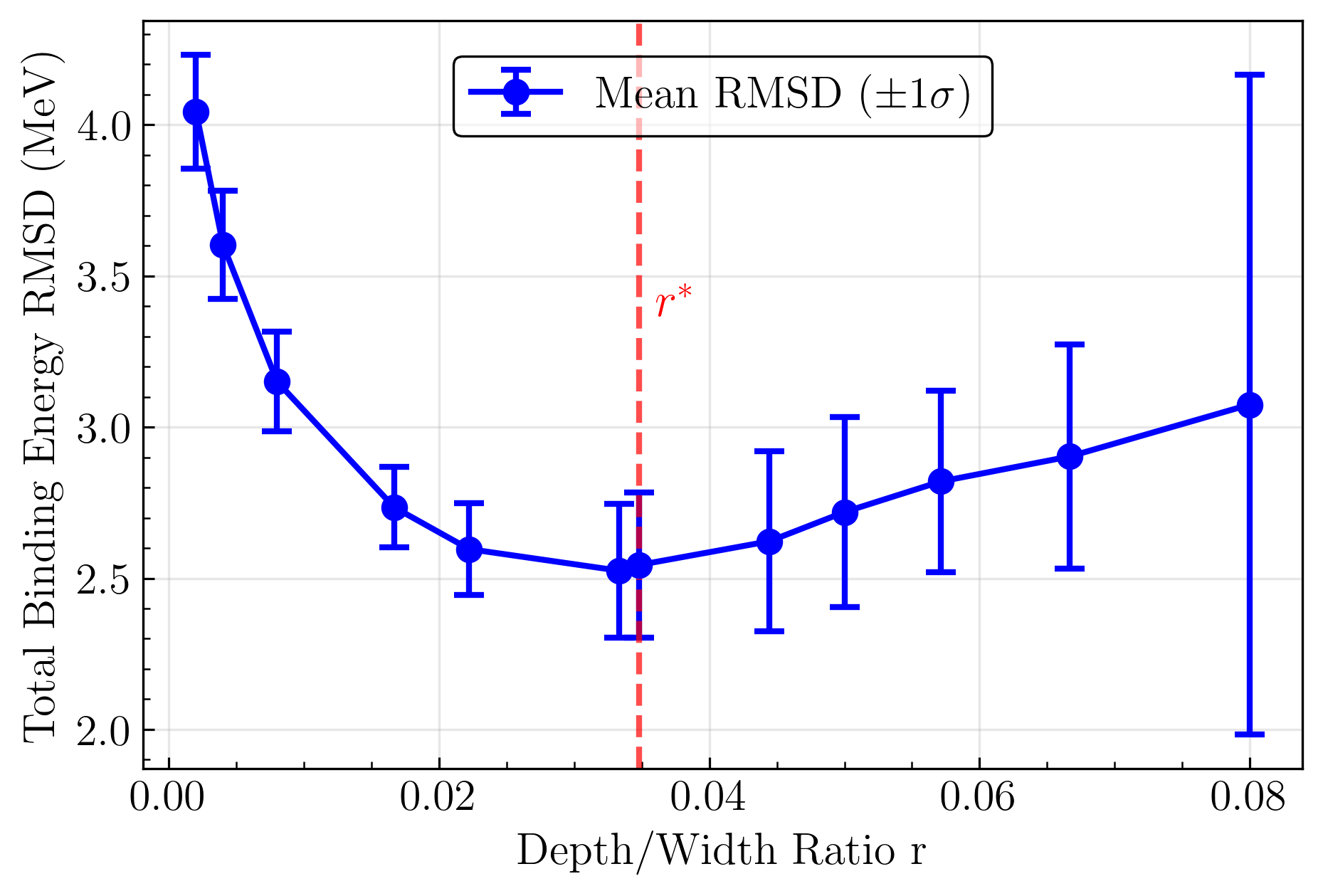}
    \caption{The mean root-mean-square (rms) deviation of the binding energy data and the two-input-network outputs  for $100$ trainings vs.\ ratio of depth-to-width $r$, with the depth $\ell_{out} = 4$. Trainings with binding-energy RMSD$\geq30$ MeV were omitted from calculation of the mean value of the binding-energy RMSD. The red horizontal line is the $r^*$ value for this network, calculated from Eq.~\eqref{eq:rstar} to be $r^*=0.034$. This labels the cutoff where the effectively deep regime begins to transition into the chaotic regime with increasing $r$.  
    \label{fig:RMSD_r_regimes}}
    
\end{figure}

\begin{figure*}[htb!]
    \centering
    \includegraphics[width=0.65\linewidth]{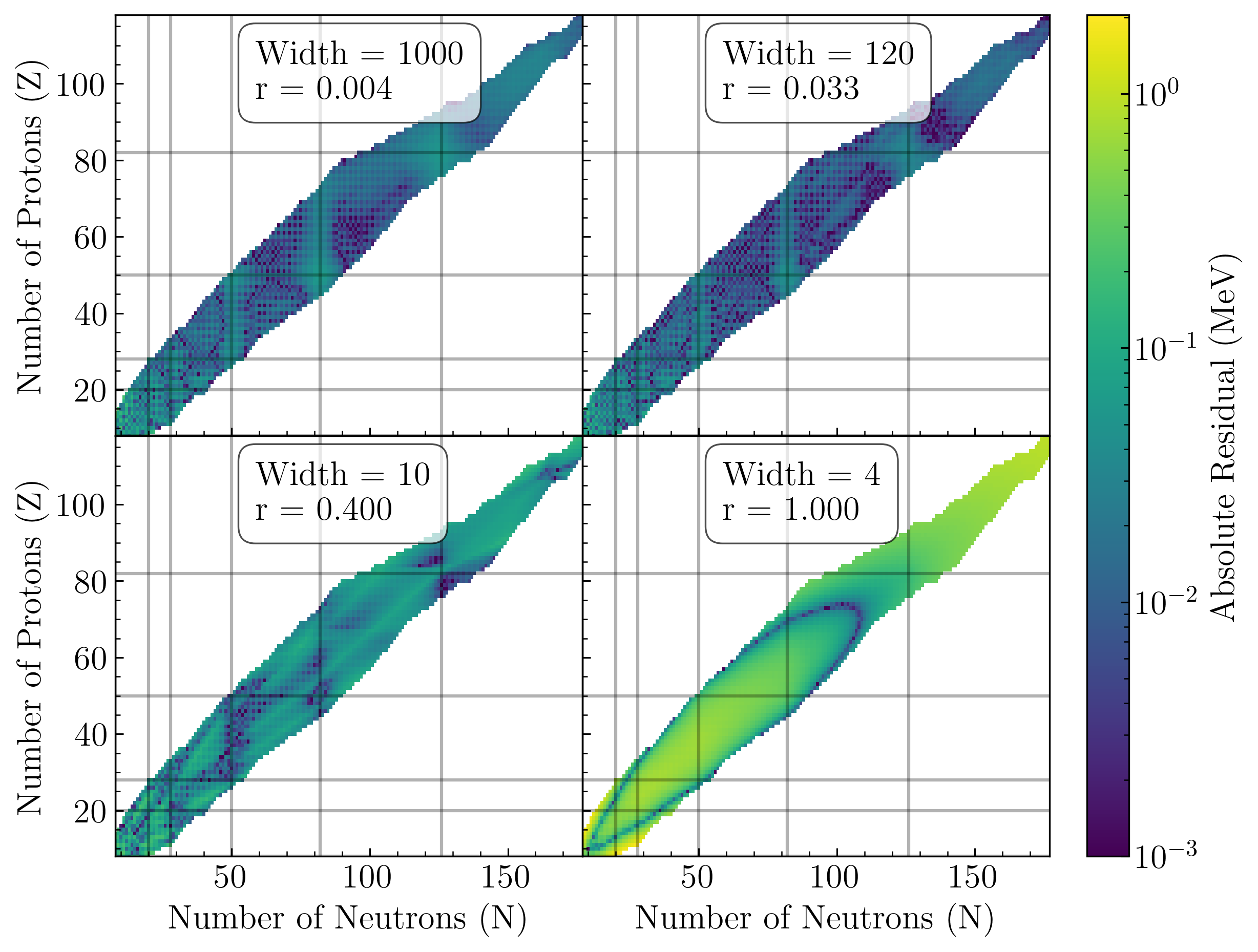}
        \phantomsublabel{-2.7}{2.2}{fig:CICT_ReLU_depth_4_residuals_grid_a}
        \phantomsublabel{-1.0}{2.2}{fig:CICT_ReLU_depth_4_residuals_grid_b}
        \phantomsublabel{-2.7}{0.65}{fig:CICT_ReLU_depth_4_residuals_grid_c}
        \phantomsublabel{-1.0}{0.65}{fig:CICT_ReLU_depth_4_residuals_grid_d}
    \caption{Residual plots for a trained 2 input binding energy network with a fixed depth of 4, and widths of (a) 4, (b) 10, (c) 120, and (d) 1000 to demonstrate the different learning regimes in ANNFT. \label{fig:Residuals_r_regimes}}
    
\end{figure*}

Different ANN learning regimes associated with the value $r$ were identified in Sec.~\ref{subsec:recursion} (see \cite{Roberts:2021fes}).
When $r\geq1$, correlations between neurons are too strong to learn features, and the networks can be considered ``overly deep''. 
When $r \ll 1$, the networks are ``effectively deep'', and can learn. When $r\rightarrow0$, the neuron correlations turn off, and features cannot be learned. 
This implies an optimal $r$ (which will be denoted $r^*$ below).
The behavior with $r$ is illustrated in Figs.~\ref{fig:RMSD_r_regimes} and~\ref{fig:Residuals_r_regimes}.
    
The binding-energy RMSD decreases with $r$ in Fig.~\ref{fig:RMSD_r_regimes} until it reaches a minimum at $r=0.033$. 
The distribution of RMSD values is roughly a normal distribution up to this point, with widths indicated by the error bars.
The decreasing values of the binding-energy RMSD demonstrate the improvement in feature learning as the network moves from the Gaussian limit ($r\rightarrow 0$) towards the effectively deep region. 
Past the mininum value, the distribution of binding-energy RMSD values becomes bimodal, with a second mode centered around about 72\,MeV (here the error bars indicate the standard deviation, but a normal distribution is not implied).
This other mode is associated with networks that fail to train, corresponding to the transition from effectively deep to overly deep (chaotic), where the network becomes incapable of learning any features.
For $r\geq1$, the distributions become entirely centered around 72\,MeV.

The role of $r$ in determining learning capability is visualized in Fig.~\ref{fig:Residuals_r_regimes}, which shows binding-energy residuals across the nuclear landscape as heat maps for a wide range of $r$ values. 
In Fig.~\subref*{fig:CICT_ReLU_depth_4_residuals_grid_a} with $r = 0.004$ the residuals are generally small, but an increase to $r=0.033$ in Fig.~\subref*{fig:CICT_ReLU_depth_4_residuals_grid_b} shows improved learning. A further increase to $r=0.4$ in Fig.~\subref*{fig:CICT_ReLU_depth_4_residuals_grid_c} shows significant degradation and $r=1$ in Fig.~\subref*{fig:CICT_ReLU_depth_4_residuals_grid_c} shows poor learning.

Given the minimum binding-energy RMSD seen in Fig.~\ref{fig:RMSD_r_regimes}, the existence of an optimal value $r$ for a neural network is implied. 
Roberts and Yaida discuss methods utilizing information theory for calculating $r^*$, the optimal aspect ratio, in appendix A of Ref.~\cite{Roberts:2021fes}.
$r^*$ defines the boundary where the effectively deep regime ends, and the chaotic regime begins, and they calculate it (to $\mathcal{O}(r^3)$) to be 
\begin{align}
    r^* = \biggl(\frac{4}{20+3n_{\ell_{out}}}\biggr)\frac{1}{\nu}.
    \label{eq:rstar}
\end{align}
Here $n_{\ell_{out}}$ is the size of the output layer ($n_{\ell_{out}}=1$ for the binding-energy network) and $\nu$ is equal to the slope factor of the critical linear EK; 
$\nu=5$ for the ReLU network EK at criticality and $r^*=0.034$.
For our $\ell_{out} = 4$ case, this suggests the optimal width of $n=115$.
This is consistent with Fig.~\ref{fig:RMSD_r_regimes}, which has a minimum binding-energy RMSD when $r=0.033$, and growing values of the binding-energy RMSD past the $r^*$ cutoff. 
As noted above, the distributions of outputs for runs with $r>r^*$ feature more non-converged runs.
This suggests that while architectures with $r^*<r<1$ can result in successfully trained networks, they have a greater chance of failing as $r$ increases.
Table~\ref{tab:rmsd_analysis} displays the width $n$, ratio of depth to width $r$, the mean and std of the binding-energy RMSD, and the number of runs used in calculating the statistics for the RMSD. 
Runs where the RMSD was greater than $100$ MeV, or produced a NaN value are excluded from calculation of the statistics, and 100 runs were used at most for each $r$ value.
The mean and std have an asterisk for $r$ values where $r^*<r<1$, where the distribution of RMSD values are split between the effectively deep and chaotic regimes as noted before.
Indeed, one can see the mean and std values transition between $\mathcal{O}(1)$ means and $\mathcal{O}(10^{-1})$ std to much larger means and variances before settling entirely at around $72$ MeV when $r=4$.

\begin{table*}[htb!]
\centering
\caption{RMSD mean and standard deviation (std) $\sigma$, and the number of runs used for the distributions, for a large range of network widths and ratio of depth to width ($r$). For this table, the RMSD cutoff whenever averaging over runs was increased to 100 MeV. Values of the mean and std where $r^* < r < 1$ are marked with an asterisk; as $r$ increases above $r^*$ the distribution of RMSDs becomes increasingly bimodal.}
\label{tab:rmsd_analysis}
\begin{tabular}{ccccc}
\hline
Width & Ratio ($r = \ell_{out}/n$) & Mean RMSD (MeV) & $\sigma$ RMSD (MeV) & Number of Runs \\
\hline
1 & 4.0000 & 71.9549 & 0.0000 & 100 \\
4 & 1.0000 & 68.9716 & 11.2120 & 100 \\
10 & 0.4000 & 55.9824* & 25.3620* & 99 \\
20 & 0.2000 & 35.2627* & 31.6383* & 99 \\
30 & 0.1333 & 14.9702* & 24.1069* & 98 \\
40 & 0.1000 & 6.0500* & 12.0067* & 99 \\
50 & 0.0800 & 3.0745* & 1.0921* & 100 \\
60 & 0.0667 & 2.9027* & 0.3699* & 100 \\
70 & 0.0571 & 2.8203* & 0.3005* & 100 \\
80 & 0.0500 & 2.7189* & 0.3137* & 100 \\
90 & 0.0444 & 2.6223* & 0.2980* & 100 \\
115 & 0.0348 & 2.5392 & 0.2326 & 100 \\
120 & 0.0333 & 2.5250 & 0.2221 & 100 \\
180 & 0.0222 & 2.5964 & 0.1523 & 100 \\
240 & 0.0167 & 2.7347 & 0.1328 & 100 \\
500 & 0.0080 & 3.1506 & 0.1641 & 100 \\
1000 & 0.0040 & 3.6029 & 0.1780 & 100 \\
2000 & 0.0020 & 4.0420 & 0.1850 & 84 \\
\hline
\end{tabular}
\end{table*}

To summarize, we find activation functions should be chosen such that critical points exist. 
Initialization widths for weights and biases should be set to criticality to avoid gradient problems during training.
Learning rates must be critically scaled to improve performance and offer further gradient stability.
The ratio of depth to width $r$ should be small ($0 <r <r^*\ll 1$) to optimally learn the features of the system.
Small $r$ also means ANNs can be treated as field theories with running couplings that run with depth, and can be tuned to criticality. 
$r^*$ provides a more specific constraint for the boundary between the overly deep $r\geq1$ regime, and the effectively deep regime beyond the condition that $r\ll 1$~\cite{Roberts:2021fes}.

We now return to the use of adaptive optimization as opposed to SGD.
To make a comparison, the SGD optimization in the ReLU and Tanh networks is replaced by an Adam optimizer using the same training hyperparameters as Ref.~\cite{Zeng:2022azv}, with the exception of the learning rate.
The learning rate was calculated instead using the same averaging over the layer-dependent critical learning rates used for the non-critical training systems (CINT and NINT).
At present, there is no critical training prescription for the Adam optimizer, and so the only cases tested and compared with adaptive Adam optimizers were CINT and NINT.

The bottom line of the comparison is that
when the SGD optimizer was replaced by Adam, both CINT and NINT networks no longer show criticality, but obtain similar binding-energy RMSD values to the original Zeng network of about 1.2 MeV~\cite{Zeng:2022azv}. 
Thus the adaptive algorithm apparently compensates for non-critical initialization and training.
The (non-adaptive) CICT network gets a best binding-energy RMSD of about $1.9$ MeV, which is only a small factor worse than the result from the Adam optimizer.  
Nevertheless, the improved adaptive optimizers offer improved training performance.
However, this is at the cost of obscuring the analysis of the neural network.
A critical training prescription consistent with adaptive algorithms would be most welcome.

Additional trials were performed for a four-input network, which included the inputs $Z_0$ and $N_0$ to account for pairing effects, following Ref.~\cite{Zeng:2022azv}.
$Z_0$($N_0$) is equal to $0$ when $Z$($N$) is even and $1$ when $Z$($N$) is odd, and as such these inputs were unscaled by the MinMax scaler.
The four-input network exhibited the same initialization and training behavior presented in Secs.~\ref{sec:ANNFTvalidation} and ~\ref{sec:binding_energy_results}.
In particular, we find that an increase in input size still possesses the essential properties of ANNFT, and allows for additional features to be added to networks if desired.

Unlike the two-input case, however, the total-dataset binding-energy RMSD values of the four-input network using critical prescriptions were not able to match the RMSD value of Ref.~\cite{Zeng:2022azv}'s optimized four-input network.
This will require further investigation.
We note that even with a four-input network using the exact architecture and training hyperparameters in Ref.~\cite{Zeng:2022azv} and trained with an adaptive optimizer (Adam) on the same dataset, we were unable to reproduce the same total-dataset binding-energy RMSD values found.

\section{Conclusion and Outlook}
\label{sec:conclusion}

Neural network criticality offers a means to understand neural network behavior, as well as providing prescriptions for architecture and training hyperparameters. 
In this work, we have explored the non-empirical approach to understanding ANNs using a field-theory-based criticality analysis (ANNFT)~\cite{Roberts:2021fes}. 
We used a protoypical nuclear physics ANN as a testbed, namely a two-input/one-output feed-forward network trained from the AME2020 dataset to fit nuclear binding energies~\cite{Zeng:2022azv}.

We first confronted ANNFT predictions of initialization distribution behavior in Sec.~\ref{sec:ANNFTvalidation}.
This is pre-training; we simply collect statistics from repeated random initializations of the weights and biases.
We found the expected approach to Gaussian distributions for the final layer output distributions with fixed depth but increasing width.
Within small fluctuations, normal distributions were seen by a width of 120 neurons for both ReLU and Softplus activations (Fig.~\ref{fig:dist_10_ReLU}) and for every other activation function we have tried.
The correlations between pairs of input values with a width of 120 and increasing depths showed Gaussian process (GP) correlations with the predicted kernel at depth 1 and increasing correlations with additional hidden layers (Fig.~\ref{fig:corner_plot_ReLU}).

A comparison of measured variance and excess kurtosis (difference of the fourth moment from Gaussian expectations) from many initializations for a fixed width of 240 and increasing depths validates the theoretical predictions of criticality (Figs.~\ref{fig:variance_above_ReLU} and \ref{fig:kurtosis_above_ReLU}).
In particular, for the ReLU activation function the variance grows or shrinks with depth if the initialization hyperparameters are tuned above or below the predicted critical value.
At criticality the variance is flat.
In contrast, the Softplus activation function does not have a critical fixed point for initialization.
In accordance with theory, the variance either grows or asymptotes to a constant value with depth.
Non-zero excess kurtosis (EK) indicates non-Gaussian four-point correlations.
For both activation functions, the observed behavior of the EK with depth is in good accord with theoretical expectations.

Next we considered training of the ANN in Sec.~\ref{sec:binding_energy_results}.
We followed the mean absolute error (MAE) loss for a large number of epochs with several different depths, using combinations of critical and non-critical initialization and training.
The networks with critical initialization and critical training (CICT) reached the lowest loss in all cases both for ReLU and Tanh activation functions, with particular contrast to when non-critical hyperparameters were used for both initialization and training (Figs.~\ref{fig:BE_loss_ReLU} and \ref{fig:BE_loss_Tanh}).

Most practitioners do not encounter these issues even though they rarely pay attention to details of initialization and training.
This is because the critically tuned networks are still outperformed by adaptive optimizers.
Nevertheless, the results are comparable, and differ by a much smaller factor than when untuned.

Of particular interest is the prediction by ANNFT of different training regimes, characterized by the ratio of depth to width $r$. 
The predictive analyses of ANNFT are based on the validity of a Taylor series expansion about the initialization.
By considering the mean hidden-layer rms deviation and Frobenius norms of the weight matrices
before and after training, we can see that the conditions of small deviations from initialization are achieved with sufficiently small $r$ (Figs.~\ref{fig:param_change_RMSD}, \ref{fig:param_change_width}, and \ref{fig:param_change_depth}).
The different learning regimes are manifested by looking at the rms deviation of the binding energy data and ANN outputs as a function of $r$ (Fig.~\ref{fig:RMSD_r_regimes}), achieved by varying the width at fixed depth, and through heatmaps of global residuals across all $N$ and $Z$ for four different $r$ values (Fig.~\ref{fig:Residuals_r_regimes}).

In future work, we will
continue to investigate criticality in binding-energy fitting with more features/inputs and work toward extending criticality to other training algorithms/hyperparameters.
At the same time, we will explore more nuclear systems with critical networks to further understand learning behavior in different problems.
The field-theory-based framework offers the possibility of applying approximations and techniques from many-body field theory to extend ANNFT beyond simple perturbation theory.
Finally, we will consider an alternative formulation of ANNFT that organizes an expansion with non-Gaussianity not around $1/n$ corrections to the infinite-width limit but through parametric violations of the necessary conditions for the CLT (e.g., independence).


\vspace*{.5in}

\begin{acknowledgments}
We thank Pranav Sharma for useful feedback on the manuscript. We also thank Dan Roberts and Pranav Sharma for useful discussions.
The work of SAS and RJF was supported in part by the 
STREAMLINE collaboration under award DE-SC0024509 and the
National Science Foundation Award No.\ PHY-2209442.
RJF also acknowledges support from the ExtreMe Matter Institute EMMI
at the GSI Helmholtzzentrum für Schwerionenforschung GmbH, Darmstadt, Germany.
\end{acknowledgments}

\bibliography{bayesian_refs}

\end{document}